\documentclass[journal,final,twocolumn,10pt,twoside]{IEEEtran}

\normalsize

\ifCLASSINFOpdf

\else

\fi

\usepackage[usenames,dvipsnames]{xcolor}
\usepackage[cmex10]{amsmath}
\usepackage{amsthm}
\usepackage{amssymb}
\usepackage{enumitem}

\newtheorem{lemma}{Lemma}

\newtheorem{remark}{Remark}

\newtheorem{auxiliary code}{Auxiliary Code}

\newtheorem{definition}{Definition}

\usepackage{algorithm}
\usepackage[noend]{algpseudocode}

\algrenewcommand\algorithmicindent{1em}%

\usepackage{tikz}
\usepackage{graphicx}
\usetikzlibrary{positioning}
\usepackage[font={small}]{caption}
\usepackage{subcaption}
\usepackage{multirow}
\usepackage{setspace}%
\usepackage{optidef}
\usepackage[scientific-notation=true]{siunitx}
\usepackage{amsbsy}

\def \tG {\mathcal{G}}
\def \tS {\mathcal{S}}
\def \td {\zeta}
\def \tH {\mathcal{H}}

\def \ttc {\Lambda}
\def \ttcsmall {\lambda}
\def \tcn {c}
\def \ttg {g_c}
\def \tGEC {\mathcal{\widetilde{G}}}
\def \teta {\kappa}
\def \tK {\mathcal{K}}
\def \talpha {\alpha_{\ttg}}
\def \tV {\mathcal{V}}
\def \tr {R}
\def \tL {\Pi}
\def \tx {\mathrm{x}}
\def \tnu {\theta}

\def \tSall {\psi}
\def \tA {\mathrm{A}}
\def \tC {\mathrm{C}}
\def \tO {\mathcal{O}}

\def \tCMT {\mathcal{T}}

\def \piA {\Delta}
\def \tsj {\Psi}

\def \tpfsa {P^{S,A}_f}
\def \tlambda {\rho}
\def \tm {M}

\def \tzero {0}

\def \trnj {\tr n_j}
\def \tnj {n_j}
\def \tleq {<}

\usepackage{multicol}
\usepackage[T1]{fontenc}
\newcommand\lev[1]{{\color{black}#1}}
\newcommand\deb[1]{{\color{black}#1}}

\newcommand\bluetext[1]{{\color{black}#1}}

\newcommand\editor[1]{{\color{black}#1}}
\newcommand\revone[1]{{\color{black}#1}}
\newcommand\revtwo[1]{{\color{black}#1}}
\newcommand\revthree[1]{{\color{black}#1}}
\newcommand\trim[1]{{\color{black}#1}}

\newcommand\editortwo[1]{{\color{black}#1}}
\newcommand\revonetwo[1]{{\color{black}#1}}
\newcommand\revfourtwo[1]{{\color{black}#1}}

\newcommand\revonethree[1]{{\color{black}#1}}

\usepackage{anyfontsize}

\begin{document}

\title{Overcoming Data Availability Attacks in Blockchain Systems: Short Code-Length LDPC Code Design for Coded Merkle Tree\vspace{-0.1cm}}

\author{Debarnab~Mitra,~\IEEEmembership{Student Member,~IEEE,}
Lev~Tauz,~\IEEEmembership{Student Member,~IEEE}
        and~Lara~Dolecek,~\IEEEmembership{Senior Member,~IEEE\vspace{-0.6cm}}%
\thanks{D.~Mitra, L.~Tauz, and L.~Dolecek  are with the ECE Department, UCLA, Los Angeles, CA 90095 USA (e-mail: debarnabucla@ucla.edu, levtauz@ucla.edu, dolecek@ee.ucla.edu). A part of this paper was presented at the IEEE Information Theory Workshop 2020 \cite{SSskewITW}. Research supported in part by the Guru Krupa Foundation and NSF-BSF grant no. 2008728.}%
}

\markboth{IEEE Transactions on Communications}%
{Submitted paper}

\maketitle

\begin{abstract}
\trim{
Light nodes in blockchains improve the scalability of the system by storing a small portion of the blockchain ledger.
In certain blockchains, light nodes are vulnerable to a \emph{data availability} (DA) attack where a malicious node makes the light nodes accept an invalid block by hiding the invalid portion of the block from the nodes in the system.}
 Recently, a technique based on LDPC codes called Coded Merkle Tree (CMT) was proposed by Yu \emph{et al.} that enables light nodes to detect a DA attack by randomly requesting/sampling portions of the block from the malicious node. However, light nodes fail to detect a DA attack with high probability if a malicious node hides a small stopping set of the LDPC code.
 To mitigate this problem, Yu \emph{et al.} used random LDPC codes that achieve large minimum stopping set size with high probability.
 Although effective, these codes are not necessarily optimal for this application, \editor{especially at short code lengths, which are relevant for low latency systems, IoT blockchains, etc.}. In this paper, \editor{we focus on short code lengths} and demonstrate that a suitable co-design of specialized LDPC codes and the light node sampling strategy can improve the probability of detection of DA attacks.
We consider different adversary models based on their computational capabilities of finding stopping sets in LDPC codes. 
For a weak adversary model, we devise a new LDPC code construction termed as the \emph{entropy-constrained} PEG (EC-PEG) algorithm which \emph{concentrates} stopping sets to a small group of variable nodes. 
We demonstrate that the EC-PEG algorithm coupled with a greedy sampling strategy improves the probability of detection of DA attacks. For stronger adversary models, we provide a co-design of a sampling strategy called \emph{linear-programming-sampling} (LP-sampling) and an LDPC code construction called \emph{linear-programming-constrained} PEG (LC-PEG) algorithm.
The new co-design demonstrates a higher probability of detection of  DA attacks compared to \trim{approaches in earlier literature.
} 
\end{abstract}


\IEEEpeerreviewmaketitle

\vspace{-0.5cm}
\section{Introduction}\label{sec:intro}
\vspace{-0.2cm}
Blockchains are tamper-proof ledgers of transaction data maintained by a network of nodes in a decentralized manner. They were initially proposed in the field of cryptocurrencies like Bitcoin  and Ethereum.
However, the decentralized nature of blockchains has lead to their application in  fields such as
supply chains \cite{supplychain},
Internet of Things \cite{IoTsurvey}, and healthcare \cite{healthcare}.

\trim{A blockchain is a collection of transaction blocks arranged in the form of a hash-chain. Full nodes in the blockchain network store the entire blockchain ledger and operate on it to validate transactions.
However, storing the entire ledger requires a significant storage overhead\footnote{At the time of writing, the size of the Bitcoin and Ethereum ledgers are around 400GB \cite{bitcoinsize} and 650GB \cite{ethereumsize}, respectively.
} which prevents resource limited nodes from joining the blockchain system.
To alleviate this problem, some blockchain systems also run light nodes \cite{Bitcoin}. These are nodes that only store the headers corresponding to each block of the blockchain.
The header for each block contains a field called a Merkle root which is constructed from the block transactions \cite{Bitcoin}.
Using the Merkle root, light nodes can verify the
inclusion of a given transaction in a block via a technique called a Merkle proof. However, 
they cannot verify the correctness of the transactions in the block.}

\begin{figure*}[t]
    \centering
    \begin{subfigure}{0.5\linewidth}
\begin{minipage}{0.99\linewidth}
\begin{tikzpicture}
  \node (img)
  {\includegraphics[scale=0.49]{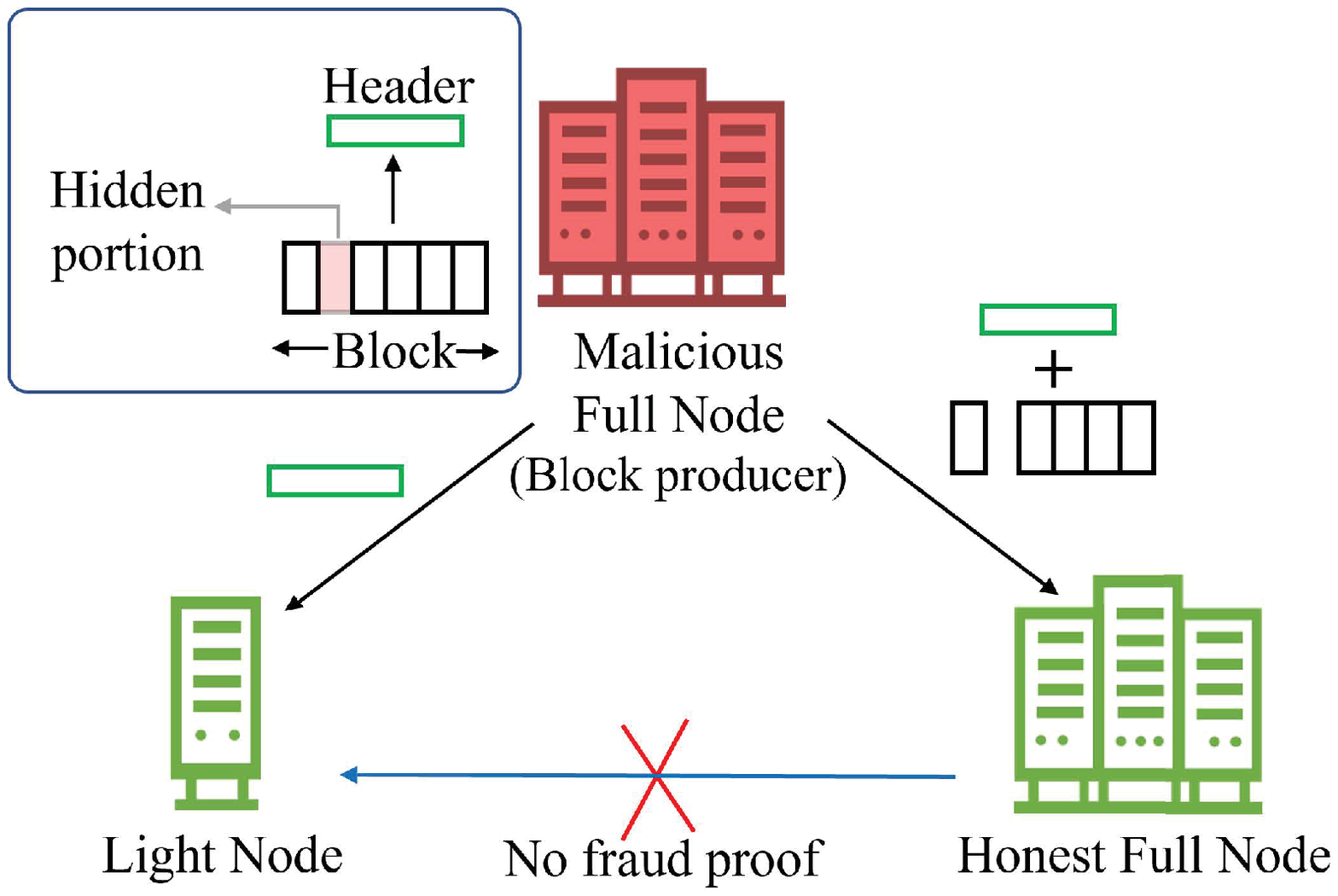}};
 \end{tikzpicture}
 \end{minipage}
    \end{subfigure}%
\begin{subfigure}{0.5\linewidth}
\begin{minipage}{0.99\linewidth}
\begin{tikzpicture}
  \node (img)
  {\includegraphics[scale=0.49]{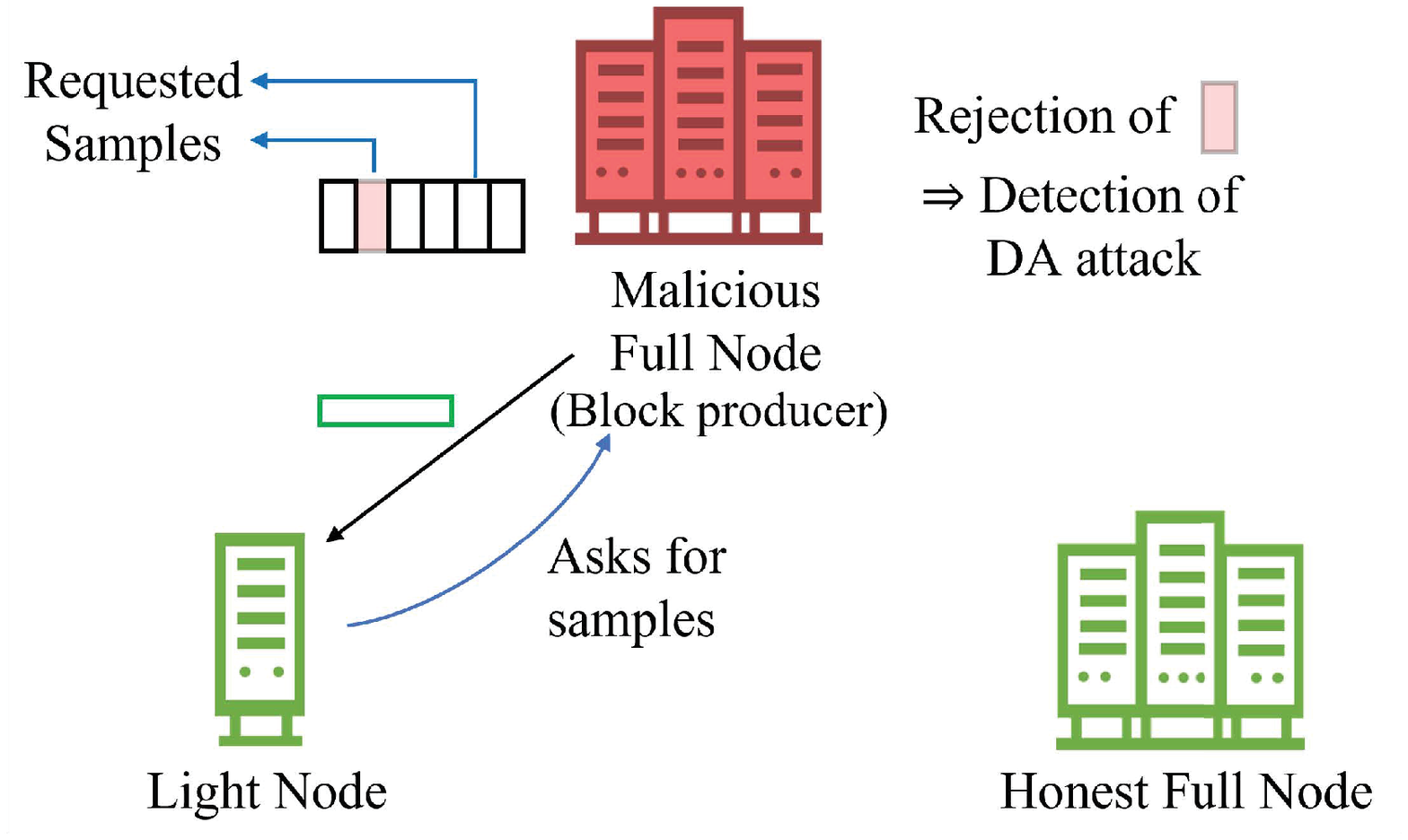}};
 \end{tikzpicture}
 \end{minipage}
\end{subfigure}
     \vspace{-10pt}
    \caption{Left: Data Availability (DA) attack; Right: Detection of DA attack via light node sampling}
    \label{fig:da_attack}
    \vspace{-10pt}
\end{figure*}

\trim{
Assuming that the system has a majority of honest full nodes, light nodes simply accept headers that are a part of the longest header chain because honest full nodes will not mine blocks on chains containing fraudulent transactions (i.e., a longest chain  consensus protocol \cite{Bitcoin} is used).
However, when the honest majority assumption is removed, the longest chain protocol becomes insecure for light nodes. 
As such, researchers were prompted to find methods to provide security even under a dishonest majority of full nodes.  One such research endeavor was \cite{dataAvailOrg} where authors provided protocols for honest full nodes to broadcast verifiable fraud proofs of invalid transactions.}  The mechanism allows light nodes, even in the presence of a majority of malicious full nodes, to reject headers of invalid blocks on receiving fraud proofs from an honest full node. \deb{However, with a majority of malicious full nodes,} the light nodes are still susceptible to data availability (DA) attacks \revone{\cite{dataAvailOrg,CMT}}. In this attack, as illustrated in Fig. \ref{fig:da_attack} left panel, a malicious full node generates
a block with invalid transactions, publishes the header of the invalid block to the light nodes, and hides the invalid portion of the block from the full nodes. Honest full nodes cannot validate the missing portion of the block and hence are unable to generate fraud proofs to be sent to the light nodes.  Since the absence of a fraud proof also corresponds to the situation that 
the block is valid, light nodes accept the invalid header\footnote{
\trim{In this system, there is no way of identifying honest alarm messages sent by full nodes about block unavailability \cite{CMT}, \cite{DAnote}.}
}.

Light nodes can independently detect a DA attack if a request for a portion of the block is rejected by the full node that generates the block. As such, as illustrated in Fig. \ref{fig:da_attack} right panel, light nodes randomly sample the block, i.e., randomly request for different portions of the block
transactions and accept the header if all the requested portions are returned.
In this paper, we are interested in reducing the probability of failure for a light node to detect a DA attack for a given sample size, thus improving the security of the system.
\editor{Since the size of individual transactions is much smaller compared to the entire block, an adversary can hide a very small portion of the block corresponding to the invalid transactions. Such a hiding will result in a high probability of failure for the light nodes using random sampling.
To alleviate this problem, authors in \cite{dataAvailOrg} proposed coding the block using erasure codes\footnote{\editortwo{As with all applications of channel coding, coded redundancy results in a rate penalty, which in this case is a storage overhead at the full nodes. In this work, we improve the trade-off between the storage overhead and the probability of failure of detecting DA attacks by providing better codes, thus, making channel coding a more viable solution despite the overhead.}}.}
When the block is erasure coded, to make the invalid portion of the block unavailable, the malicious block producer must prevent honest full nodes from decoding back the original block. \trim{They do so %
by either} 1) hiding a larger portion of the coded block (more than the erasure correcting capability of the code). This hiding
can be detected with a high probability by the light nodes
using random sampling; 2) incorrectly generating the coded data. In this case, honest full nodes can broadcast verifiable \emph{incorrect-coding (IC) proofs} \cite{dataAvailOrg}, \cite{CMT} allowing light nodes to reject the header. To keep the IC proof size small, authors in \cite{dataAvailOrg} used 2D Reed-Solomon (RS) codes. 2D-RS codes result in an IC proof size of $O(\sqrt{b}\log b)$, where $b$ is the size of the block. Work in \cite{CMT} extends the idea into a
technique called Coded Merkle Tree (CMT).
\trim{A CMT uses Low-Density Parity-Check (LDPC) codes for encoding a Merkle tree and it provides the following benefits:}
1) small check node (CN) degrees in the LDPC codes reduce the IC proof size to $O(\log b)$ \cite{CMT}; 2) LDPC codes can be decoded using a linear time peeling decoder \cite{ModernCodingTheory}, thus reducing the decoding complexity compared to Reed-Solomon codes. Despite these benefits, an LDPC code with a peeling decoder leads to certain problematic objects, called stopping sets \cite{ModernCodingTheory}  \deb{that allow malicious nodes to successfully hide a smaller portion of the block compared to Reed-Solomon codes.} 
A stopping set of an LDPC code is a set \deb{of variable nodes (VNs)} that if erased prevents a peeling decoder from fully decoding the original block.  If a malicious node hides coded symbols corresponding to a stopping set of the LDPC code, full nodes will not be able to decode the CMT. 
Since the malicious node can hide the smallest stopping set, the best code design strategy to reduce the probability of failure using random sampling is to construct deterministic LDPC codes with large minimum stopping set size. Constructing such LDPC codes is considered a hard problem \cite{SSelim}.

\editor{Another important coding parameter for  the CMT is the length of the LDPC codes which affects the encoding/decoding complexity and Merkle proof sizes. \revonetwo{Similar to applications such as wireless systems, short code lengths are beneficial in CMT applications (like low latency blockchains \cite{LowLatency} or resource limited IoT blockchains \cite{IoTsurvey}) since they keep the above quantities small.} Previous work in \cite{CMT} have focused on using codes from an LDPC ensemble to construct the CMT. At large code lengths, the LDPC ensemble guarantees, with high probability, a large stopping ratio (the smallest stopping set size divided by the code length \cite{CMT}) and hence a low probability of failure. However, at short code lengths,  the LDPC ensemble is unable to provide good guarantees on the minimum stopping set size. Authors in \cite{CMT} combat this issue through the use of \emph{bad-code} proofs when  codes with a smaller stopping ratio (\emph{bad-codes}) than guaranteed by the ensemble get used. A bad-code proof triggers all nodes in the system to use a newly sampled code from the ensemble.
However, at short code lengths, this approach requires many rounds of bad codes until a good code has been found which undermines the security of the system.
Thus, the LDPC code design of \cite{CMT} is inappropriate for short CMT code lengths. Hence, in this paper, we focus on short CMT code lengths and provide deterministic LDPC codes that allow for good detection of DA attacks. 
\revonetwo{Due to our focus on short code-lengths, we do not make guarantees for the extension of the techniques proposed in this paper to longer code lengths.}
For various adversary models, we provide a co-design of specialized LDPC codes and sampling strategies that reduce the probability of failure compared to techniques used in earlier literature.}

\deb{We can broadly categorize all possible adversaries into three types based on their computational capabilities. The computational complexity is based on how hard it is for a malicious node to find the minimum stopping set in the LDPC code (which is known to be an NP-hard problem \cite{SSNP-hard}). Note that the light node sampling strategy is known by all entities in the system.
The first adversary type is termed as a \emph{weak adversary}.  A weak adversary does not have the resources to find a large number of stopping sets. It settles for hiding a random one it finds and is unable to take advantage of the light node sampling strategy. The second type is a \emph{medium adversary}.  A medium adversary, using more computational resources, can find all stopping sets up to a certain size and select the stopping set that performs the worst under the posted light node sampling strategy. While the medium adversary has more computational capability than a weak adversary, a medium adversary represents a malicious node with bounded resources and can only find stopping sets up to a certain size  within a reasonable time frame. The final type is a \emph{strong adversary} which we assume has unlimited resources and can find all stopping sets \lev{(of any size)} and hide one among them that performs the worst. These three models represent how much resources we assume an adversary possesses to disrupt our system. As such, our modeling encompasses everything from a single hacker with a standard computer to a small group of hackers with a cluster of computers to a large organization with unlimited resources.}\\

\vspace*{-0.75cm}
\subsection{Contributions}\label{sec:contributions}
\vspace{-0.1cm}
Our main contributions in this paper are co-design techniques for LDPC codes and coupled light node sampling strategies that result in a low probability of failure under the different adversary models described above. \revone{In LDPC codes with no degree-one VNs, all stopping sets are made up of cycles \cite{WesselSS}. Since working with stopping sets directly is  computationally difficult, in this paper we design LDPC codes by optimizing cycles to indirectly optimize stopping sets.
We show that our LDPC codes result in the desired stopping set properties and produce  low  probability of failures for the different adversary models.} The contributions are listed as follows:
\begin{enumerate}[wide, labelwidth=!, labelindent=10pt]

    \item For the weak adversary, we demonstrate that \emph{concentrating} stopping sets in LDPC codes to a small set of VNs and then greedily sampling this small set of VNs results in a low probability of light node failure. We then provide a specialized LDPC code construction technique called the \emph{entropy-constrained}  Progressive Edge Growth (EC-PEG) algorithm that is able to concentrate stopping sets in the LDPC code to a small set of VNs.
\trim{We provide a greedy sampling strategy for the light nodes to sample this small set of VNs.} We demonstrate that for a weak adversary, LDPC codes constructed using the EC-PEG algorithm along with the greedy sampling strategy result in a significantly lower probability of failure compared to techniques used  in earlier literature.

\item To secure the light nodes against a medium and a strong adversary, we provide a co-design
of a light node sampling strategy called \emph{linear-programming-sampling} (LP-sampling) and an LDPC code construction called \emph{linear-programming-constrained} PEG (LC-PEG) algorithm. LP-sampling is tailor-made for the particular LDPC codes used to construct each layer of the CMT. \trim{It is designed by solving a linear program (LP) based on the knowledge of the small stopping sets in the LDPC codes to minimize the probability of failure. 
We demonstrate that, for a medium and a strong adversary, LDPC codes designed by the LC-PEG algorithm coupled with LP-sampling result in a lower probability of failure compared to techniques used  in earlier literature.}

\end{enumerate}

\vspace{-0.4cm}
\subsection{Previous Work}\label{sec:previous_work}
\vspace{-0.1cm}

\revonetwo{In \cite{dataAvailOrg}, authors proposed to solve DA attacks by encoding the block using 2D-RS codes. Their
approach was optimized in \cite{RSoptimize}.} However, 2D-RS codes results in an IC proof size of $O(\sqrt{b}\log b)$.
In \cite{CMT}, authors proposed the CMT and demonstrated that encoding the CMT using LDPC codes results in a small IC proof size of $O(\log b)$. Authors in \cite{CMT} used codes from a random LDPC ensemble of \cite{SSemsemble} to construct the CMT to result in a low probability of failure. \editor{However, random LDPC ensembles used in \cite{CMT} were originally designed for other types of channels (i.e., BSC) and we show that they are not the best choice for this specific application at short CMT code lengths. At the same time, as described before, random LDPC ensembles undermine the security of the system, especially at short CMT code lengths.
In this work, we demonstrate that the presented co-design techniques result in a lower probability of failure compared to using codes from a random LDPC ensemble and random sampling. Furthermore, to alleviate the security problem, we provide deterministic LDPC code design algorithms in this paper.}
In \cite{Cover}, authors provide a protocol called \emph{CoVer} based on CMT, which allows light nodes to collectively validate blocks.
However,  \cite{Cover} still uses
  random sampling and random LDPC ensembles to mitigate DA attacks.

\trim{DA attacks are possible in other blockchain systems as well. \emph{Sharded blockchains} where each node stores
a fraction of the entire block
are vulnerable to DA attacks that can be solved using the CMT \cite{Trifecta}.
The LDPC co-design techniques described in this paper can also be used in sharded blockchains.
\emph{Side Blockchains} \cite{AceD} that improve the throughput of block transactions  are also vulnerable to DA attacks. The vulnerability is mitigated in \cite{AceD} by introducing a \emph{DA oracle} that uses the CMT. A similar idea as this paper of co-design to construct specialized LDPC codes to improve the performance of the DA oracle was demonstrated in \cite{DE-PEG}.} 

\editor{While this paper focuses on designing codes to mitigate DA attacks, channel coding has been extensively used to mitigate other scalability issues in blockchain systems: \cite{networkcodingstorage} uses network codes to reduce the storage cost associated with full nodes; 
\cite{downsampling} combines downsampling and erasure coding to reduce the storage cost while allowing nodes to directly use the stored data without decoding; \cite{SeF} proposes secure fountain codes to reduce the storage and bootstrapping communication cost of full nodes;  \cite{polyshard} uses Lagrange coding in sharded blockchains to simultaneously improve storage, computation, and security;
\cite{erasurelowstorage} proposes using erasure codes to allow light nodes to contribute in storing the blockchain. The proposal in \cite{erasurelowstorage} can be combined with techniques proposed in this paper to enable light nodes to ensure data availability.}

The rest of this paper is organized as follows. \trim{In Section \ref{sec:preliminaries}, we provide the preliminaries and system model. 
In Section \ref{sec:EC-PEG}, we describe the greedy sampling strategy and the EC-PEG algorithm and how they overcome DA attacks against the weak adversary. 
 In Section \ref{sec:LC-PEG}, we present our approach for the medium and strong adversary where we describe the LP-sampling strategy and the LC-PEG algorithm.}
\editor{We discuss system aspects of our co-design in Section \ref{sec:system_aspects}.}
We provide simulation results 
in Section \ref{sec:simulations} and concluding remarks in Section \ref{sec:conclusion}.

\vspace{-0.35cm}
\section{Preliminaries and System Model}\label{sec:preliminaries}
\vspace{-0.1cm}
 In this section, we first look at the preliminaries of the CMT
 
 \noindent
 and LDPC notation.
We then present our system, network, and threat model. We use the following notation in the rest of the paper.  For $\mathrm{p} = (p_1, \ldots, p_t)$ such that $p_i \geq 0$,   $\sum_{i=1}^{t}p_i = 1$, we \deb{use} the entropy function $\mathcal{H}(\mathrm{p}) = -\sum_{i=1}^{t}p_i\log(p_i)$. For a vector $\mathrm{a}$, let $\max(\mathrm{a})$ ($\min(\mathrm{a})$) denote the largest (smallest) entry of $\mathrm{a}$ and let $\mathrm{a}_i$ denote the $i^{th}$ element of $\mathrm{a}$. For a matrix $\mathrm{M}$ of size $c \times d$, let
  $\mathrm{M}_{ki}$ denote the element of $\mathrm{M}$ on the $k^{th}$ row and $i^{th}$ column, $1 \leq k \leq c$, $1 \leq i \leq d$. Define $x \bmod {p} := (x)_p$.

\begin{figure*}[t]
    \centering
    \begin{subfigure}{0.52\linewidth}
\begin{minipage}{0.99\linewidth}
\begin{tikzpicture}
  \node (img)
  {\includegraphics[scale=0.34]{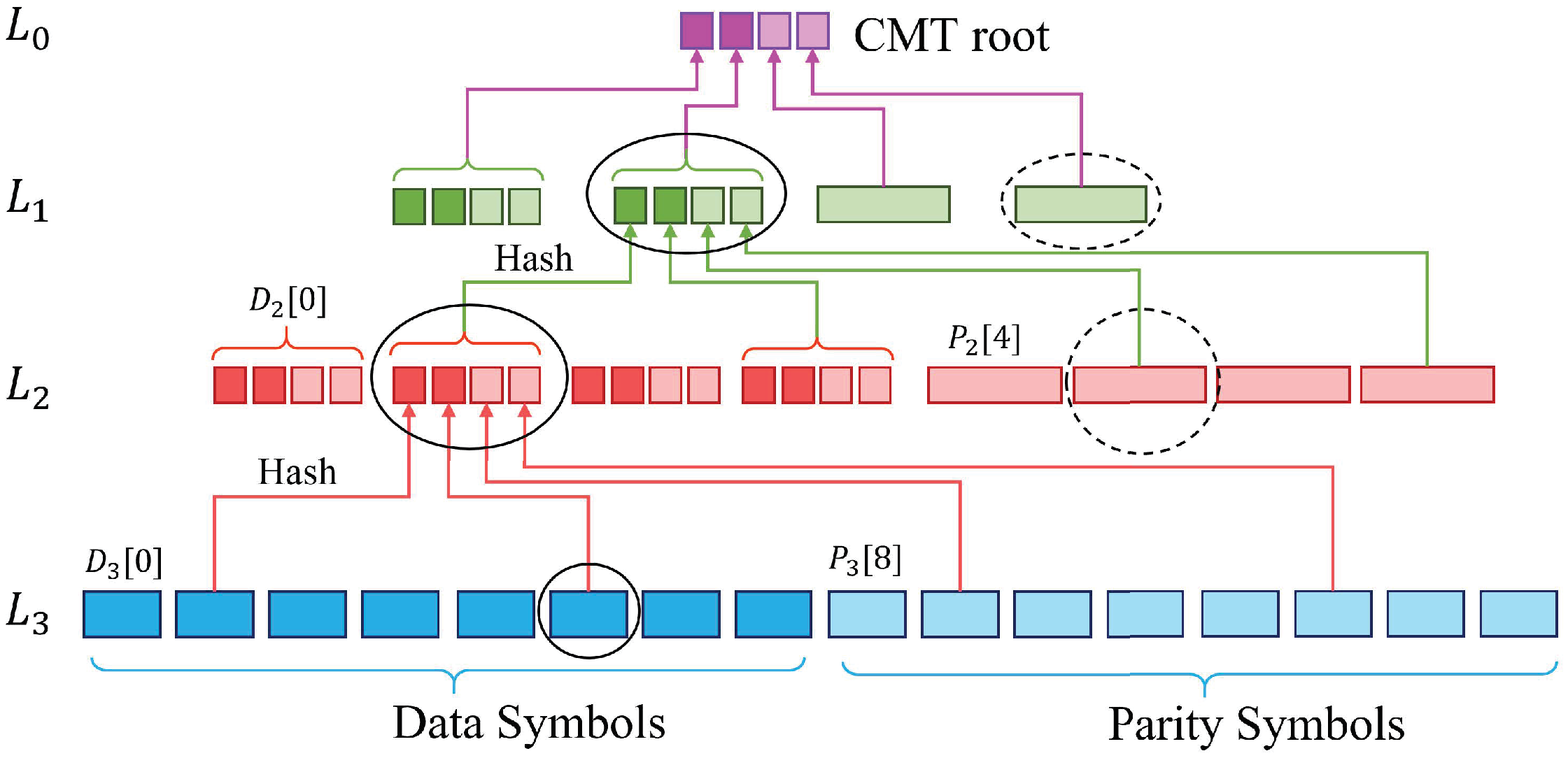}};
 \end{tikzpicture}
 \end{minipage}
    \end{subfigure}%
\begin{subfigure}{0.5\linewidth}
\begin{minipage}{0.99\linewidth}
\begin{tikzpicture}
  \node (img)
  {\includegraphics[scale=0.34]{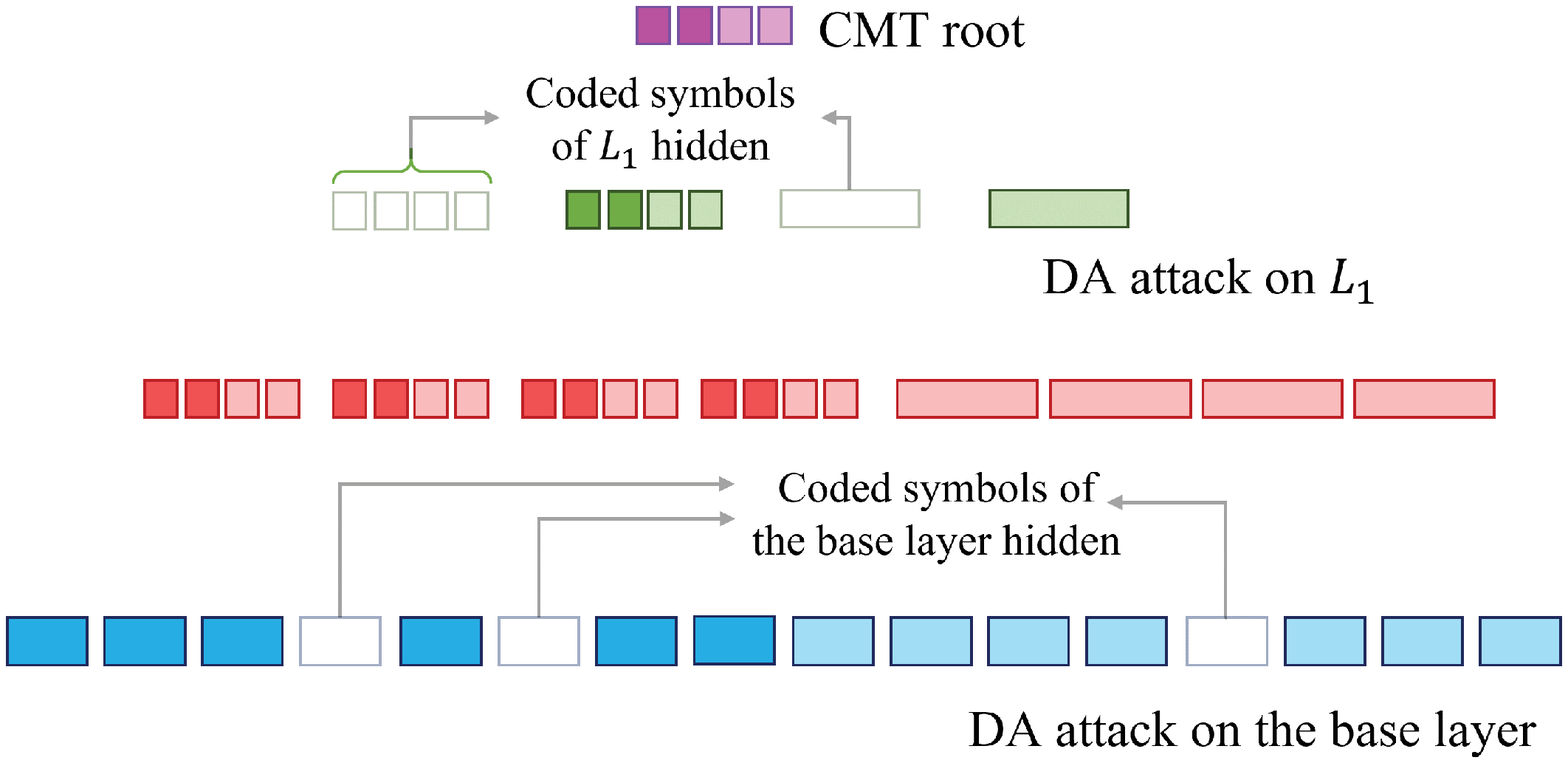}};
 \end{tikzpicture}
 \end{minipage}
\end{subfigure}
     \vspace{-10pt}
    \caption{\deb{Left Panel: Construction process of a CMT. A block of size $b$ is partitioned into $k$ data chunks (data symbols) each of size $\frac{b}{k}$ and a rate $\tr$ systematic LDPC code is applied to generate $n$ coded symbols. These $n$ coded symbols form the base layer of the CMT. The $n$ coded symbols are then hashed using a hashing function and the hashes of every
$q$ coded symbols are concatenated to get one data symbol of the parent layer. The data symbols of this layer are again coded using a rate $\tr$ systematic LDPC code and the coded symbols are further hashed and concatenated to get the data symbols of its parent layer. This iterative process is continued until there are only $t$
$(t > 1)$ hashes in a layer which form the CMT root. Left panel shows a CMT with $n = 16$, $q = 4$, $\tr = 0.5$ and $t = 4$. The circled symbols in $L_1$ and $L_2$ are the Merkle proof of the circled symbol in $L_3$.  Right panel: DA attack on the CMT.}}
    \label{fig:CMT}
     \vspace{-15pt}
\end{figure*}

\vspace{-0.4cm}
\subsection{Coded Merkle Tree (CMT) }\label{sec:CMT_construction}
\vspace{-0.1cm}
\subsubsection{CMT construction} A CMT of a block is built using the block transactions as leaf nodes and the CMT root is included in the block header. It is constructed by encoding each layer of the Merkle tree \cite{Bitcoin} with an LDPC code and then hashing the layer to generate its parent layer. A simplified description of the CMT construction is shown in Fig. \ref{fig:CMT} left panel.
As shown in Fig. \ref{fig:CMT} left panel, coded symbols of a layer are interleaved into the data symbols of the parent layer\footnote{
\revonetwo{In this paper, we refer to chunks of a fixed length as symbols of a field. A symbol of $c$ bits is represented as an element in $\mathbb{F}^{c}_2$ and encoding and decoding are performed using bitwise XOR operations over the bitwise representation of the symbols (similar to \cite{SeF}). 
\revonethree{Thus, the complexity of encoding and decoding depends on the size of the chunks (i.e., symbols) $c$ which is calculated as $c = \frac{b}{nR}$ where $b$ is the block size, and $n$ and $R$ are the length and rate of the LDPC code in the CMT base layer.}
}
}.
In this paper, we adopt the interleaving technique introduced in \cite{AceD}.
Let the CMT have $l$ layers (except the root), $L_1, L_2, \ldots, L_l$, where $L_l$ is the base layer. The root of the CMT is referred to as $L_0$.
\trim{For $1 \leq j \leq l$, let $L_j$ have $n_j$ coded symbols and let the LDPC code used in  $L_j$ have a parity check matrix $H_j$.}
\editor{Let $N_j[i]$, $ \tzero \leq i \tleq \tnj$, be the $(i+1)^{th}$ symbol of the $j^{th}$ layer $L_j$\footnote{\editor{Due to modulo operations, we define $N_j[i]$ starting with index 0 for $i$. All other variables in the paper start with index 1.}}. Also, let $D_j[i] = N_j[i], \tzero \leq i \tleq \trnj$ and $P_j[i] = N_j[i]$, $ \trnj \leq i \tleq \tnj$, be the systematic (data) and parity symbols of $L_j$, respectively.
Coded symbols $P_j[i]$, $  \trnj \leq i \tleq \tnj$ are obtained from $D_j[i]$, $\tzero \leq i \tleq \trnj$ using a rate $\tr$ systematic LDPC code $H_j$.}  \trim{In the above CMT, hashes of every $q$ coded symbols of $L_j$ are concatenated to form a data symbol of $L_{j-1}$.} Hence, $n_j = \frac{n_l}{(q\tr)^{l-j}}$, $j = 1, \ldots, l$. The CMT root has $t = n_1$ hashes.
 Let the number of systematic and parity symbols in $L_j$ be denoted by $s_j = \tr n_j$ and $p_j = (1-\tr)n_j$, respectively.
\trim{For $1 \leq j \leq l$, the data symbols of  $L_{j-1}$ are formed from the coded symbols of $L_j$ as follows:}
\begin{align*}
    D_{j-1}[i] = N_{j-1}[i] =  &{\fontfamily{qcr}\selectfont \text{concat}}(\{ {\fontfamily{qcr}\selectfont \text{Hash}}(N_{j}[x]) \:\vert \:\tzero\leq x \tleq \tnj,\\& i = (x )_{s_{j-1}} \})\; \forall \; \tzero  \leq i \tleq s_{j-1},
\end{align*}

\noindent
where ${\fontfamily{qcr}\selectfont \text{Hash}}$ and ${\fontfamily{qcr}\selectfont \text{concat}}$ represent the hash and the string concatenation functions, respectively. 

\subsubsection{Merkle Proof for CMT symbols}
The Merkle
proof of a symbol in $L_j$ consists of a data symbol and a parity symbol from each intermediate layer of the tree that is above $L_j$ \cite{AceD}. An illustration of a Merkle proof is shown in Fig. \ref{fig:CMT} left panel.
In particular, the Merkle proof of the symbol $N_j[i]$, $1 < j \leq l$, is the set of symbols 
\editor{\{$N_{j'}[\;(i)_{s_{j'}}]$, $N_{j'}[\;s_{j'}+ (i)_{p_{j'}}] \:\vert\: 1 \leq j' \leq j-1$\}}.
Detailed discussion on the properties of Merkle proofs\footnote{The data part of the Merkle proof of $N_j[i]$ from each layer lie on the path of $N_j[i]$ to the CMT root and can be used to check the integrity of $N_j[i]$ in a manner similar to regular Merkle trees in \cite{Bitcoin}. The parity symbols in the Merkle proof are only for sampling purposes and the information provided in the Merkle proof of $N_j[i]$ are sufficient to check their integrity \cite{AceD}.%
} can be found in \cite{AceD}.

\vspace{-0.01cm}
\subsubsection{Hash-Aware Peeling decoder}
Using the CMT root and the available  symbols of each layer of the CMT, the original block can be decoded using a hash-aware peeling decoder described in \cite{CMT}. The hash-aware peeling decoder decodes each layer of the CMT (from top to bottom) like a conventional peeling decoder \cite{ModernCodingTheory}. However, after decoding a symbol in layer $j$, the decoder matches its hash with the corresponding hash present in layer $j-1$. Matching the hashes allows the decoder to detect IC attacks and generate IC proofs as described in \cite{CMT}. \trim{The IC proof size is proportional to the degree of CNs in the LDPC codes used to build the CMT.
}

\vspace{-0.3cm}
\subsection{Stopping sets and LDPC notation}

\vspace{-0.05cm}
A stopping set of an LDPC code is a set of VNs such that every CN connected to this set is
connected to it at least twice \cite{ModernCodingTheory}. A stopping set is hidden (made unavailable) by a malicious node if all VNs present in it are hidden. The hash-aware peeling decoder fails to successfully decode layer $j$ of the CMT \trim{if a stopping set of $H_j$} is unavailable.
\revfourtwo{Let the Tanner graph (TG) \cite{ModernCodingTheory} representation of $H_j$ be denoted by $\tG_j$ such that  $\tG_j$ has $n_j$  VNs $\{v^{(j)}_1, \ldots, v^{(j)}_{n_j}\} $. VN $v^{(j)}_i$ corresponds to the $i^{th}$ column of $H_j$ and  CNs in $\tG_j$ correspond to the  rows of $H_j$. Let $H_j[v^{(j)}_{i}]$ denote the column of the parity check matrix corresponding to VN $v^{(j)}_{i}$. CMT symbol $N_{j}[i]$, $\tzero \leq i \tleq \tnj$, corresponds to VN $v^{(j)}_{i+1}$ of $\tG_j$.}
  A cycle of length $g$ is called a $g$-cycle.  For a set $\mathcal{S}$, let $|\mathcal{S}|$ denote its cardinality. For a cycle (stopping set) in the TG $\mathcal{G}$, we say that a VN $v$ \emph{touches} the cycle (stopping set) iff $v$ is part of the cycle (stopping set). Define the weight of a stopping set as the number of VNs touching it. Let $\omega^{(j)}_{\min}$ denote the minimum stopping set size of $H_j$, $1 \leq j \leq l$. 
  The girth of a TG is defined as the length of the smallest cycle present in the graph.

\vspace{-0.3cm}
\subsection{System and Network Model}\label{sec:network_nodes}
\vspace{-0.05cm}
\editor{We consider a blockchain system similar to \cite{dataAvailOrg} and \cite{CMT} that has full nodes and  light nodes. 
One of the full nodes acts as a \emph{block producer} of a new block.} \revthree{
We consider the same blockchain network model as \cite{CMT}. In particular, we assume a synchronous network where
\revonetwo{the subgraph of honest full nodes is connected}\footnote{\revonetwo{The connected subgraph of honest full nodes ensures that a message broadcasted by a honest node reaches all honest nodes.}} and the messages sent on the network are anonymous.
The network can have a dishonest majority of full nodes, but each light node is connected to at least one honest full node (thus preventing eclipse attacks \cite{dataAvailOrg}). Nodes broadcast a message (fraud proofs, IC proofs, and CMT symbols) by sending the message to all its connected nodes. The connected nodes check the message correctness (Merkle proofs) and forward valid messages to their neighbors}\footnote{\revthree{ Since messages are communicated only to connected nodes, the cost of broadcasting is not high. Moreover, honest nodes prevent fake communication from malicious nodes by  forwarding only valid messages. }}.
In the following, we describe actions performed by the block producer, other full nodes, and light nodes. We also mention the items included in the publicly available protocol that is designed by a blockchain system designer to be used by nodes in the system. In Section \ref{sec:system_aspects}, we provide a discussion on the 

\noindent
blockchain system designer.
\begin{enumerate}[wide, labelwidth=!, labelindent=10pt,font=\itshape]
    \item \editor{ \emph{Items included in the protocol:} Parity check matrices $H_j$, $1 \leq j \leq l$, systematic
    generator
     matrix of each $H_j$,
   and the light node sampling strategy (a rule to sample CMT symbols). \\[-3mm]

    \item \emph{Block Producer:} A full node that produces (mines) a new block (see Fig. \ref{fig:da_attack}).
    On producing a new
 block, the block producer encodes the block to construct its CMT using the systematic generator matrices specified in the protocol. It then
    broadcasts all the coded symbols in the CMT (including the root) to other full nodes and the root of the CMT to the light nodes. 
    On receiving  a sampling request from the light nodes, it returns the requested symbols along with their Merkle proofs. The block producer can be malicious and can act arbitrarily. \\[-3mm]
    
    \item \emph{Full nodes that are not the block producer:}
    These nodes perform Merkle proof checks on the coded symbols of the CMT that they receive from a block producer, other full nodes, or light nodes (see Fig. \ref{fig:da_attack}).
    They forward  symbols that satisfy the Merkle proofs to other connected full nodes. Using the symbols that they received, they decode each layer of the CMT with a hash-aware peeling decoder using the parity check matrices $H_j$, $1\leq j\leq l$, specified in the protocol. 
After decoding the base layer of the CMT, which contains transaction data, they verify all the transactions. 
They store a local copy of all blocks (i.e., its CMT) that they verify to be valid (i.e., fully available, having no fraudulent transactions and no incorrect-coding at any layer). They declare the availability of this valid block to all other nodes and respond to sample requests from the light nodes. If they find a certain block to be invalid, either due to fraudulent transactions  or incorrect coding, they broadcast a fraud proof or an IC proof for other nodes to reject the block. If they find a certain layer of the CMT to be unavailable (i.e., having coded symbols missing that prevent decoding), they reject the block.
A malicious full node need not follow the above protocol and can act arbitrarily. \\[-3mm]

    \item \emph{Light nodes:} These nodes are storage constrained and only store the CMT
    root corresponding to each block (see Fig. \ref{fig:da_attack}). They download only a small portion of the block and perform tasks like fraud and IC proof checks. Additionally, light nodes check the availability of each layer of the CMT. They do so by making sampling requests for coded symbols of the CMT base layer from the block producer (or any other full node that declares the block to be available). They make sample requests using the sampling strategy specified in the protocol.
    They perform Merkle proof checks on the returned symbols and broadcast symbols that satisfy the Merkle proofs to other connected full nodes.
    Upon receiving all the requested symbols and verifying their Merkle proofs, light nodes accept the block as available and store the block header. 
    On receiving fraud proofs or IC proofs sent out by a full node, light nodes verify the proof and reject  the header if the proof is correct. We assume that each light node is honest.}
\end{enumerate}

\vspace{-0.2cm}
\begin{remark}\label{remark:comparisons}
\trim{In this paper, we provide co-design of LDPC codes and sampling strategies (that are included in the protocol) to reduce the probability
of failure. As such, we do not compromise on other performance metrics considered in \cite{CMT}:} the CMT root has a fixed size $t$ which does not grow with the blocklength; the hash-aware peeling decoder has a decoding complexity linear in the blocklength;  we empirically show that the IC proof size for our codes is similar to \cite{CMT}.
  \end{remark}

\vspace{-0.4cm}
\subsection{Threat Model}\label{sec:adv_models}
\vspace{-0.05cm}

\revthree{A blockchain system involves  two aspects: block generation and block verification. The block
generation depends on the consensus algorithm used in the blockchain e.g., Proof of Work (PoW) \cite{Bitcoin}, Proof of Stake (PoS) \cite{SnowWhite}, etc.. However, a DA attack caused by an adversary with dishonest majority (in terms of work, stake, etc.) affects the block verification process. Hence, the exact consensus algorithm used by the blockchain is not relevant to our work. Similar to \cite{dataAvailOrg} and \cite{CMT}, we focus on block verification and propose LDPC codes to mitigate DA attacks\footnote{\revthree{Note that forking-based double spending attacks (related to block generation) where an adversary generates an invalid longest chain are still possible with a dishonest majority of full nodes \cite{dataAvailOrg} but are not necessary to launch a DA attack.}}.
}

\editor{Similar to \cite{dataAvailOrg} and \cite{CMT}, we model our system security in terms of two properties:  \emph{i) Soundness:} If a light node thinks that a block is available and accepts the block, then at least one honest full node in the system will be able to fully decode all layers of the CMT corresponding to the block;  \emph{ ii) Agreement:} \revonethree{If a light node} determines that a block is available, all light nodes in the system determine that the block is available. \revonethree{Similar to \cite{dataAvailOrg}, we analyse probability of soundness or agreement failure per light client. Let $\tpfsa$ be the probability that soundness or agreement fails for a single light client due to a DA attack.} In Section \ref{sec:system_aspects}, we show that in our proposed co-design, $\tpfsa$ is reduced by reducing the probability of failure of a single light node to detect DA attacks when there is a sufficiently large number of light nodes in the system. Thus, in the rest of the paper, we focus on reducing the probability of failure of a single light node.}

We consider an adversary  that conducts a DA attack by hiding coded symbols of the CMT.
\deb{ An illustration of a DA attack is shown in Fig. \ref{fig:CMT} right panel.} On receiving sampling requests from the light nodes, the adversary only returns coded symbols that it has not hidden and ignores other requests.
The adversary conducts a DA attack at layer $j$ of the CMT by 1) generating coded symbols of layer $j$, that satisfy their Merkle proof, for the light nodes to accept these coded symbols as valid, and 2) hiding a small portion of the coded symbols of layer $j$, corresponding to a stopping set of $H_j$, such that honest full nodes are not able to decode the layer. A DA attack at layer $j$ prevents an honest full node from generating a fraud proof of fraudulent transactions (if $j =l$) or an IC proof for incorrect coding at layer $j$. 
 Since an incorrect coding can occur at any layer, for the full nodes to be able to send IC proofs, light nodes must detect a DA attack at any layer $j$ that the  adversary may perform. They do so by sampling few base layer coded symbols. For each intermediate layer $j$, $1 \leq j < l$, the symbols of layer $j$ collected as part of
 the Merkle proofs of the base layer samples are used to check  the availability of layer $j$.

Light nodes fail to detect a DA attack if none of the base samples requested or the symbols in their Merkle proofs are hidden. 
Let $P^{(j)}_f(s)$, $1 \leq j \leq l$, be the probability of failure of detecting a DA attack at layer $j$ by \editor{a single light node when it samples $s$ base layer coded symbols.} Also, let $J^{\max} = \underset{1 \leq j \leq l}{\operatorname{argmax}} \: P^{(j)}_f(s)$. To maximize the probability of failure,  we assume that the adversary is able to perform a DA attack at layer $J^{\max}$.  We now provide precise mathematical definitions of the three adversary models discussed in Section \ref{sec:intro} based on their computational capabilities:
\subsubsection{Weak Adversary} For each layer $j$, $1 \leq j \leq l$, they hide stopping sets of  size $< \mu_j$ for the parity check matrix $H_j$ (for some integer $\mu_j$). Moreover, they do not exhaustively find all stopping sets of a particular size of a given parity check matrix or perform a tailored search for stopping sets. Instead, we assume that to conduct a DA attack at layer $j$, for all stopping sets of $H_j$ of a particular size, they randomly choose one of them to hide.
\subsubsection{Medium Adversary} For each layer $j$, $1 \leq j \leq l$, they hide stopping sets of size $< \mu_j$ for the parity check matrix $H_j$. However, they use the knowledge of the sampling strategy employed by the light nodes to hide the worst case stopping set that has the lowest probability of being sampled by the light nodes. Let $\tsj_j$ be set of all stopping sets of $H_j$ of size $< \mu_j$. Also, let $P_f^{(j)}(s) = \underset{\tSall \in \tsj_j}{\max}\; P_f^{(j)}(s;\tSall)$, where $P_f^{(j)}(s;\tSall)$ is the probability of failure for the light nodes to detect a DA attack at layer $j$ under the light node sampling strategy when the adversary hides the stopping set $\tSall$ of $H_j$. For $J^{\max} = \underset{1 \leq j \leq l}{\operatorname{argmax}} \: P^{(j)}_f(s)$, the medium adversary conducts a DA attack at layer $J^{\max}$ by hiding a stopping set $\tSall$ from $\tsj_{J^{\max}}$ with the highest $P_f^{J^{\max}}(s;\tSall)$. 
 \subsubsection{Strong Adversary} 
   They can find the worst case stopping sets of any size of $H_j$, $1 \leq j \leq l$. Let $\tsj^{\infty}_{j}$ be the set of all stopping sets of $H_j$. Similar to the medium adversary, define $P_f^{(j)}(s) = \underset{\tSall \in \tsj^{\infty}_j}{\max} P_f^{(j)}(s;\tSall)$ and   $J^{\max} = \underset{1 \leq j \leq l}{\operatorname{argmax}} \: P^{(j)}_f(s)$. The strong adversary conducts a DA attack at layer $J^{\max}$ by hiding a stopping set $\tSall$ from $\tsj^{\infty}_{J^{\max}}$ with the highest $P_f^{J^{\max}}(s;\tSall)$. 

\vspace{-0.05cm}

 The co-design that we provide to mitigate DA attacks against weak adversaries, i.e., the EC-PEG algorithm and the greedy sampling strategy, has the advantage of being computationally cheap and does not involve finding stopping sets.
In order to mitigate DA attacks against a medium and a strong adversary we provide  LP-sampling and the LC-PEG algorithm. LP-sampling uses  stopping sets of size $< \mu_j$ from layer $j$ of the CMT and is more computationally expensive. It is an overkill for the weak adversary which can be mitigated using cheaper techniques. \revtwo{
Factors such as the choice of the consensus algorithm, area of deployment, etc. can give an idea about the expected computational capabilities of full nodes in the system and allow the system designer to choose the adversary model. For example, in PoS \cite{SnowWhite} and PoSpace \cite{PoSpace} consensus blockchains, full nodes need not have a high computational power and a weak adversary would be a reasonable model to follow.  For PoW blockchains \cite{Bitcoin}, full nodes are expected to have high computational power and a strong and medium adversary model would be a suitable design choice. Another example is small scale IoT-blockchains where the blockchain nodes are IoT devices \cite{IoTsurvey}. Here, full nodes have low computational power and a weak adversary model would be appropriate. }
    
In our co-design to mitigate a DA attack against a medium and a strong adversary, we assume that a blockchain system designer decides the value of $\mu_j$, $1 \leq j \leq l$, and is able to find all stopping sets of $H_j$ of size $< \mu_j$, that is used to design LP-sampling. 
Although finding all stopping sets of $H_j$ of size $< \mu_j$ is NP-hard, since we focus on short code lengths
in this paper, the set of stopping sets can be found in a reasonable amount of time using Integer Linear Programming (ILP) methods demonstrated in \cite{ILPsearch}.
\editor{
Note that $\mu_j$ and the set of all stopping sets of $H_j$ of size $< \mu_j$ that the designer uses to design LP-sampling is not publicly released. Only the final design output, i.e., the LP-sampling strategy is included in the protocol. 
Here, we have made a trusted set up assumption of a blockchain system designer to design the items included in the protocol.
In Section \ref{sec:system_aspects}, we will discuss potential ways to prevent security attacks by a malicious designer and how some attacks are naturally handled by our co-design method.} 

\editor{Given the above adversary models, we provide LDPC code construction and sampling strategies to minimize the probability of failure for a single light node  to detect DA attacks. In the next section, we discuss the techniques to mitigate DA attacks conducted by a weak adversary.  }

 \vspace{-0.2cm}
\section{LDPC code and sampling co-design for Weak Adversary}\label{sec:EC-PEG}
\vspace{-0.1cm}
In this section, we demonstrate our novel design idea of \deb{\emph{concentrating}} stopping sets in LDPC
codes to reduce the probability of failure against a weak adversary. 
\revone{Since working with stopping sets directly is  computationally difficult, we focus on \emph{concentrating} cycles to indirectly \emph{concentrate} stopping sets.} It is well known that codes with irregular VN degree distributions are prone to small stopping sets. Thus, we consider VN degree regular LDPC codes of VN degree $d_v \geq 3$ in this paper.
\trim{In the following, we first look at the effect of the light node sampling strategy on the probability of failure when a DA attack occurs on the base layer of the CMT.}
This will motivate the LDPC code construction for the base layer. 
Later, we demonstrate how the LDPC code construction for the base layer can be used in all layers by aligning the columns of the parity check matrices before constructing the CMT. For simplicity of notation, we denote
$H_l$ by $H$ having $n$ VNs $\tV =$ $\{v_1, v_2, \ldots, v_n\}$ and TG $\mathcal{G}$.  Consider the following definition.

\vspace{-0.1cm}
\begin{definition}\label{remark:skewingSS}
For a parity check matrix $H$, let $ss^{\teta} = (ss^{\teta}_1, ss^{\teta}_2, \ldots, ss^{\teta}_n)$ denote the VN-to-stopping-set of weight $\teta$ distribution where $ss^{\teta}_i$ is the fraction of stopping sets of $H$ of weight  $\teta$ touched by $v_i$.  Similarly, for a parity check matrix $H$, let $\td^g = (\td^g_1, \td^g_2, \ldots, \td^g_n)$ be the VN-to-$g$-cycle distribution where $\td^g_i$ is the fraction of $g$-cycles of $H$ touched by $v_i$. 
\end{definition} 

\vspace{-0.2cm}
We informally say that distribution $ss^{\teta}$ ($\td^g$) is concentrated if a small set of VNs have high 
corresponding stopping set ($g$-cycle) fractions $ss^{\teta}_i$ ($\td^g_i$). The following lemma demonstrates that LDPC codes with concentrated $ss^{\teta}$ results in a smaller probability of light node failure when a weak adversary conducts a DA attack (on the base layer).
\revone{The proof is straightforward and we omit it due to space limitations. It can be found in \cite{SSskewITW} and references therein.}

\vspace{-0.1cm}
\begin{algorithm*}
{
\caption{Light node sampling strategy for weak adversary: {\fontfamily{qcr}\selectfont \text{greedy-set}}($\mathcal{G},g_{\min},g_{\max},s$)}\label{alg:LN-greedysamp}
\begin{algorithmic}[1]
\State \textbf{Inputs:} TG $\mathcal{G}$, $g_{\min}$, $g_{\max}$, $s$, \textbf{Output:} \trim{$S^{(s)}_{greedy}$}, \textbf{Initialize:} $S^{(s)}_{greedy} = \emptyset$, $g = g_{\min}$, $\widehat{\mathcal{G}} = \mathcal{G}$
\While{$\vert S^{(s)}_{greedy} \vert < s$}
\State $v$ = VN that touches the maximum number of $g$-cycles in $\widehat{\mathcal{G}}$ (ties broken randomly)
\State $S^{(s)}_{greedy} = S^{(s)}_{greedy} \cup \{v\}$,\;
Purge $v$ and all its incident edges from $\widehat{\mathcal{G}}$ 

\If{$\widehat{\mathcal{G}}$ has no $g$-cycles} $g = g+2$
\EndIf
\If{ $g \geq g_{\max}$}
\State $\mathcal{V}_r$ = randomly select $s - \vert S^{(s)}_{greedy} \vert$ VNs from $\widehat{\mathcal{G}}$ (ordered arbitrarily);
$S^{(s)}_{greedy} = S^{(s)}_{greedy} \cup \mathcal{V}_r$
\EndIf
\EndWhile
\end{algorithmic}
}
\end{algorithm*}
\vspace*{-0.05cm}

\begin{lemma}\label{lemma:average_pf}
Let $\mathcal{SS}_{\teta}$ denote the set of all weight $\teta$ stopping sets of $H$. For a weak adversary that randomly hides a stopping set from  $\mathcal{SS}_{\teta}$, the probability of failure at the base layer, $P^{(l)}_f(s)$, when the light nodes  use $s$ samples and any sampling strategy satisfies $P^{(l)}_f(s) \geq 1 - \max_{\tS \subseteq \tV, |\tS| = s}\tau(\tS,\teta)$. Here, $\tau(\tS,\teta)$ is the fraction of stopping sets of weight $\teta$ touched by the subset of VNs  $\tS$ of $H$. 
The lower bound in the above equation is achieved when light nodes sample, with probability one, the set $\tS^{opt}_{\teta} = \mathrm{argmax}_{\tS \subseteq \tV, |\tS| = s}\tau(\tS,\teta)$.
\vspace{-0.2cm}
\end{lemma}

\vspace{-0.01cm}
Lemma \ref{lemma:average_pf} suggests that for a sample size $s$, the lowest probability of failure is $1 - \tau(\tS^{opt}_{\teta},\teta)$ and is achieved when the light nodes sample the set $\tS^{opt}_{\teta}$.
Now, $\tau(\tS^{opt}_{\teta},\teta)$ is large if a majority of stopping sets of weight $\teta$ are touched by a small subset of VNs. This goal is achieved if the distributions $ss^{\teta}$ are concentrated towards a small set of VNs.
Thus, designing LDPC codes with  concentrated $ss^{\teta}$ increases $\tau(\tS^{opt}_{\teta},\teta)$ and reduces the probability of failure. 
In Section \ref{sec:alg:EC-PEG}, we 
design the EC-PEG algorithm that achieves concentrated stopping set distributions.

We are unaware of an efficient method to find $\tS^{opt}_{\teta}$.
 Instead, we use a greedy algorithm using cycles to find the light node samples, provided in Algorithm \ref{alg:LN-greedysamp}. 
Algorithm \ref{alg:LN-greedysamp} takes as input the TG $\mathcal{G}$, its girth $g_{\min}$, an upper bound cycle length $g_{\max}$, and the sample size $s$. It outputs a set of VNs $S^{(s)}_{greedy}$ that the light nodes will sample, which we call \emph{greedy samples}.
The probability of failure using this strategy when a weak adversary randomly hides a stopping set of size $\teta$ from the base layer is $P^{(l)}_f(s) = 1 - \tau(S^{(s)}_{greedy},\teta)$
\trim{(see proof of Lemma \ref{lemma:average_pf} in \cite{SSskewITW})}.
At the end of this section, we empirically show that concentrating the cycle distributions $\td^{g}$ also concentrates the stopping set distributions. Thus, the EC-PEG algorithm aims to concentrate the cycle distributions to improve the probability of failure. \editor{It is easy to see that the complexity of Algorithm \ref{alg:LN-greedysamp} is dominated by the complexity of finding cycles (of worst case length $g_{\max}$) and is $O(n^{g_{\max}/2})$ using brute force.}

\vspace{-0.1cm}
\editor{\begin{remark}\label{remark:overall_strategy}
(Overall Greedy Sampling Strategy)
In the above sampling strategy,  some coded symbols may never get sampled which can affect the soundness of the system. We alleviate this problem without affecting the probability of failure by modifying the sampling strategy as follows: Let  $\tlambda$ be a fixed parameter where $0 < \tlambda < 1$. For a total of $s$ samples, the light nodes select $\tlambda s$ greedy samples $S^{(\tlambda s)}_{greedy}$ =  {\fontfamily{qcr}\selectfont \text{greedy-set}}($\mathcal{G},g_{\min},g_{\max},\tlambda s$) and randomly select $s - \tlambda s$ base layer coded symbols for the remaining samples. We discuss the soundness and agreement of this modified strategy in Section \ref{sec:system_aspects}.
For this strategy, $P^{(l)}_f(s) = [1 - \tau(S^{(\tlambda s)}_{greedy},\teta)] \big(1 - \frac{\omega^{(l)}_{\min}}{n_l}\big)^{(s - \tlambda s)}$. 
\end{remark}}

\vspace{-0.24cm}

\vspace{-0.3cm}
\subsection{Aligning the parity check matrices of the CMT}\label{sec:colm_align}
\vspace{-0.05cm}
In the above discussion, we demonstrated how to mitigate a DA attack conducted by a weak adversary on the base layer of the CMT using greedy sampling. Now, we extrapolate the idea of greedy sampling to the intermediate layers. \trim{Since the  intermediate layers are sampled via the Merkle proofs of the base layers samples, we align the base and intermediate layer symbols such that the intermediate layers are also sampled greedily. We do so by aligning (permuting) the columns of the parity check matrices used in different CMT layers. We align the columns such that the samples of an intermediate layer $j$ collected from the Merkle proofs of the base layer samples coincide with the greedy samples for layer $j$ provided by {\fontfamily{qcr}\selectfont \text{greedy-set}}($\mathcal{G}_j,g^{(j)}_{\min},g^{(j)}_{\max},\tilde{s}$). Here, $g^{(j)}_{\min}$ is the girth of $\mathcal{G}_j$ and $g^{(j)}_{\max}$ is the upper cycle length for layer $j$.}

We assume that the output $S^{(s)}_{greedy}$ of Algorithm \ref{alg:LN-greedysamp} is ordered according to the order
VNs were added to $S^{(s)}_{greedy}$. 
 Let $S^{(j)}_{ordered} =$ {\fontfamily{qcr}\selectfont \text{greedy-set}}($\mathcal{G}_j,g^{(j)}_{\min}, g^{(j)}_{\max},n_j$), $1 \leq j \leq l$.
  VNs in $S^{(j)}_{ordered}$ are all the VNs of $H_j$ ordered (permuted) according to the order they were added to $S^{(j)}_{ordered}$. Hence, we denote $S^{(j)}_{ordered}[i]$ as the $i^{th}$ VN in this ordered list of VNs. The procedure to align the columns of the parity check matrices of different layers of the CMT is provided in Algorithm \ref{alg:layer-greedy-align}.
 In the algorithm, we first permute the columns of the base layer parity check matrix $H_l$ (to obtain $\widetilde{H}_l$) such the the VNs in $S^{(l)}_{ordered}$ appear as columns $1, 2, \ldots, n_l$ in $\widetilde{H}_l$ (line 3).
 \editor{Recall that when the base layer symbol corresponding to $v^{(l)}_i$ is sampled, then for every intermediate layer $j$, the VNs with with subscript indices $\{1+(i-1)_{s_j}, 1 + s_j + (i-1)_{p_j}\}$ get sampled.} We assign columns of $\widetilde{H}_j$ at these indices (starting from $i=1$) the columns of $H_j$ correspond to the greedy samples in $S^{(j)}_{ordered}$ from start to end (lines 4-6).  
We continue this process until all columns of
$\widetilde{H}_j$ have been assigned.
 \editor{The complexity of Algorithm \ref{alg:layer-greedy-align} is dominated by the complexity of finding $S^{(j)}_{ordered}$ using Algorithm \ref{alg:LN-greedysamp} and has a complexity of $O(\sum_{j=1}^{l}n^{g^{(j)}_{\max}/2}_j)$.
 }

\vspace{-0.2cm}
\begin{algorithm*}
{
\caption{Aligning parity check matrices of CMT for greedy sampling}\label{alg:layer-greedy-align}
\begin{algorithmic}[1]
\State \textbf{Inputs:} $H_j$, $S^{(j)}_{ordered}$, $1 \leq j \leq l$, \textbf{Outputs:} $\widetilde{H}_j$, $1 \leq j \leq l$
\State \textbf{Initialize:} $\widetilde{H}_j$: matrix with unassigned columns, $1 \leq j \leq l$, $\mathrm{counter} = 1$ %
\State $\widetilde{H}_l[i]$ =  $H_l[S^{(l)}_{ordered}[i]]$, $1 \leq i \leq n_l$
\For{$j = 1, 2, \ldots, l-1$}
\textbf{for}{ $i = 1, 2, \ldots, n_l$ \textbf{do}} $d = 1+(i-1)_{s_j}$,\; $p = 1 + s_j + (i-1)_{p_j}$
\If{$\widetilde{H}_j[d]$ is not assigned before} $\widetilde{H}_j[d] = H_j[S^{(j)}_{ordered}[\mathrm{counter}]]$,\;
$\mathrm{counter}$ $\mathrel{+}=1$
\EndIf
\If{$\widetilde{H}_j[p]$ is not assigned before} 
$\widetilde{H}_j[p] = H_j[S^{(j)}_{ordered}[\mathrm{counter}]]$,
$\mathrm{counter}$ $\mathrel{+}=1$
\EndIf
\If{all columns of $\widetilde{H}_j$ have been assigned} Break $i$ for loop
\EndIf
\EndFor
\end{algorithmic}
}
\end{algorithm*}

\vspace{-0.01cm}
\begin{remark}\label{remark:systematic_generator} \editor{ The
 parity check matrices $\widetilde{H}_j$, $1 \leq j \leq l$, after the alignment are included in the protocol.
  Recall that a CMT is built using systematic LDPC codes. Under the assumption of full rank, for the parity check matrices $\widetilde{H}_j$, $1 \leq j \leq l$, the corresponding generator matrices are constructed in a systematic form which are then included in the protocol for CMT construction.} \editor{Also, after the alignment, the overall greedy sampling strategy as described in Remark \ref{remark:overall_strategy} becomes: sample the first $\tlambda s$ coded symbols of the base layer of the CMT and then randomly sample $s - \tlambda s$ base layer coded symbols. This sampling rule is included in the protocol.}
\end{remark}

For a CMT built using $\widetilde{H}_j$, $1 \leq j \leq l$, provided by Algorithm \ref{alg:layer-greedy-align}, greedy sampling of the base layer of the CMT according to Algorithm \ref{alg:LN-greedysamp} ensures that all intermediate layers of the CMT are greedily sampled  according to Algorithm \ref{alg:LN-greedysamp} through the Merkle proofs of the base layer samples. 
Next, we provide a design strategy to construct LDPC codes with concentrated stopping set distributions that result in a low probability of failure under greedy sampling. \revthree{Note that codes produced in the next subsection are aligned by Algorithm \ref{alg:layer-greedy-align} and then included in the protocol.}

\vspace{-0.3cm}
\subsection{Entropy-Constrained PEG (EC-PEG) Algorithm}\label{sec:alg:EC-PEG}
\vspace{-0.02cm}
The EC-PEG algorithm is based on
minimizing the entropy of cycle distribution  $\td^{g}$. \revone{The intuition behind our algorithm is using the fact that uniform distributions have high entropy and distributions that are concentrated have low entropy.} \trim{Thus, we construct LDPC codes using the PEG algorithm \cite{PEG} by making CN selections that minimize the entropy of the cycle distributions.
 Algorithm \ref{alg:EC-PEG} presents the EC-PEG algorithm
 for constructing a TG $\tGEC$ with $n$ VNs, $m$ CNs, and VN degree $d_v$  that concentrates distributions $\td^{g'}$, $\forall g'< \ttg$. Choice of $\ttg$ is a complexity constraint
 of how 
 many cycles we keep track in the algorithm.
 All ties in the algorithm are broken randomly.}

  The PEG algorithm builds a TG by iterating over the set of VNs and for each VN $v_j$ in the
  TG, establishing $d_{v}$ edges to it.   For establishing the $k^{th}$ edge to VN $v_j$, the PEG  algorithm encounters two situations: i) addition of the edge is possible without creating cycles; ii) addition of the edge creates cycles.
  In both situations, the PEG algorithm finds a set of \emph{candidate} CNs that it proposes to connect to $v_j$, to maximize the girth. We abstract out the steps followed in \cite{PEG} to find the set of \emph{candidate CNs} by a procedure {\fontfamily{qcr}\selectfont PEG}$(\tGEC,v_j)$. The procedure returns the set of \emph{candidate} CNs $\tK$
  for establishing a new edge to VN $v_j$ under the TG setting $\tGEC$ according to the PEG algorithm in \cite{PEG}.
  \trim{For ii), the procedure returns the cycle length $g$ of the smallest cycles formed when an edge is added between any CN in $\tK$ and $v_j$. For i), it returns $g = \infty$. $\tK$ is the set of all CNs in $\tGEC$ that create new $g$-cycles when an edge is added between the CN and $v_j$. 
    When $g = \infty$, $\tK$ is the set of all CNs in $\tGEC$ that if connected to $v_j$ create no cycles.}

    \vspace{-0.1cm}
  \begin{algorithm*}
  {
\caption{EC-PEG Algorithm}\label{alg:EC-PEG}
\begin{algorithmic}[1]
\State \textbf{Inputs:} $n$, $m$ , $d_v$, $\ttg$, \textbf{Outputs:} $\tGEC$, $g_{\min}$, \textbf{Initialize} $\tGEC$ to $n$ VNs, $m$ CNs and no edges
\State \textbf{Initialize} $\ttc^{(g')}_i = 0$, for all $g'  < \ttg$ and $1\leq i \leq n$, $T = |\{4, 6, \ldots, \ttg-2\}|$
\For{$j=1$ to $n$}
    \For{$k=1$ to $d_{v}$}
    \State [$\tK , g$] = {\fontfamily{qcr}\selectfont \text{PEG}}$(\tGEC, v_j)$
        \If{ $g \geq \ttg$} 
        \State
    $\tcn^{sel}$ = Select a CN from $\tK$ with the minimum degree under the current TG setting $\tGEC$
        \Else\Comment{\textit{(\bluetext{$g$-cycles}, $g < \ttg$, are created)}}
        \For{each $c$ in $\tK$}
        \State $\ttcsmall^{(g',c)}_i = \ttc^{(g')}_i$, $g' < \ttg$, $1 \leq i \leq n$
        \State $\mathcal{L}_{cycles} =$ new $g$-cycles formed in $\tGEC$ due to the addition of edge between $c$ and $v_j$ 
        \For{all $v$ in $\tGEC$}
         $\ttcsmall^{(g,c)}_v$ = $\ttcsmall^{(g,c)}_v$ + $\vert \{\mathcal{O} \in \mathcal{L}_{cycles}\; |\; v \text{ is part of } \mathcal{O} \} \vert$
        \EndFor
        \State  $\alpha^{(g')} = (\alpha^{(g')}_1, \alpha^{(g')}_2, \ldots, \alpha^{(g')}_n)$, where $\alpha^{(g')}_i = \frac{\ttcsmall^{(g',c)}_i}{\sum_{i=1}^{n}\ttcsmall^{(g',c)}_i}$, $g' < \ttg$ (define $\frac{0}{0} = 0$)
        \State $\talpha = (\sum_{g'<\ttg}\frac{\alpha^{(g')}_1}{T}, \sum_{g'<\ttg}\frac{\alpha^{(g')}_2}{T}, \ldots, \sum_{g'<\ttg}\frac{\alpha^{(g')}_n}{T})$; $\mathrm{Entropy}[c]  = \mathcal{H}(\talpha)$ 
        \EndFor
        
        \State
            $\tcn^{sel}$ = CN in $\tK$ with minimum $\mathrm{Entropy}[c]$;\; 
         $\ttc^{g}_i = \ttcsmall^{(g, \tcn^{sel})}_i$,  $1 \leq i \leq n$
        \EndIf
    \State  $\tGEC = \tGEC \cup \mathrm{edge}\{\tcn^{sel},v_j\}$

    \EndFor
\EndFor
\end{algorithmic}
}
\end{algorithm*}

Thus, when the {\fontfamily{qcr}\selectfont PEG}$(\tGEC,v_j)$ procedure returns $g \geq \ttg$, either no new cycles are created or the
cycles created have length $\geq \ttg$. In both these situations, similar to the original PEG algorithm in \cite{PEG}, we select a CN from $\tK$ with the minimum degree under the current TG setting $\tGEC$ (line 7). When {\fontfamily{qcr}\selectfont PEG}$(\tGEC,v_j)$ returns $g < \ttg$, we modify the CN selection procedure so that the resultant
cycle distributions get concentrated. We explain the modified CN selection procedure next. 

While progressing through the EC-PEG algorithm, for all $g'$-cycles, $g' < \ttg$, we maintain
\textit{VN-to-$g'$-cycle} counts $\ttc^{(g')} = (\ttc^{(g')}_1, \ttc^{(g')}_2, \ldots, \ttc^{(g')}_n)$, where $\ttc^{(g')}_i$ is the number of \bluetext{$g'$-cycles} that are touched by VN $v_i$. When the {\fontfamily{qcr}\selectfont PEG}$(\tGEC,v_j)$ procedure returns $g < \ttg$, for each candidate CN $\tcn \in \tK$, new $g$-cycles are formed in the TG when an edge is established between $\tcn$ and $v_j$. These cycles are listed in $\mathcal{L}_{cycles}$ (line 11). 
\trim{For these new $g$-cycles, we calculate the resultant \emph{VN-to-$g$-cycle} counts $\ttcsmall^{(g,c)}_i$, $1 \leq i \leq n$, if an edge is established between $\tcn$ and $v_j$ (line 12).}
Using $\ttcsmall^{(g,c)}_i$, we calculate the  \textit{VN-to-$g'$-cycle} normalized counts $\alpha^{g'} = (\alpha^{g'}_1, \alpha^{g'}_2, \ldots, \alpha^{g'}_n)$ (line 13) and then the joint normalized cycle counts $\talpha$ for $g'$-cycles, $g' < \ttg$ (line 14). The joint normalized cycle counts $\talpha$ is simply the average of the normalized cycle counts across all the cycle lengths.
Using $\talpha$, we calculate the entropy $\mathcal{H}(\talpha)$ for each CN $c$ in $\tK$ (line 14). Our modified CN selection procedure is to select a CN from $\tK$ with minimum $\mathrm{Entropy[]}$ (line 15). We then update the \emph{VN-to-$g$-cycle} counts for the new $g$-cycles that get created (line 15) to be used in future iterations. Minimizing the entropy of the joint normalized cycle counts ensures that the different cycle distributions are concentrated towards the same set of VNs.

\begin{figure*}[t]
    \centering
    \begin{subfigure}{0.33\linewidth}
\begin{minipage}{0.99\linewidth}
\begin{tikzpicture}
  \node (img) {\includegraphics[scale=0.38]{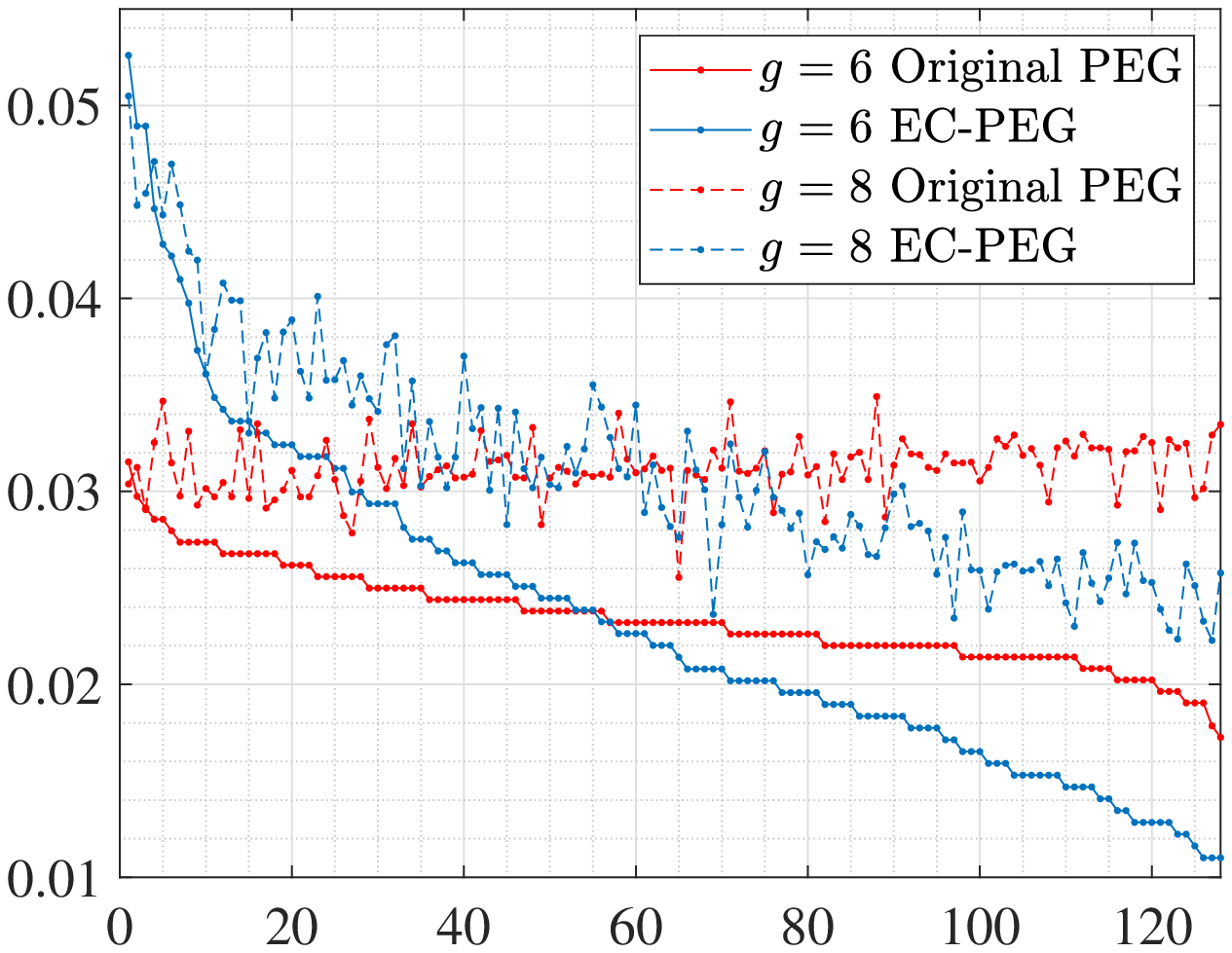}};
  \node[below=of img, node distance=0cm, yshift=1.3cm,font=\color{black}] {\small VN index};
  \node[left=of img, node distance=0cm, rotate=90, anchor=center,yshift=-1.1cm,font=\color{black}] {\footnotesize{$\td^g$}};
 \end{tikzpicture}
 \end{minipage}
\label{fig:cycle_dist}
    \end{subfigure}%
    \begin{subfigure}{0.33\linewidth}
\begin{minipage}{0.99\linewidth}
\begin{tikzpicture}
  \node (img) {\includegraphics[scale=0.38]{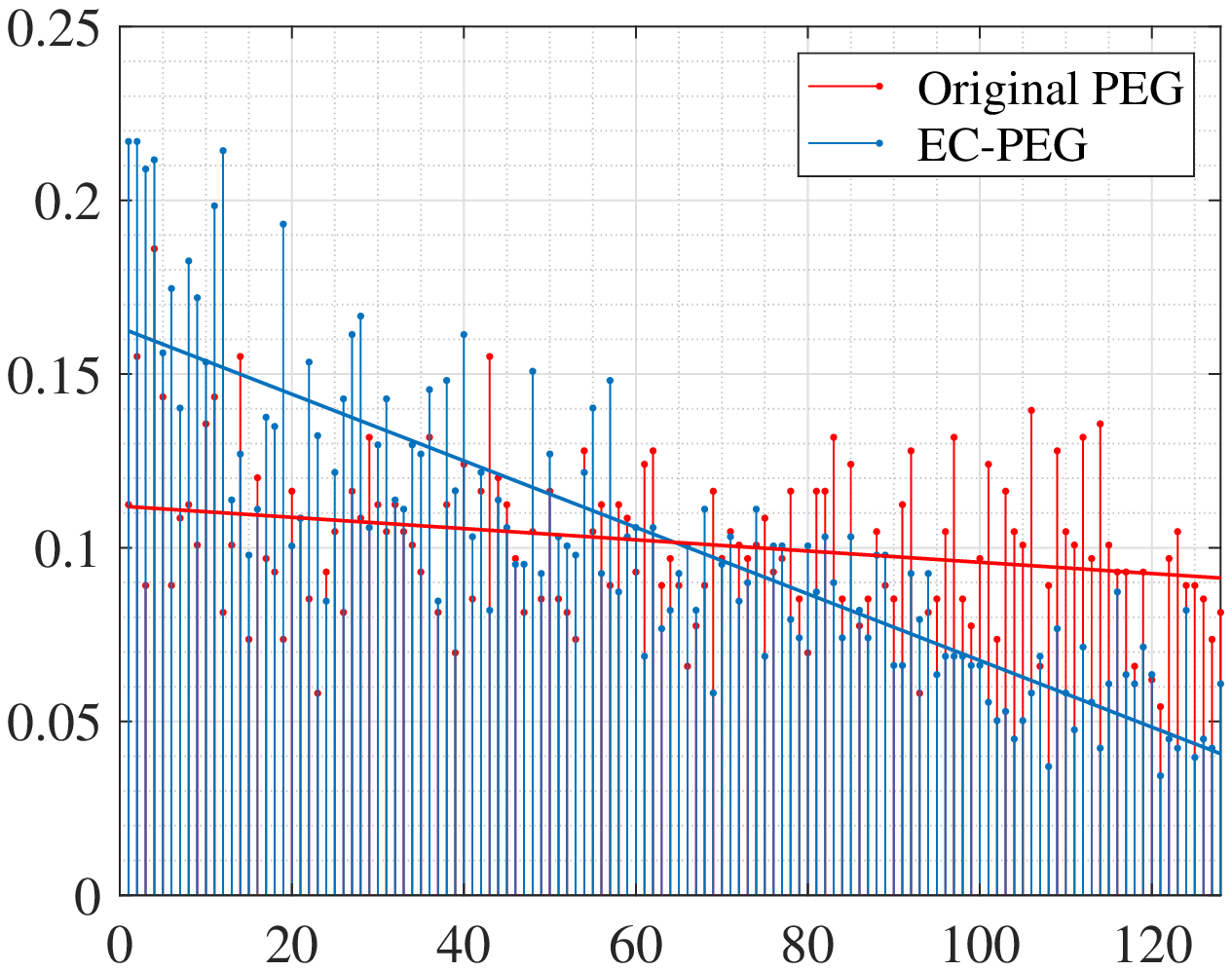}};
  \node[below=of img, node distance=0cm, yshift=1.3cm,font=\color{black}] {\small VN index};
  \node[left=of img, node distance=0cm, rotate=90, anchor=center,yshift=-1.15cm,font=\color{black}] {$ss^{13}$};
 \end{tikzpicture}
 \end{minipage}
    \end{subfigure}%
\begin{subfigure}{0.33\linewidth}
\begin{minipage}{0.99\linewidth}
\begin{tikzpicture}
  \node (img) {\includegraphics[scale=0.38]{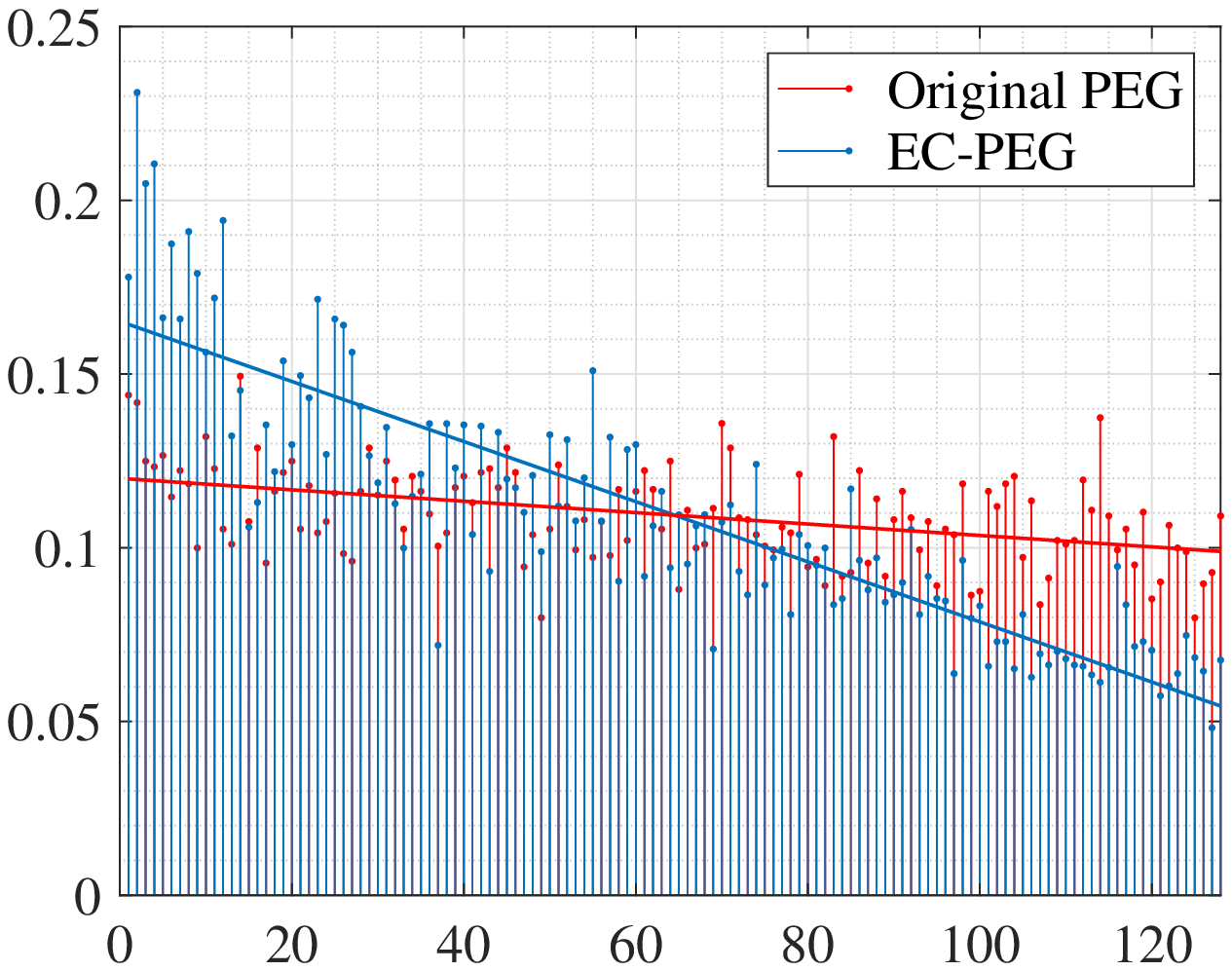}};
  \node[below=of img, node distance=0cm, yshift=1.3cm,font=\color{black}] {\small VN index};
  \node[left=of img, node distance=0cm, rotate=90, anchor=center,yshift=-1.15cm,font=\color{black}] {$ss^{14}$};
 \end{tikzpicture}
 \end{minipage}
\end{subfigure}
     \vspace{-2pt}
    \caption{\revone{Results for LDPC codes with $\tr=0.5$, $d_v = 4$, $n = 128$ using  different  PEG  algorithms. The x-axis in all the plots are the VN indices $v_i$ in the decreasing order of the 6-cycle fractions $\td^6_i$ (for the respective codes); Left panel:
    cycle distributions $\td^6$ and $\td^8$;
    Middle Panel: stopping set distribution $ss^{13}$; Right Panel: stopping set distribution $ss^{14}$. The lines in the middle and right panels are the best fit lines for $ss^{\teta}$  indicating the graph slope.}}
    \label{fig:EC-PEG-distribution}
     \vspace{-10pt}
\end{figure*}

\editor{We now mention the complexity of the EC-PEG algorithm. Note that the complexity of the original PEG algorithm is $O(mn)$ \cite{PEG}. The EC-PEG algorithm differs from the original PEG algorithm in steps 8-15. Step 14 has the largest complexity which results in the complexity of the EC-PEG algorithm to be at most $O(mn^2) = O(n^3)$. Note that $\mathcal{L}_{cycles}$ in step 11 is obtained during step 5 as a by-product and does not incur additional complexity.}

Fig. \ref{fig:EC-PEG-distribution} demonstrates the effectiveness of the EC-PEG algorithm in concentrating the stopping set distribution. In Fig. \ref{fig:EC-PEG-distribution} left panel, we plot the cycle distributions generated by the PEG and EC-PEG algorithms. From the figure, we see that the EC-PEG algorithm
generates significantly concentrated distributions $\td^6$ and $\td^8$ compared to the original PEG algorithm. \trim{
Fig. \ref{fig:EC-PEG-distribution} middle and right panels show the corresponding stopping set distributions $ss^{\teta}$.
We see that for the EC-PEG algorithm, the VNs towards the left (right) on the x-axis have high (low) stopping set fraction. Thus, concentrating the cycle distributions concentrates the stopping set distributions towards the same set of VNs as the cycles. In Section \ref{sec:simulations}, \deb{we demonstrate that such concentrated distributions result in a low probability of failure using the greedy sampling strategy in Algorithm \ref{alg:LN-greedysamp}.} }

\vspace{-0.2cm}
\section{LDPC code and sampling co-design for Medium and Strong Adversary}\label{sec:LC-PEG}
\vspace{-0.1cm}
\deb{
\trim{For the medium and strong adversary, the EC-PEG algorithm and greedy sampling is insufficient to secure the system and requires stronger code and sampling design. In this section, we focus on overcoming these stronger adversaries}
that hide the worst case stopping set.}
Similar to Section \ref{sec:EC-PEG}, we first look at a medium and a strong adversary who conduct a DA attack on the base layer of the CMT and propose a sampling strategy for the light nodes to sample the base layer to minimize the probability of failure. This will motivate the construction of LDPC codes for the base layer.
Finally, we will generalize the sampling strategy and LDPC construction for the situation when the adversary conducts a DA attack at any layer of the CMT.   

Recall that for each layer $j$, $1 \leq j \leq l$, the medium adversary hides stopping sets of $H_j$ of
size $< \mu_j$. Let $\tsj_j = \{\tSall^{(j)}_1, \tSall^{(j)}_2, \ldots, \tSall^{(j)}_{\vert\tsj_j\vert}\}$ be the set of all stopping sets of $H_j$ of size $< \mu_j$, $1 \leq j \leq l$. For $\tsj_j$, let $\tL^{(j)}$ denote the \emph{VN-to-stopping-set} adjacency matrix of size $\vert \tsj_j \vert \times n_j$,
where $\tL^{(j)}_{ki} = 1$ iff $v^{(j)}_{i}$ touches  stopping set $\tSall^{(j)}_k$, else $\tL^{(j)}_{ki} = 0$, $1 \leq i \leq n$, $1 \leq k \leq \vert \tsj_j\vert $.

\vspace{-0.1cm}
\begin{definition}
A sampling (with replacement) strategy $\left(\tx\;,\;\beta^{(l)}\right)$ is a $n_l \times 1$ vector 
$\tx = \left[\tx_1\;\cdots\; \tx_{n_l}\right]^T$,
where $\tx_i$ is the probability that a light node requests for the $i^{th}$ base layer symbol \editor{(i.e., $v^{(l)}_i$)} for every sample request and $\beta^{(l)}$ controls the minimum probability of requesting a given CMT base layer symbol. $\left(\tx\;,\;\beta^{(l)}\right)$ satisfy $0\leq \beta^{(l)} \leq \tx_i \leq 1$, $\sum_{i=1}^{n}\tx_i = 1$.
\end{definition}

\vspace{-0.1cm}
\revthree{Let $P^{(l)}_{(f,med)}(s)$ ($P^{(l)}_{(f,str)}(s)$) be the probability of failure against a medium (strong) adversary for a DA attack on the base layer of the CMT. (Define similarly $P^{(j)}_{(f,med)}(s)$ and $P^{(j)}_{(f,str)}(s)$ for DA attack on layer $j$). 
We have the following lemma (all proofs are deferred to the Appendix). }

\vspace{-0.1cm}
\revthree{\begin{lemma}\label{eqn:p_f_str}
 For a sampling strategy $\left(\tx\;,\;\beta^{(l)}\right)$, $P^{(l)}_{(f,med)}(s) = \left[\max(1 - \tL^{(l)}\tx)\right]^s$. Define $P^{(l)}_{(f,str\text{-}bnd)}(s):= \left(1 - \beta^{(l)}\mu_l\right)^s$.
 Then, $P^{(l)}_{(f,str)}(s) \leq \max\left(P^{(l)}_{(f,med)}(s),P^{(l)}_{(f,str\text{-}bnd)}(s)\right)$.
\end{lemma}}

\vspace*{-0.1cm}

In the rest of the paper, we assume that $P^{(l)}_{(f,str)}(s)$ is equal to the upper bound provided in
Lemma \ref{eqn:p_f_str}.
We find the light node sampling strategy by formulating a linear program (LP) in $\left(\tx\;,\;\beta^{(l)}\right)$ to minimize the probabilities in Lemma \ref{eqn:p_f_str}.
The optimization problem (which can be
easily converted into an LP by introducing additional variables) is provided below:

\vspace{-0.1cm}
\begingroup
\allowdisplaybreaks
\begin{align}\label{eqn:LP_base}
& \underset{\tx\;,\;\beta^{(l)}}{\text{minimize}}
& & \max\left( \max(1 - \tL^{(l)}\tx), \tnu\times\left[1 - \beta^{(l)}\mu_l\right]\right) \\
& \text{subject to}
& & \beta^{(l)}\leq \tx_i \leq 1,\; i = 1, \ldots, n_l;\;  \beta^{(l)} \geq 0;\; \sum_{i=1}^{n_l}\tx_i = 1,\nonumber
\end{align}%
\endgroup
\trim{where $\tnu$, $0 \leq \tnu \leq 1$, is a parameter that controls the trade-off between $P^{(l)}_{(f,med)}(s)$ and $P^{(l)}_{(f,str)}(s)$.}

\vspace{-0.3cm}
\subsection{Linear-programming-sampling (LP-sampling) for DA attacks on any layer of the CMT}\label{sec:LP-sampling-anylayer}
\vspace{-0.1cm}

\trim{In this subsection, we modify LP \eqref{eqn:LP_base} to take into effect a DA attack conducted on any layer
  of the CMT and derive the sampling strategy based on the modified LP. We first align the columns of the parity check matrices of all the CMT layers as described in Section \ref{sec:colm_align}. Assume that the stopping sets and VNs in the following are based on the aligned parity check matrices.}

Since a base layer symbol samples, via its Merkle proof, two symbols from every intermediate layer of the CMT, the events of sampling intermediate layer symbols are not disjoint.
To calculate the probability that each intermediate layer symbol is sampled, we define for each $j$,  $1 \leq j \leq l-1$, a matrix $\tA^{(j)}$ of size $n_j \times n_l$ whose entries are as follows: 1)   if ($k \leq s_j$ and $1+ (i-1)_{s_j}=k$) then $\tA^{(j)}_{ki} =1$; 2) if ($k > s_j$ and $1+ s_j+ (i-1)_{p_j}=k$) then $\tA^{(j)}_{ki} =1$;
3) $\tA^{(j)}_{ki} =0$ for all other cases. For simplicity, assume that $\tA^{(l)}$ is an $n_l \times n_l$ identity matrix. Also, define for $1 \leq j \leq l$, the matrices $\piA^{(j)} = \min(\tL^{(j)}\tA^{(j)},1)$ where the minimum is element wise. \trim{Using the above matrices,  we calculate $P^{(j)}_{(f,med)}(s)$ and $P^{(j)}_{(f,str)}(s)$ in Lemma \ref{lemma:p_f_med_str_any_layer}. 
First, consider the following definition.}

\vspace{-0.1cm}
\begin{definition}
A sampling (with replacement) strategy $\left(\tx\;,\;\beta^{(1)}\;,\;\beta^{(2)}\;,\ldots,\;\beta^{(l)}\right)$ is a sampling
strategy $\left(\tx\;,\;\beta^{(l)}\right)$, such that for $\tx^{(j)} = \tA^{(j)}\tx$, $1 \leq j \leq l-1$, $\tx^{(j)}_i$'s satisfy $\tx^{(j)}_i \geq \beta^{(j)}$, $1 \leq i \leq n_j$, $1 \leq j \leq l-1$.
Parameter $\beta^{(j)}$, $1 \leq j \leq l$, is a non-negative real number and controls the minimum
 probability of requesting a given symbol from layer $j$ of the CMT. 
\end{definition}

\vspace{-0.25cm}
\revthree{\begin{lemma}\label{lemma:p_f_med_str_any_layer}
For a sampling strategy $\left(\tx\;,\;\beta^{(1)}\;,\ldots,\;\beta^{(l)}\right)$, let $\tx^{(j)} = \tA^{(j)}\tx$, $1 \leq j \leq l$. $\tx^{(j)}_k$  is the probability that $v^{(j)}_k$ is sampled and
$P^{(j)}_{(f,med)}(s) = \left[\max(1 - \piA^{(j)}\tx)\right]^s$. Also, for $1 \leq j < l$, let $P^{(j)}_{(f,str\text{-}bnd)}(s) := \left(1 - \frac{1}{2}\beta^{(j)}\mu_j\right)^s$. Then, $P^{(j)}_{(f,str)}(s) \leq \max(P^{(j)}_{(f,str\text{-}bnd)}(s),  P^{(j)}_{(f,med)}(s))$. 
\end{lemma}}

\vspace{-0.1cm}
 
 \trim{Using Lemmas \ref{eqn:p_f_str} and \ref{lemma:p_f_med_str_any_layer}, we formulate the following LP to find the light node sampling strategy:}

\vspace{-0.25cm}
\begin{mini!}<b>
{{\small \tx,\beta^{(1)},\ldots,\beta^{(l)}}}
{\max\left(\max_{1\leq j\leq l}\max(1 - \piA^{(j)}\tx), \right.\label{objective}}
{\label{eqn:LP_full}}
{}
\breakObjective{\max_{1 \leq j \leq l}\tnu^{(j)} \times \left. \left[1 - \xi^{(j)}\beta^{(j)}\mu_j\right] \right)}
\addConstraint{ \beta^{(l)}\leq \tx_i\leq 1, \; i = 1, \ldots, n_l,\quad\sum_{i=1}^{n_l}\tx_i = 1}{}
\addConstraint{\beta^{(j)}\leq \min(\tA^{(j)}\tx) ,\; j = 1, \ldots, l-1 }{}
\addConstraint{\beta^{(j)} \geq 0,\; j = 1, \ldots, l,}{}
\end{mini!}

{\raggedright where $\xi^{(j)} = \frac{1}{2}$ for $1 \leq j < l$ and $\xi^{(l)} = 1$.
The first and second term in the outer maximum}
above corresponds to the probability of failure against the medium and strong adversary for a DA attack on different layers of the CMT. $\tnu^{(j)}$'s are trade-off parameters and control the
importance  given to a strong adversary on layer $j$ of the CMT compared to a medium adversary.

\revthree{The sampling strategy $\left(\tx\;,\;\beta^{(1)},\ldots,\beta^{(l)}\right)$ obtained as the optimal solution of LP \eqref{eqn:LP_full} is called LP-sampling and is included in the protocol.}
\trim{To reduce the probability of failure against a medium and a strong adversary under LP-sampling, we design LDPC codes aimed towards minimizing the probability for each layer.} \editor{The complexity of LP-sampling is determined by the complexity of finding all stopping sets of $H_j$ of size $< \mu_j$. Although stopping set enumeration is NP-hard, they can be found in a reasonable time for small code lengths using an ILP \cite{ILPsearch}. However, it is difficult to obtain an analytical complexity expression for stopping set enumeration using ILP.
}
\vspace{-0.5cm}\subsection{Linear-programming-Constrained PEG (LC-PEG) Algorithm}\label{subsec:LC-PEG}
\vspace{-0.1cm}
In this section, we design LDPC codes that perform well under LP-sampling. \trim{
We design such codes 
by modifying the CN selection procedure in the PEG algorithm. We call our construction} \emph{linear-programming-constrained} PEG or LC-PEG algorithm since it is trying to minimize the optimal objective value of an LP. Codes designed in this section are aligned by Algorithm 2 and 
then included in the protocol.
Similar to the EC-PEG algorithm, we optimize cycles instead of stopping sets. The motivation for focusing on cycles is the following: for lists $\mathcal{C}$ and $\tsj$ of cycles and stopping sets, respectively, such that for every $\psi \in \tsj$, there exists a $\mathcal{O} \in \mathcal{C}$ which is part of $\psi$, we have
$\max_{\psi \in \tsj}\left(1 - \sum_{v_i : v_i \in \psi}\tx_i\right) \leq \max_{\mathcal{O} \in \mathcal{C}}\left(1 - \sum_{v_i : v_i \in \mathcal{O}}\tx_i\right)$. Thus, the optimal objective value of LP \eqref{eqn:LP_base} can be upper bounded by the optimal objective value of a modified version of LP \eqref{eqn:LP_base} which is based on cycles. We select CNs in the PEG algorithm depending on the optimal objective value they produce on the modified LP. Algorithm \ref{alg:LC-PEG} presents our LC-PEG algorithm
 for constructing a TG $\tGEC$ with $n$ VNs, $m$ CNs, and VN degree $d_v$. \trim{All ties are broken randomly.}

  \begin{algorithm*}
    {
\caption{LC-PEG Algorithm}\label{alg:LC-PEG}
\begin{algorithmic}[1]
\State \textbf{Inputs:} $n$, $m$ , $d_v$, $\ttg$, $T_{th}$, $\hat{\theta}$, $\hat{\mu}$; \textbf{Outputs:} $\tGEC$, $g_{\min}$
\State \textbf{Initialize} $\tGEC$ to $n$ VNs, $m$ CNs and no edges, $\mathcal{L} = \emptyset$
\For{$j=1$ to $n$}
    \For{$k=1$ to $d_{v}$}
    \State [$\tK , g$] = {\fontfamily{qcr}\selectfont \text{PEG}}$(\tGEC, v_j)$;\; $\tK_{mindeg}$ = CNs in $\tK$ with the minimum degree under the TG setting $\tGEC$
        \If{ $g \geq \ttg$} 
    $\tcn^{sel}$ = Select a CN randomly from $\tK_{mindeg}$
        \Else\Comment{\textit{(\bluetext{$g$-cycles}, $g < \ttg$, are created)}}
        \State $\tK_{mincycles}$ = CNs in $\tK_{mindeg}$ that result in the minimum number of new $g$-cycles due to the addition of edge \hspace*{3cm} between the CN and $v_j$
        \For{each $c$ in $\tK_{mincycles}$}
        \State $\mathcal{L}^{\tcn}_{cycles} =$ new $g$-cycles formed in $\tGEC$ due to the addition of edge between $c$ and $v_j$
        \State $\mathrm{cost}[\tcn] =$ {\fontfamily{qcr}\selectfont \text{LP-objective}}$(\mathcal{L}\cup\mathcal{L}^{\tcn}_{cycles}, \tGEC)$
        \EndFor
        
        \State
            $\tcn^{sel}$ = CN in $\tK_{mincycles}$ with minimum $\mathrm{cost}[\tcn]$ 
            \State $\mathcal{L}^{sel} = $ cycles in $\mathcal{L}^{\tcn^{sel}}_{cycles}$ that have EMD $\leq T_{th}$;\;
           $\mathcal{L} = \mathcal{L} \cup \mathcal{L}^{sel}$
        \EndIf
    \State  $\tGEC = \tGEC \cup \mathrm{edge}\{\tcn^{sel},v_j\}$

    \EndFor
\EndFor
\end{algorithmic}
}
\end{algorithm*}

 \deb{In the LC-PEG algorithm, we use the concept of the extrinsic message
degree (EMD) of a
set of VNs that  allows us to rank the harm a cycle may have in creating stopping sets. EMD of a set of VNs is  the number of CN neighbors singly connected to the set \cite{WesselSS} and is calculated using the method in \cite{EMDcalculation}. EMD of a cycle is the EMD of the VNs involved in the cycle. Low EMD cycles are more likely to form stopping sets and we term cycles with EMD below a threshold
$T_{th}$ as \emph{bad cycles}. We use bad cycles to form the modified linear program below:}
\vspace{-0.05cm}
\begin{equation}\label{eqn:LP_LC_PEG}
\begin{aligned}
& \underset{\hat{\tx},\;\hat{\beta}}{\text{min}}
\; \max\left(\max(1 - \tC\hat{\tx}), \hat{\theta}[1 - \hat{\beta}\hat{\mu}]\right) 
\;\\& \text{s.t.}\;
 \hat{\beta}\leq \hat{\tx}_i \leq 1,\;i = 1, \ldots, \hat{n};\; \hat{\beta} \geq 0;\;\sum_{i=1}^{\hat{n}}\hat{\tx}_i = 1.
\end{aligned}
\vspace{-0.05cm}
\end{equation}

The LC-PEG algorithm uses LP \eqref{eqn:LP_LC_PEG} via the procedure  {\fontfamily{qcr}\selectfont \text{LP-objective}}$(\widehat{\mathcal{L}}, \widehat{\mathcal{G}})$ which outputs its
optimal objective value. The procedure has inputs of a list $\widehat{\mathcal{L}} = \{\tO_1, \ldots, \tO_{\vert \widehat{\mathcal{L}}\vert} \}$ of cycles and a TG $\widehat{\mathcal{G}}$.
Let $\widehat{\mathcal{G}}$ have $\hat{n}$ VNs $\{\hat{v}_1, \ldots, \hat{v}_{\hat{n}}\}$. Here, $\tC$ is a matrix of size $\vert \widehat{\mathcal{L}}  \vert \times \hat{n}$, such that $\tC_{ki} = 1$ if $\hat{v}_{i}$ touches  $\tO_k$, else $\tC_{ki} = 0$, $1 \leq i \leq \hat{n}$, $1 \leq k \leq \vert \widehat{\mathcal{L}}\vert $. Also, $\hat{\theta}$, $0 \leq \hat{\theta}\leq 1$, is a parameter. 

\vspace{-0.01cm}
In the LC-PEG algorithm, we use the procedure {\fontfamily{qcr}\selectfont \text{PEG}}$()$ defined in Section \ref{sec:alg:EC-PEG} for the EC-
PEG
algorithm. 
The LC-PEG algorithm proceeds exactly as the EC-PEG algorithm when the {\fontfamily{qcr}\selectfont \text{PEG}}$()$ procedure returns cycle length $g \geq \ttg$.
When the {\fontfamily{qcr}\selectfont \text{PEG}}$()$ procedure returns cycle length $g < \ttg$, we select a CN from the set of \emph{candidate} CNs $\tK$ such that the resultant LDPC codes have a low optimal objective value of LP \eqref{eqn:LP_base}. We explain the CN selection procedure next.

While progressing through the LC-PEG algorithm, we maintain a list $\mathcal{L}$ of cycles. $\mathcal{L}$  contains cycles of length $g < \ttg$ that had EMD less than or equal to threshold $T_{th}$ when they were formed. Cycles in  $\mathcal{L}$ are considered \emph{bad} cycles and we base our CN selection procedure on these cycles. When the {\fontfamily{qcr}\selectfont \text{PEG}}$()$ procedure returns candidate CNs $\tK$, we first select the set of CNs $\tK_{mindeg}$ that have the minimum degree under the current TG setting $\tGEC$ (line 5). Of the CNs in $\tK_{mindeg}$, we select the set of CNs $\tK_{mincycles}$ that form the minimum number of new $g$-cycles if an edge is established between the CN and $v_j$ (line 8). Now for every CN $\tcn$ in $\tK_{mincycles}$, we find the list $\mathcal{L}^{\tcn}_{cycles}$ of new $g$-cycles formed due to the addition of an edge between $\tcn$ and $v_j$ (line 10) and compute {\fontfamily{qcr}\selectfont \text{LP-objective}}$(\mathcal{L}\cup\mathcal{L}^{\tcn}_{cycles}, \tGEC)$ to get $\mathrm{cost}[\tcn]$ (line 11). Our modified CN selection procedure is to select a CN in  $\tK_{mincycles}$ that has the minimum $\mathrm{cost}[c]$ (line 12). After selecting $\tcn^{sel}$ using the above criteria, we update $\mathcal{L}$ as follows: let $\mathcal{L}^{sel}$ be the list of $g$-cycles in $\mathcal{L}^{\tcn^{sel}}_{cycles}$ that have EMD $\leq T_{th}$. We add $\mathcal{L}^{sel}$ to $\mathcal{L}$ (line 13). Finally, we update the TG $\tGEC$ (line 14).

\vspace{-0.1cm}
\begin{remark}\label{remark:MC_PEG}
We empirically observed that reducing the number of cycles in the TG (and hence the number of stopping sets) reduces the probability of failure against the medium and strong adversary when LP-sampling is employed. The above holds even if the size of the smallest stopping set remains unchanged. This is in contrast to random sampling where the probability of failure only depends on the size of the smallest stopping set and is agnostic to the number of stopping sets of small size present in the code. Thus, based on this observation, we have added line 8 in our LC-PEG algorithm which selects CNs $\tK_{mincycles}$ that form the minimum number of cycles when a new edge is established. However, we further make an informed choice among the CNs in $\tK_{mincycles}$ to select a CN that has the minimum optimal objective value of LP \eqref{eqn:LP_LC_PEG}. 
\end{remark}
\vspace{-0.1cm}
\editor{ 
We now discuss the complexity of the LC-PEG algorithm. Note that it differs from the original PEG algorithm (that has complexity $O(mn)$ \cite{PEG}) in steps 7-13. Of these steps, step 11 has the largest complexity due to solving LP \eqref{eqn:LP_LC_PEG}. An LP $\min_{Az \leq b}c^Tz$ with $d$ variables and $t$ constraints can be solved with complexity $\tilde{O}((nnz(A) + d^2)\sqrt d)$ \cite{LPcomplexity} where $nnz(A)$ is the number of non-zero entries in $A$ and $\tilde{O}$ hides factors poly-logarithmic in $d$ and  $t$. In our case, LP \eqref{eqn:LP_LC_PEG} has $n$ variables and at most $mnd_v$ constraints (step 10 in the algorithm can result in at most $m$ cycles) and hence $nnz(A) \leq \ttg mnd_v$. Thus, the overall complexity of the LC-PEG algorithm is at most $\tilde{O}(mn \sqrt{n}(\ttg mnd_v + n^2)) = \tilde{O}(n^{4.5})$. 
}\revonethree{In our simulations, we were able to generate codes up to length 500 for different rates in a reasonable time frame (within a day) using the LC-PEG algorithm.}
\revfourtwo{
Note that the algorithms proposed in this paper for LDPC code construction and sampling strategy design have more complexity compared to \cite{CMT}. However, these algorithms are used offline instead of on-the-fly. The  complexity increase is still tractable for short code lengths. We demonstrate improvement in the probability of failure  using our algorithms in Section \ref{sec:simulations}.}

\vspace{-0.3cm}
\editor{\section{System Aspects}\label{sec:system_aspects}}
\vspace{-0.1cm}

\editor{\subsubsection{Security Performance} Here, we discuss how \emph{soundness} and \emph{agreement} defined in Section
\ref{sec:adv_models}
are affected by our co-design. Let $\tm$ be the total number of light nodes in the system and $\eta_{rec} = \left(\underset{1\leq j\leq l}{\max} \frac{n_j-\omega^{(j)}_{\min} + 1}{n_j}\right)$, \revfourtwo{where $\omega^{(j)}_{\min}$ is the minimum stopping set size of the LDPC code used in layer $j$ of the CMT.} We have the following lemmas (we defer the proofs to 

\noindent
the Appendix).

\vspace{-0.2cm}
\begin{lemma}\label{lemma:soundess_weak}
\revonethree{For a weak adversary, when light nodes sample according to the overall greedy
sampling strategy, the probability of soundness or agreement failure per light client $\tpfsa$ satisfies\vspace{0.08cm}}
\begin{align*}
    &\tpfsa \leq\\& \max\bigg(\max_{1 \leq j \leq l,\; \omega^{(j)} < \mu_j}\left[[1 - \tau(S^{(\tlambda s \;,\;j)}_{greedy},\omega^{(j)})]\big(1 - \frac{\omega^{(j)}}{n_j}\big)^{s - \tlambda s}\right],\\ &\hspace{3.5cm}2^{[\tH(\eta_{rec},1-\eta_{rec})n_l - \tm s(1-\tlambda) \log(\frac{1}{\eta_{rec} })]}\bigg)
\end{align*}
\noindent
Here, $S^{(\tlambda s \;,\;j)}_{greedy}$ is the samples of layer $j$, $1 \leq j \leq l$, collected  when the light nodes request for
the first $\tlambda s$ coded symbols from the base layer of the CMT.
\end{lemma}

\vspace{-0.2cm}
\begin{lemma}\label{lemma:soundness_med_strong}
\revonethree{For a medium and a strong adversary, when light nodes sample according to LP-sampling $\tx$, the probability of soundness or agreement failure per light client $\tpfsa$ satisfies\vspace{-0.02cm}}
\begin{align*}
    \tpfsa \leq \max\bigg( \max_{1\leq j\leq l}&P^{(j)}_f(s) ,\;\\ &2^{[\tH(\eta_{rec},1-\eta_{rec})n_l - \tm s \log\big(\frac{1}{\sum_{i=1}^{\eta_{rec}n_l}\tx_{[i]} }\big)]}\bigg)
\end{align*}
Here, $P^{(j)}_f(s) = P^{(j)}_{(f,med)}(s)$ and $P^{(j)}_f(s) = P^{(j)}_{(f,str)}(s)$ for the medium and strong adversary,
respectively, as defined in Section \ref{sec:LP-sampling-anylayer} and $x_{[i]}$ is the $i^{th}$ largest entry in vector $\tx$. 

\end{lemma}

\vspace{-0.1cm}
The first term in the maximum in Lemma \ref{lemma:soundess_weak} and \ref{lemma:soundness_med_strong} is the probability of failure of a single light
node against different adversaries. Thus, when the number of light nodes $M$ is large, $\tpfsa$
is affected by the probability of failure of a single light node, which we minimize in this paper. }

\vspace{-0.01cm}
\subsubsection{Blockchain System Designer}\label{sec:subsub:malicious_designer} In the system model in Section \ref{sec:network_nodes}, we have made a trusted
set up assumption of a
blockchain system designer who designs the parity checks matrices and the LP-sampling strategy. Note that for greedy sampling, after the overall sampling rule described in Remark \ref{remark:systematic_generator}, nothing more needs to be designed by the system designer.
Additionally, as previously mentioned in Section \ref{sec:adv_models}, only the final LP-sampling strategy obtained by solving LP \eqref{eqn:LP_full} is included in the protocol and inputs to LP \eqref{eqn:LP_full} ($\mu_j$ and set of all stopping sets of $H_j$ of size $< \mu_j$) are not part of the protocol. Existing examples of blockchain systems that rely on trusted set up assumptions include  \cite{SnowWhite,zerocash,InformationDispersal}.  In our system, there are two attacks possible by a compromised designer: i) incorrect protocol design (i.e., the designed sampling strategy and LDPC codes do not result in the claimed probability of failure. Here, the probability of failure can be thought of as an output of the protocol
design computation task and nodes join the system based on the published probability of failure
performance); ii) the designer acts as the adversary and launches a DA attack using the known stopping sets of $H_j$ of size $< \mu_j$.

A possible direction to remove the first attack is as follows. A cryptographic tool called zk-STARK \cite{zk-STARKS} can be used by the system designer to create verifiable proofs of correct computation of the LDPC codes, the LP-sampling strategy, and the probability of failure. This proof can be verified by nodes (full and light) before joining the blockchain system to ensure that the protocol is correctly designed. The proof created using zk-STARK has the following properties: it has a small size, it can be verified using significantly less computational complexity compared to the actual computation, it is secure against quantum computers, it reveals no information about the secrets involved in the computation (here $\mu_j$ and all stopping sets of $H_j$ of size $< \mu_j$).

\begin{figure*}[t]
    \centering
    \begin{subfigure}{0.5\linewidth}
\begin{minipage}{0.99\linewidth}
\begin{tikzpicture}
  \node (img)
  {\includegraphics[scale=0.51]{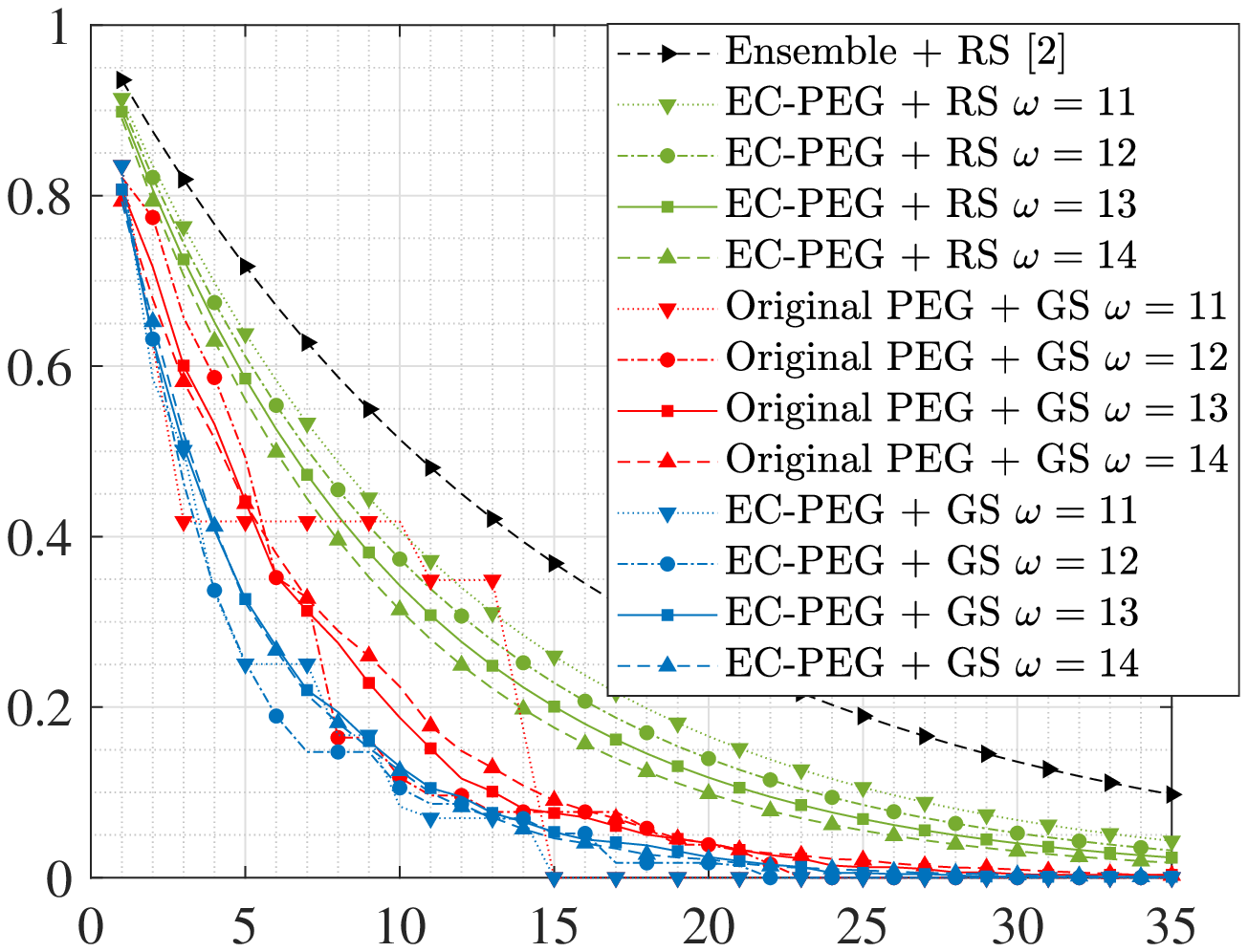}};
  \node[below=of img, node distance=0cm, yshift=1.4cm,font=\color{black}] {$s$};
  \node[left=of img, node distance=0cm, rotate=90, anchor=center,yshift=-1.1cm,font=\color{black}] {$P^{(4)}_{f,\;\omega}(s)$};
 \end{tikzpicture}
 \end{minipage}
    \end{subfigure}%
\begin{subfigure}{0.5\linewidth}
\begin{minipage}{0.99\linewidth}
\begin{tikzpicture}
  \node (img)
  {\includegraphics[scale=0.51]{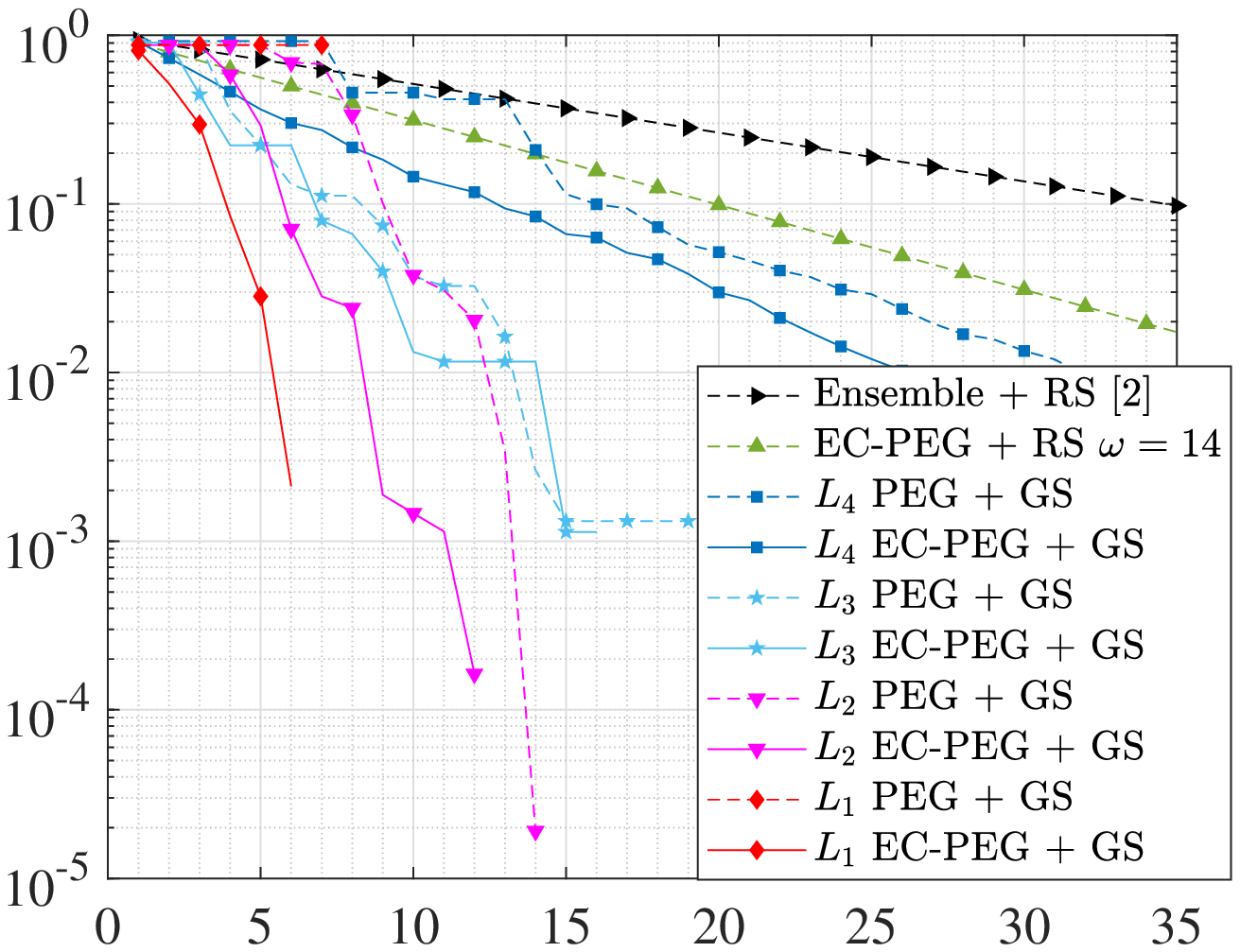}};
  \node[below=of img, node distance=0cm, yshift=1.4cm,font=\color{black}] {$s$};
  \node[left=of img, node distance=0cm, rotate=90, anchor=center,yshift=-1.1cm,font=\color{black}] {$P^{(j)}_f(s)$};
 \end{tikzpicture}
 \end{minipage}
\end{subfigure}
     \vspace{-7pt}
    \caption{\revone{The probability of light node failure for various coding schemes and sampling strategies for CMT $\tCMT_1 = (128,0.5,4,4)$ and weak adversary. \revfourtwo{In this and all other figures, RS refers to random sampling}; Left panel: probability of failure for a DA attack on the base layer for different stopping set sizes. The black curve is achieved using 
    stopping ratio $\nu^* = 0.064353$. The value $\nu^*$ is the best stopping ratio obtained for a rate $0.5$ code following the method in \cite[section 5.3]{CMT} using parameters $(c,d) = (8,16)$. 
    GS refers to the overall greedy sampling strategy described in Section \ref{sec:system_aspects} 1) where we have used $\tlambda = 0.9$. $P^{(j)}_{f,\;\omega}(s)$ for GS is calculated as $[1 - \tau(S^{(\tlambda s\;,\;j)}_{greedy},\omega)](1 - \frac{\omega}{n_j})^{s - \tlambda s}$, where $S^{(\tlambda s\;,\;j)}_{greedy}$ is described in Lemma \ref{lemma:soundess_weak}; 
    Right panel: probability of failure across different layers of the CMT.
    $P^{(j)}_f(s)$ is calculated as $P^{(j)}_f(s)  = \max_{\omega < \mu_j}P^{(j)}_{f,\;\omega}(s)$, where $\mu_j = \omega^{(j),PEG}_{\min} + 6$.
   } }
    \label{fig:EC-PEG-p_f}
     \vspace{-10pt}
\end{figure*}

In the second attack, the system designer acts as the adversary (medium) to launch a DA attack using the knowledge of the stopping sets (which it enumerated while correctly designing LP-sampling $\tx$). However, this DA attack will be detected by the light nodes with a probability of failure $P^{(j)}_{(f,med)}(s)$ which is guaranteed by the protocol. Also, to launch this DA attack, the system designer spends the same amount of computational power as a medium adversary who doesn't have the knowledge of the stopping sets and wishes to attack the system. Thus, the system designer is not at an advantage to launch DA attacks due to the knowledge of the secret.

\vspace{-0.6cm}
\section{Simulation Results}\label{sec:simulations}
\vspace{-0.1cm}
\trim{In this section, we compare the performance of our co-design techniques  with that of codes
designed by the original PEG algorithm and the performance of \cite{CMT} using random LDPC codes and random sampling (RS). \revfourtwo{Since many works e.g., \cite{Cover} \cite{Trifecta} use random LDPC codes and random sampling to mitigate DA attacks, any improvements we show in comparison to \cite{CMT} will also provide benefits in these works.}
The different CMTs used for simulation are parametrized by $\tCMT = (n_l,\tr,q,l)$ (individual parameters are defined in Section \ref{sec:CMT_construction}). For a CMT $\tCMT$, in order to compare the performance of different PEG based codes, we choose $\mu_j = \omega^{(j),PEG}_{\min} + \gamma$, $1 \leq j \leq l$, for the various adversary models described in Section \ref{sec:adv_models}. Here, $\omega^{(j),PEG}_{\min}$ is the minimum stopping set size for an LDPC code constructed using the original PEG algorithm for layer $j$ of the CMT $\tCMT$ and $\gamma$ is a parameter. We calculate the probability of failure when the light nodes request for $s$ base layer samples using random sampling for various scenarios as follows: for the base layer   when the adversary hides a stopping set of size $\omega$, $P^{(l)}_{f,\;\omega}(s) = \left(1 - \frac{\omega}{n_l}\right)^s$; for intermediate layers, we calculate the probability of failure for the medium and strong adversary by substituting $\tx = \frac{1_{n_l}}{n_l}$ in the probability of failure expressions provided in Section \ref{sec:LP-sampling-anylayer}, where $1_{n_l}$ is a vector of ones of length $n_l$; for an LDPC code with a stopping ratio $\nu^*$ we calculate the probability of failure at the base layer using random sampling as $P^{(l)}_f(s) = (1 - \nu^*)^s$. The LDPC codes at different layers of the CMTs are aligned using Algorithm \ref{alg:layer-greedy-align} where we use $g_{\max} = \ttg$
(observed cycles in the code constructions) and $g_{\min}$ is set to the girth of the respective codes.}

Fig. \ref{fig:EC-PEG-p_f} demonstrates the performance of the EC-PEG algorithm and the greedy sampling
strategy for CMT $\tCMT_1 = (128,0.5,4,4)$ and a weak adversary. For the EC-PEG algorithm, we have used the parameters: $d_v = 4$ for all layers, $\tr = 0.5$, $\ttg^{(4)} = 10$ and $\ttg^{(j)} = 8$ for $j = 1,2,3$. For the adversary model we have chosen $\gamma = 6$. Note that $\omega^{(4),PEG}_{\min} = 9$ and thus  $\mu_4 = 9 + 6 = 15$.
In Fig. \ref{fig:EC-PEG-p_f} left panel, we plot $P^{(4)}_{f,\;\omega}(s)$ for various coding algorithms and sampling strategies  when a weak adversary conducts a DA attack on the base layer of the CMT by hiding stopping sets of size $\omega < \mu_4$.
The codes designed by the original PEG and EC-PEG algorithms  have a minimum stopping set size of $9$ and $10$, respectively.
For these algorithms, $P^{(4)}_{f,\;\omega}(s)$ quickly becomes zero for $\omega =9,\;10$ using greedy sampling as $s$ increases. Hence, we have not included these stopping set sizes in Fig. \ref{fig:EC-PEG-p_f} left panel.
 The figure demonstrates three benefits of our co-design.
The first benefit is due to the use of deterministic LDPC codes that provide larger stopping set sizes than random ensembles, as can be seen when comparing the black and green curves. 
\trim{The second benefit comes from using greedy sampling as opposed to random sampling, which can be observed by comparing the green and red curves. The final benefit is provided by the EC-PEG algorithm, as can be seen by comparing the red and blue curves. These benefits combine to significantly reduce $P^{(4)}_{f,\;\omega}(s)$ compared to the black curve which was proposed  in earlier literature\footnote{\revthree{The singularities in some plots in Fig. \ref{fig:EC-PEG-p_f} (e.g., Original PEG + GS $\omega = 11$) is because $P^{(4)}_{f,\;\omega}(s)$ becomes zero after certain number of greedy samples. This situation happens when all the stopping sets of weight $\omega$ get touched by the greedy samples.}}}.

\begin{figure*}[t]
    \centering
    \begin{subfigure}{0.5\linewidth}
\begin{minipage}{0.99\linewidth}
\begin{tikzpicture}
  \node (img)
  {\includegraphics[scale=0.51]{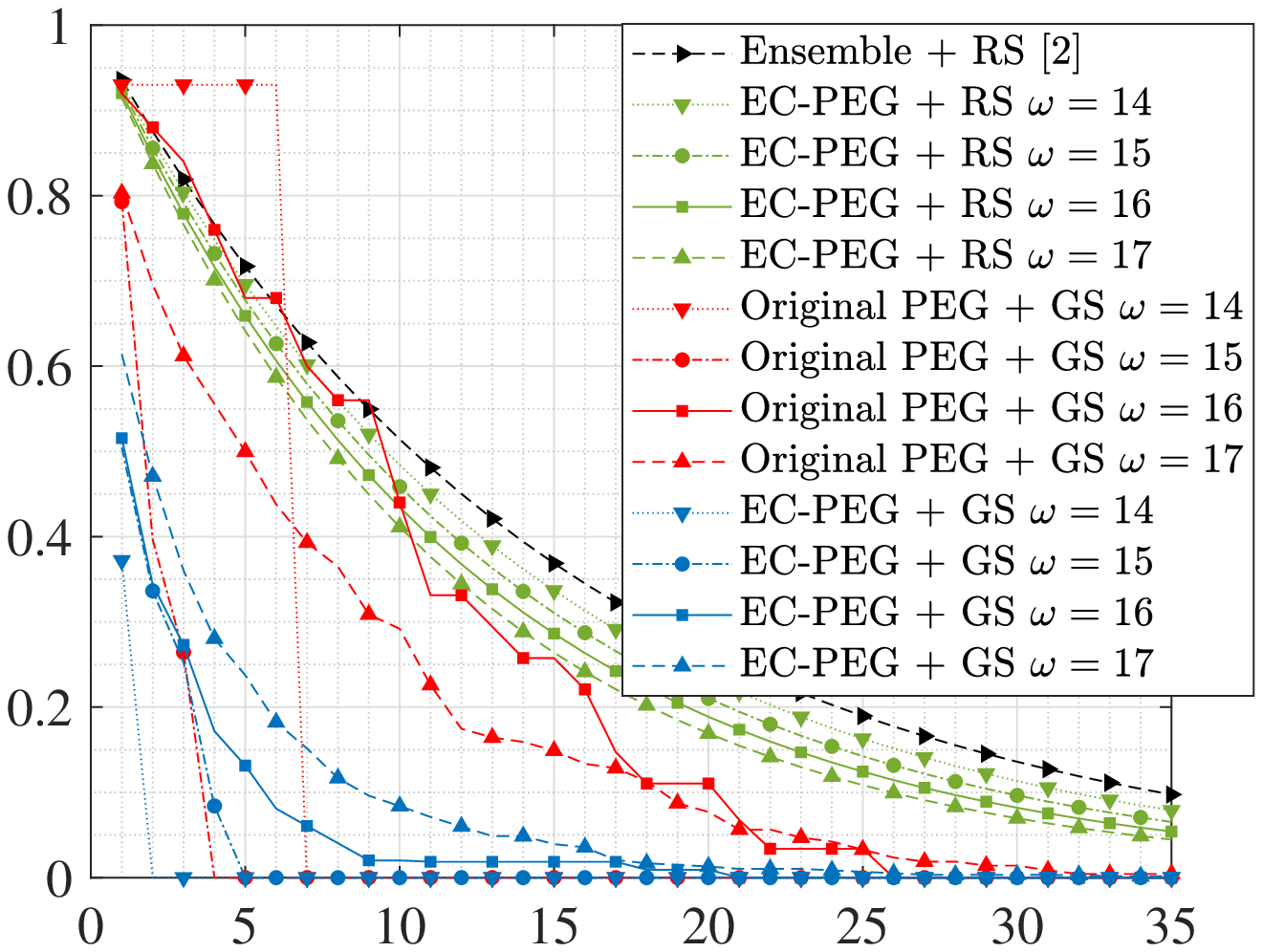}};
  \node[below=of img, node distance=0cm, yshift=1.4cm,font=\color{black}] {$s$};
  \node[left=of img, node distance=0cm, rotate=90, anchor=center,yshift=-1.2cm,font=\color{black}] {$P^{(4)}_{f,\;\omega}(s)$};
 \end{tikzpicture}
 \end{minipage}
    \end{subfigure}%
\begin{subfigure}{0.5\linewidth}
\begin{minipage}{0.99\linewidth}
\begin{tikzpicture}
  \node (img)
  {\includegraphics[scale=0.51]{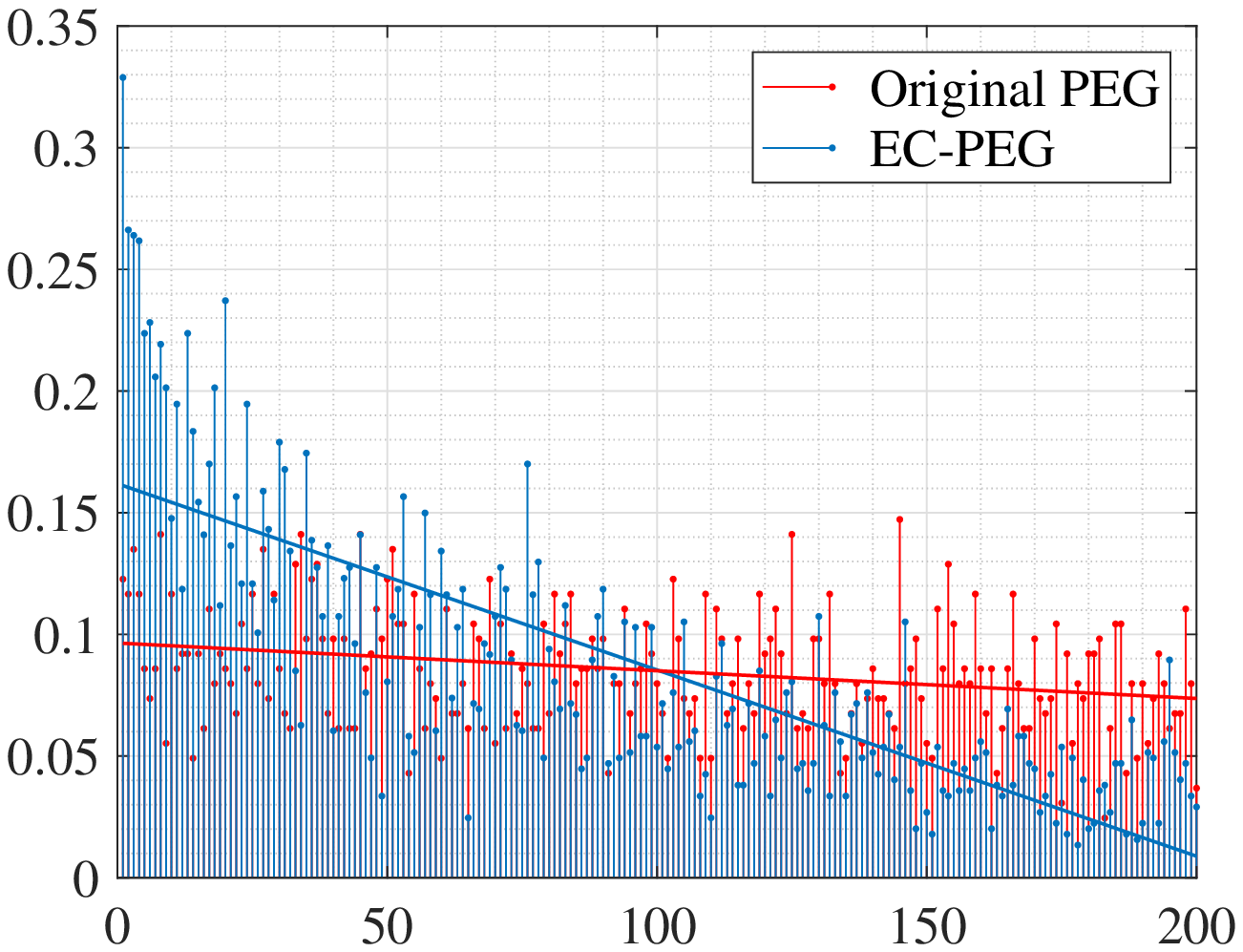}};
  \node[below=of img, node distance=0cm, yshift=1.5cm,font=\color{black}] {\footnotesize{VN index}};
  \node[left=of img, node distance=0cm, rotate=90, anchor=center,yshift=-1.1cm,font=\color{black}] {$ss^{17}$};
 \end{tikzpicture}
 \end{minipage}
\end{subfigure}
     \vspace{-5pt}
    \caption{Weak adversary performance plots for $N_l = 200$, $\tr = 0.5$, $d_v = 4$. Left panel: probability of failure for a DA attack on the base layer for different stopping set sizes (see Fig. \ref{fig:EC-PEG-p_f} left panel for plot properties). We have parameters $\omega^{PEG}_{\min} = 13$ and $\gamma = 5$, $\tlambda = 0.9$; 
    Right panel: stopping set distribution $ss^{17}$. }
    \label{fig:EC-PEG-p_f_additional}
     \vspace{-10pt}
\end{figure*}

\begin{figure*}[t]
    \centering
    \begin{subfigure}{0.33\linewidth}
\begin{minipage}{0.99\linewidth}
\begin{tikzpicture}
  \node (img) {\includegraphics[scale=0.42]{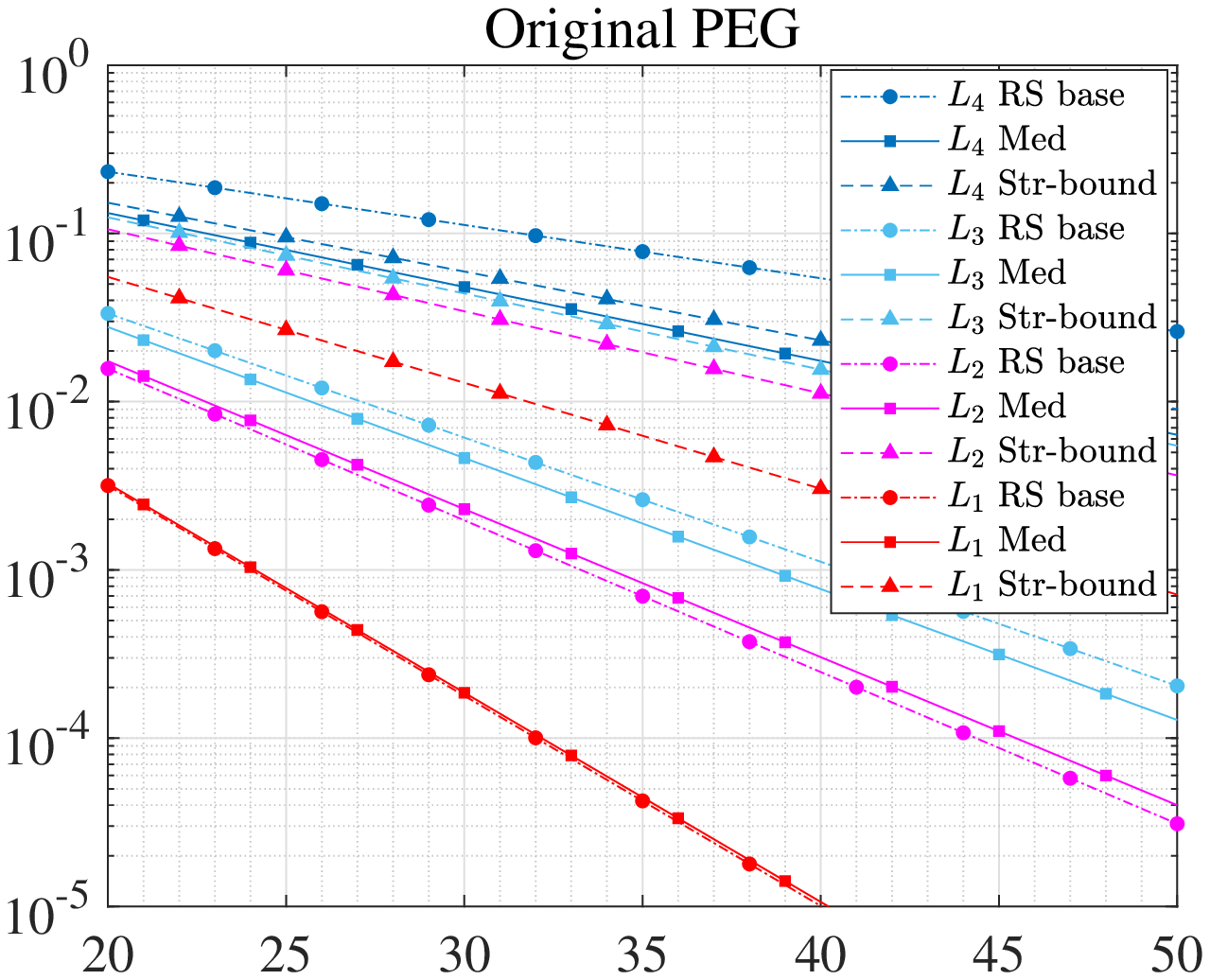}};
  \node[below=of img, node distance=0cm, yshift=1.4cm,font=\color{black}] {{ $s$}};
  \node[left=of img, node distance=0cm, rotate=90, anchor=center,yshift=-1.2cm,font=\color{black}] {\footnotesize {$P^{(j)}_f(s)$}};
 \end{tikzpicture}
 \end{minipage}
\label{fig:PEG_p_f}
    \end{subfigure}%
    \begin{subfigure}{0.33\linewidth}
\begin{minipage}{0.99\linewidth}
\begin{tikzpicture}
  \node (img) {\includegraphics[scale=0.42]{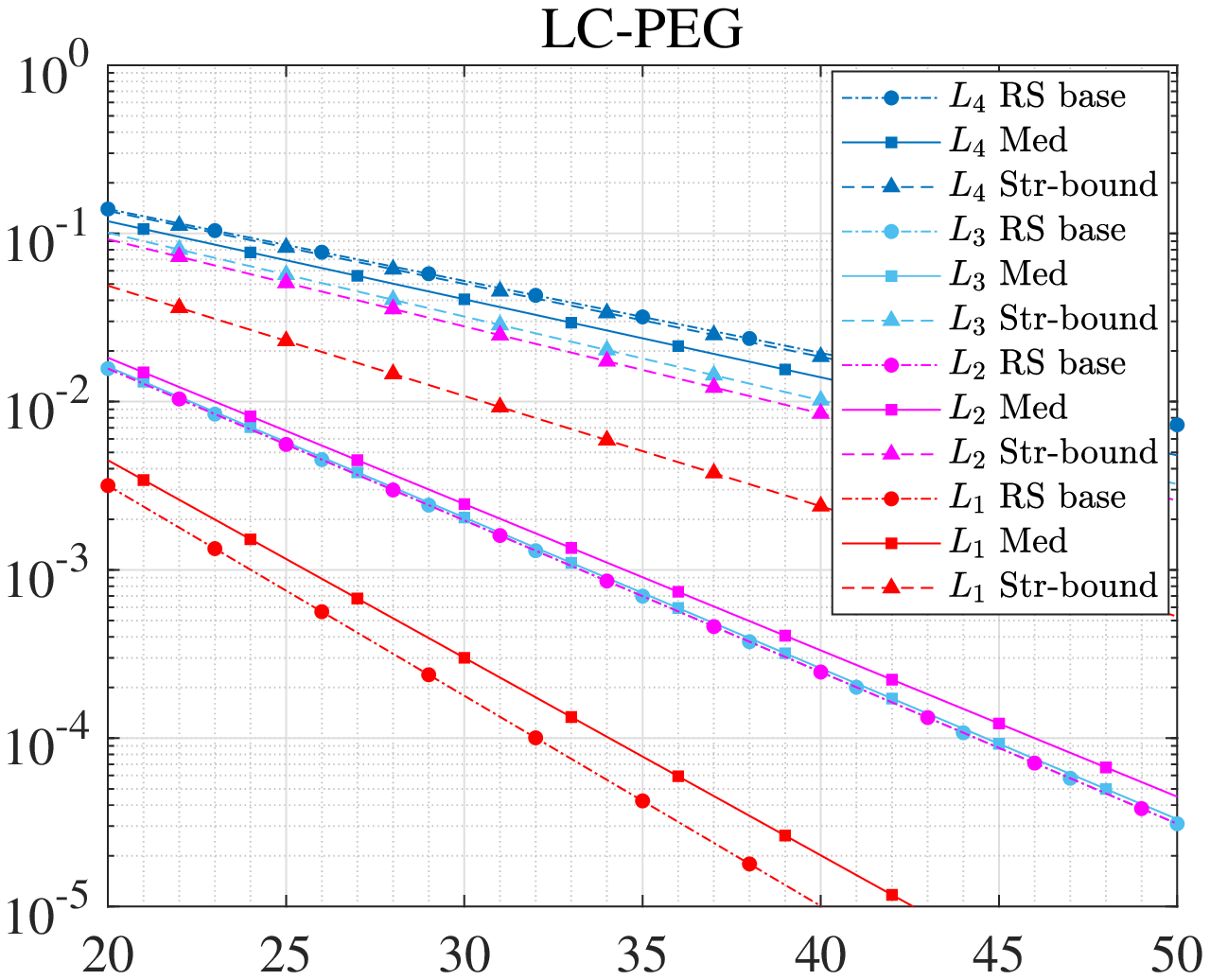}};
  \node[below=of img, node distance=0cm, yshift=1.4cm,font=\color{black}] {{$s$}};
  \node[left=of img, node distance=0cm, rotate=90, anchor=center,yshift=-1.2cm,font=\color{black}] {\footnotesize{$P^{(j)}_f(s)$}};
 \end{tikzpicture}
 \end{minipage}
    \end{subfigure}%
\begin{subfigure}{0.33\linewidth}
\begin{minipage}{0.99\linewidth}
\begin{tikzpicture}
  \node (img) {\includegraphics[scale=0.42]{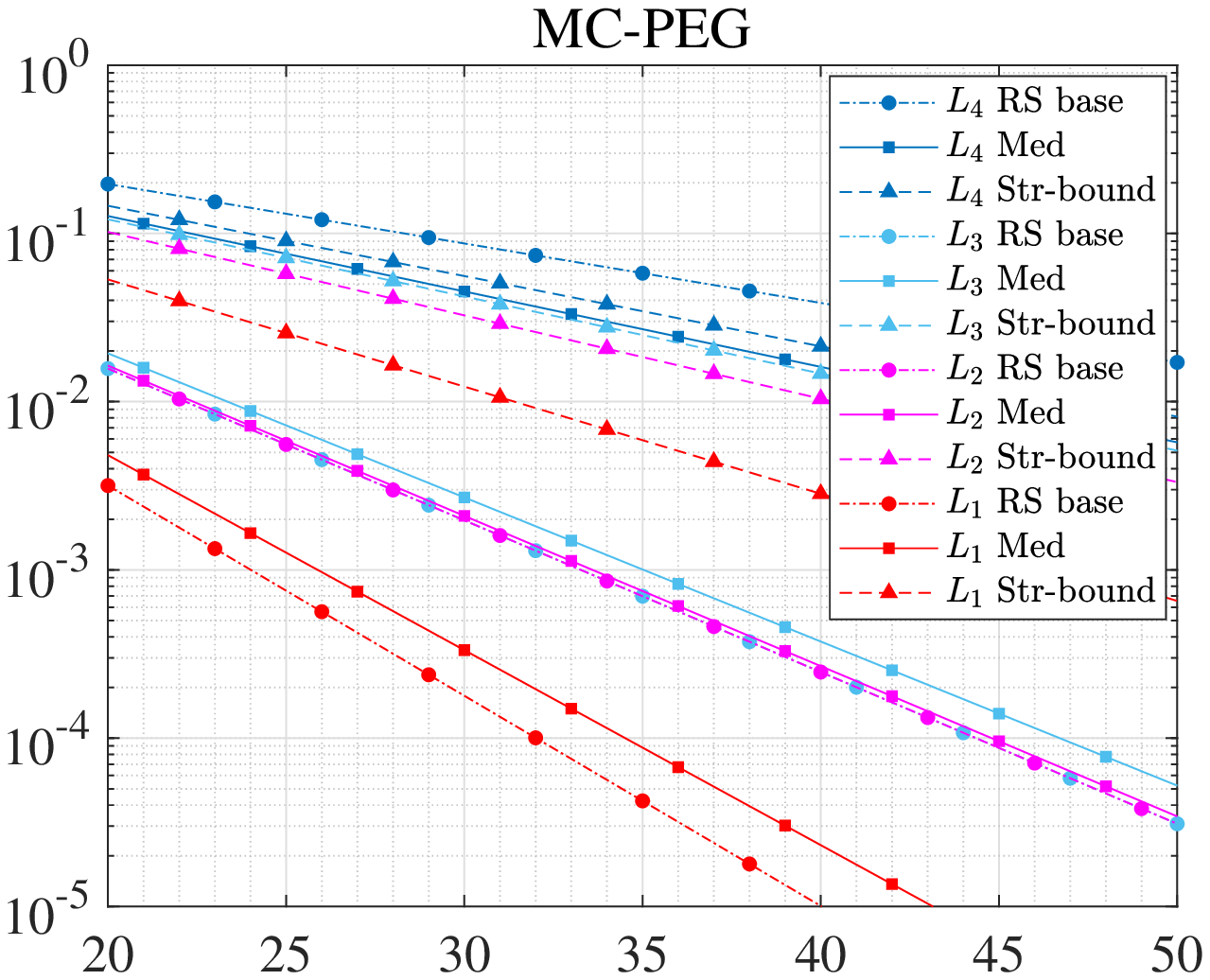}};
   \node[below=of img, node distance=0cm, yshift=1.4cm,font=\color{black}] {$s$};
  \node[left=of img, node distance=0cm, rotate=90, anchor=center,yshift=-1.2cm,font=\color{black}] {\footnotesize{$P^{(j)}_f(s)$}};
 \end{tikzpicture}
 \end{minipage}
\end{subfigure}
     \vspace{-5pt}
    \caption{\revone{
    For CMT $\tCMT_1 = (128,0.5,4,4)$, we plot the probability of failure for a DA attack at all layers for different codes under the following cases; i) Base layer is randomly sampled (RS base), ii) LP-Sampling with medium adversary (Med), iii) LP-sampling with strong adversary (Str-bound).
    For the strong adversary, we plot $P^{(j)}_{(f,str)}(s)$.
    \vspace{-10pt}}
   }
    \label{fig:LC-PEG-Full-CMT}
\end{figure*}

In Fig. \ref{fig:EC-PEG-p_f} right panel, we plot the probability of failure $P^{(j)}_f(s)$ when the weak adversary
conducts a DA attack on layer $j$ of CMT $\tCMT_1$. 
From Fig. \ref{fig:EC-PEG-p_f} right panel\footnote{\revone{The plots for $P^{(4)}_{f,\;\omega}(s)$ and $P^{(j)}_{f,\;\omega}(s)$ in Fig. \ref{fig:EC-PEG-p_f} sometimes exhibit floors (i.e., they remain constant for different values of $s$). This is due to i) the new greedy samples that are selected (on increasing $s$) do not increase the number of stopping sets that are touched; ii) the number of random samples in the overall greedy sampling strategy remain same on increasing $s$.
}}, we see that the base layer of the CMT ($L_4$) has a larger probability of failure compared to other layers and the probability of failures for the intermediate layers quickly become very small. This is due to the alignment of the columns of the parity check matrices, which ensures that each intermediate layer is greedy sampled.
We next observe that the EC-PEG algorithm with greedy sampling results in a lower $P^{(j)}_f(s)$ compared to the original PEG algorithm for all layers of the CMT. Moreover, for the base layer, $P^{(4)}_f(s)$ (for both EC-PEG and original PEG coupled with greedy sampling) is lower than  the probability of failure using random sampling for $\omega = 14$ (green curve) and the probability of failure achieved by random LDPC codes and random sampling (black curve).  Thus, in combination, the co-design of concentrated LDPC codes and greedy sampling results in a significantly lower  $P^{(4)}_{f,\;\omega}(s)$ compared to methods proposed in \cite{CMT}. \editor{To illustrate the benefits of the EC-PEG algorithm and the greedy sampling strategy, we provide plots similar to Fig. \ref{fig:EC-PEG-distribution} and Fig. \ref{fig:EC-PEG-p_f} for a different choice of code parameters in Fig. \ref{fig:EC-PEG-p_f_additional}. From the figure, we see similar stopping set concentration and probability of failure improvement as in Fig. \ref{fig:EC-PEG-distribution} and Fig. \ref{fig:EC-PEG-p_f}}.

\begin{figure*}[t]
    \centering
    \begin{subfigure}{0.5\linewidth}
\begin{minipage}{0.99\linewidth}
\begin{tikzpicture}
  \node (img)
  {\includegraphics[scale=0.52]{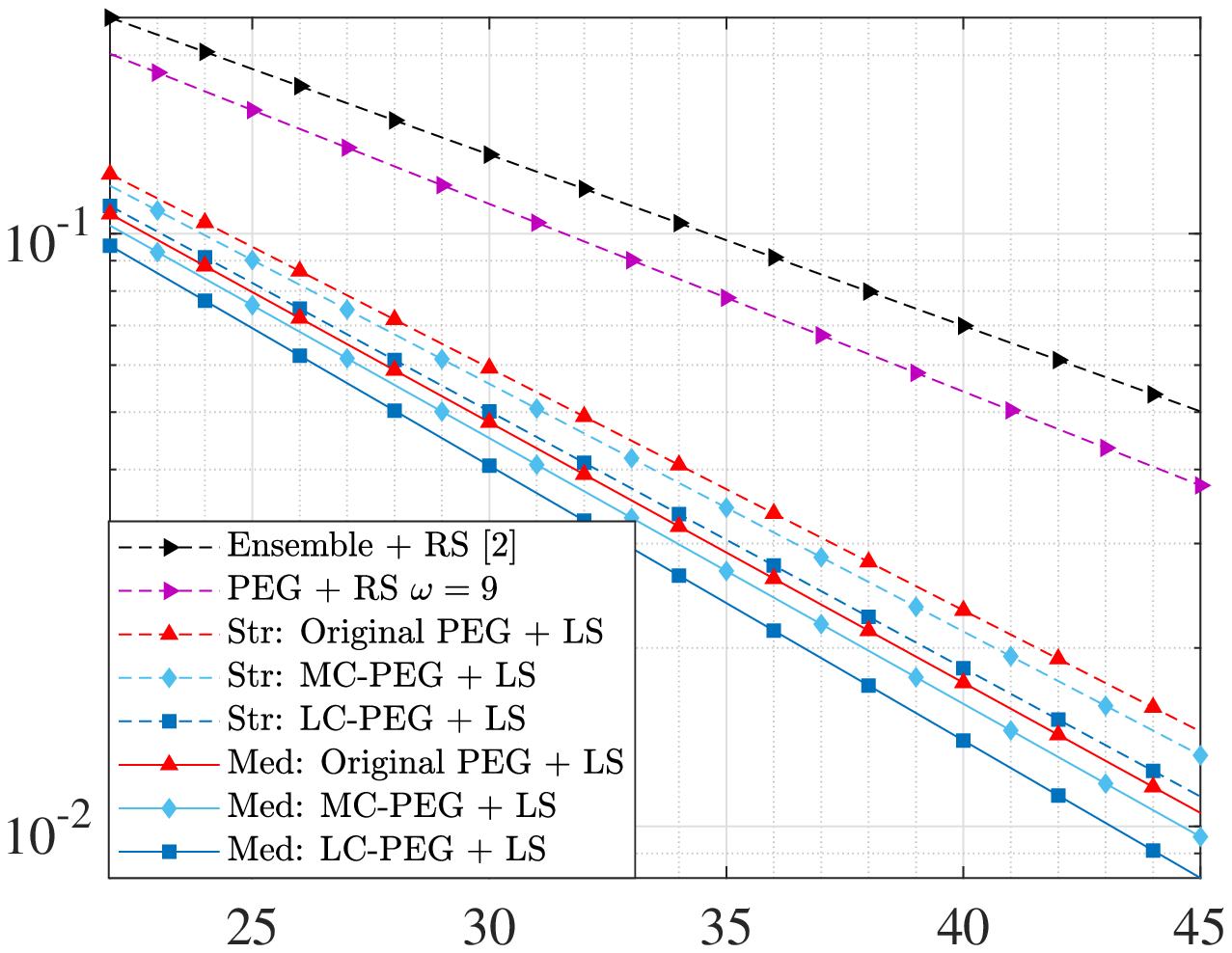}};
  \node[below=of img, node distance=0cm, yshift=1.4cm,font=\color{black}] {$s$};
  \node[left=of img, node distance=0cm, rotate=90, anchor=center,yshift=-1.2cm,font=\color{black}] {$P^{(4)}_f(s)$};
 \end{tikzpicture}
 \end{minipage}
    \end{subfigure}%
\begin{subfigure}{0.5\linewidth}
\begin{minipage}{0.99\linewidth}
\begin{tikzpicture}
  \node (img)
  {\includegraphics[scale=0.52]{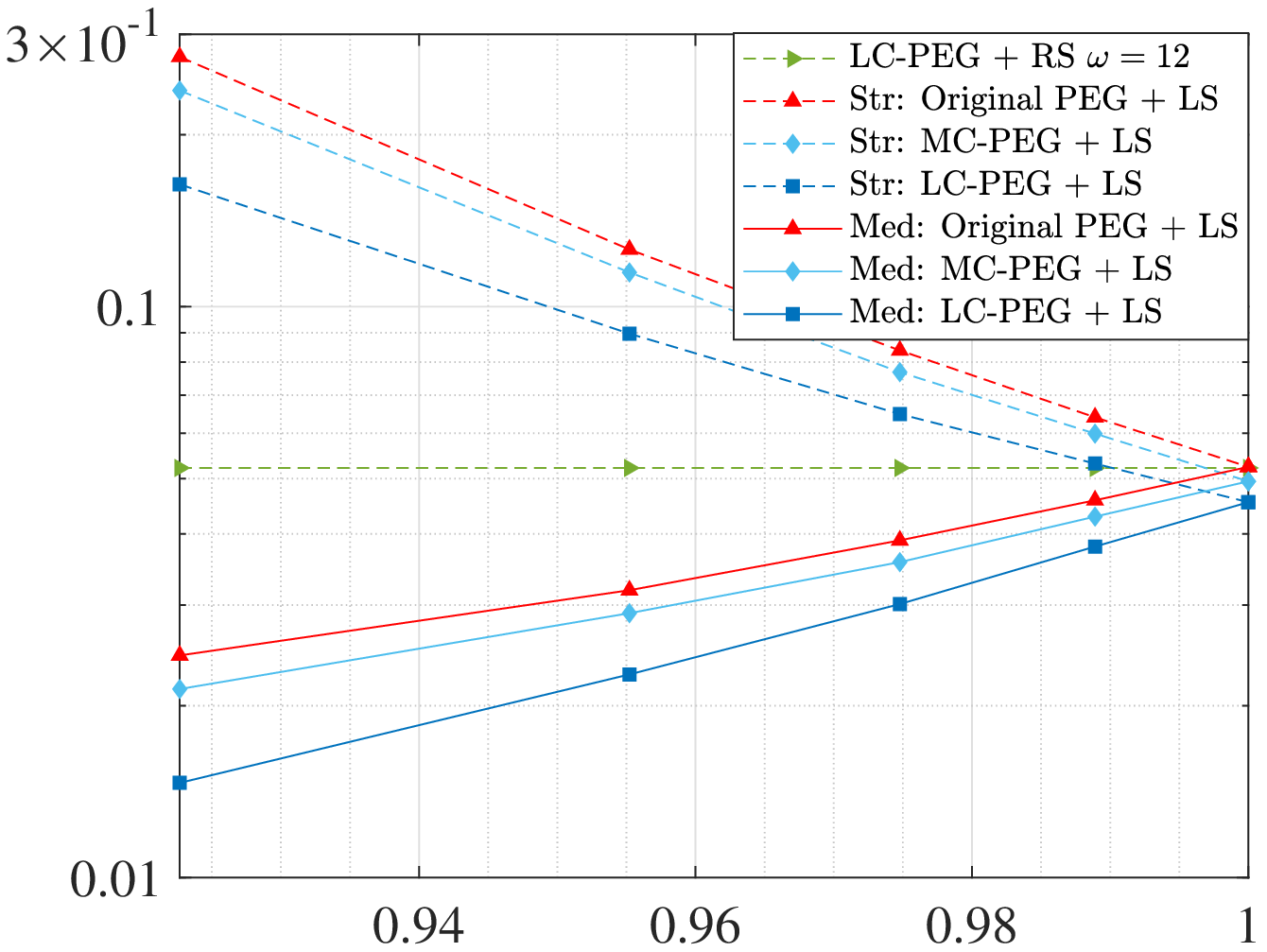}};
  \node[below=of img, node distance=0cm, yshift=1.4cm,font=\color{black}] {\small $\tnu^{(4)}$};
  \node[left=of img, node distance=0cm, rotate=90, anchor=center,yshift=-1.2cm,font=\color{black}] {$P^{(4)}_f(s=30)$};
 \end{tikzpicture}
 \end{minipage}
\end{subfigure}
     \vspace{-5pt}
    \caption{\revone{The probability of light node failure for a DA attack on the base layer of CMT $\tCMT_1 = (128,0.5,4,4)$; Left panel: comparison of different coding schemes and sampling strategies. The black curve uses $\nu^* = 0.064353$;
    Right panel: variation in $P^{(4)}_f(s=30)$ for the strong and medium adversary as a function of $\tnu^{(4)}$ for $\tnu^{(j)} = 1, j =1,2,3$.}
    }
   \label{fig:LC-PEG-baselayer}
     \vspace{-5pt}
\end{figure*}

\setlength{\tabcolsep}{2pt}
\begin{table*}[t]
\caption{ $P^{(l)}_f(s = 0.25n_l)$ for a DA attack on the base layer 
for various CMT parameters, coding schemes, and sampling strategies. The parameters used for the different CMTs is listed in Table \ref{table:all_CMTs_parameters}. 
\editortwo{For the ensemble codes, we follow the method of \cite[Section 5.3]{CMT} and for each $R$ obtain the following parameters $(R,c,d,\nu^*) = \{(0.5,8,16,0.0643),\\(0.4,6,10,0.0851),(0.8,11,55,0.0187)\}$ where $(c,d)$ are optimized to maximize the stopping ratio $\nu^*$.}
 }\label{table:all_CMTs}
 \vspace{-2pt}
\centering
\resizebox{!}{1.5cm}{
\begin{tabular}{| c | c | c | c | c | c | c | c |c | c| c|}
\hline
 &\multicolumn{4}{c|}{Random Sampling}&\multicolumn{6}{c|}{LP-Sampling}\\
\hline
CMT & \multirow{2}{*}{Ensemble} & \multirow{2}{*}{PEG} & \multirow{2}{*}{MC-PEG} & \multirow{2}{*}{LC-PEG} &
\multicolumn{3}{c|}{Strong Adversary} &
\multicolumn{3}{c|}{Medium Adversary}\\
$\tCMT = (n_l,\tr,q,l)$ & & & & & PEG & MC-PEG  & LC-PEG& PEG & MC-PEG & LC-PEG \\
\hline
$(128,0.5,4,4)$ & 0.1190 & 0.0970 &  0.0740 & 0.0428  &   0.04914 &	0.04602&	0.04106
&	0.03925&	0.03675&	0.03279
\\
$(208,0.5,4,4)$	&	0.0314 & 0.0204 & 0.0155&	0.0267& 0.0304 &	0.02427 &	0.02294
 &	0.00815&	0.00651& 0.00615
 \\
$(200,0.5,4,3)$	& 0.0356	& 0.0347&0.0202 &	0.0202 & 0.02778&	0.02028	& 0.01781
 &	0.00606&	0.00447&	0.00388
\\
$(200,0.4,5,4)$	& 0.0117	&0.0089 & 0.0067& 0.0052& 0.00558	&0.00513 &	0.00484&	0.00225&	0.00207&	0.00195
	\\
$(200,0.8,5,2)$ & 0.3891 & 0.4697 &	0.3641 &	0.2820 & 0.2827 &	0.256	&0.2332&	0.171&	0.1549&	0.1411
 \\
\hline
\end{tabular}
}
\vspace{-3pt}
\end{table*} 

\setlength{\tabcolsep}{3pt}
\begin{table*}[t]
\caption{Parameters used for LP-sampling and LC-PEG code construction for various CMTs in Table \ref{table:all_CMTs}.  For all LDPC codes we use $d_v = 4$, $\ttg^{(j)} =  g^{(j)}_{\max} = g^{(j)}_{\min}+4$, $j = 1, \ldots, l$. For LC-PEG algorithm, we use $\hat{\mu}^{(j)} = \mu^{(j)} = \omega^{(j),PEG}_{\min}+\gamma$, $j = 1,\ldots, l$. Under each variable that depends on the layer, we enumerate the layer numbers.}
 \label{table:all_CMTs_parameters}
 \vspace{-2pt}
\centering
\resizebox{!}{1.4cm}{
\begin{tabular}{| c | c | c | c | c | c | c | c | c | c |c | c| c|c|c|c|c|c|c|c|c|c|}
\hline
CMT &\multicolumn{4}{c|}{$T^{(j)}_{th}$}&\multicolumn{4}{c|}{$\widehat{\theta}^{(j)}$}&\multicolumn{4}{c|}{$g^{(j)}_{\min}$}&\multicolumn{4}{c|}{$\omega^{(j),PEG}_{\min}$}&\multirow{2}{*}{$\gamma$}&\multicolumn{4}{c|}{$\tnu^{(j)}$}\\
$\tCMT = (n_l,\tr,q,l)$ & 1& 2& 3& 4& 1& 2& 3 & 4  & 1& 2 & 3 &4 &1&2&3&4&&1&2&3&4\\
\hline
$(128,0.5,4,4)$& 3	&3	&4	&4	&0.997	&0.997&	0.997&	0.997 &	4&	4	&4&	6&	2&	4&	5&	9 & 4 & 1 & 1 & 1 &0.993\\
$(208,0.5,4,4)$ & 3	&4	&4	&5	&0.997	&0.997&	0.997&	0.9959&	4	&4	&6	&6	&4	&6	&10	&15 & 3  & 1 & 1 & 0.997 & 0.975\\
$(200,0.5,4,3)$ & 4 &	4	&5	& - &	0.997	&0.997&	0.998	& - &	4	& 6	& 6	& - &	6	&8	&13	& - & 3 & 1 & 0.99 & 0.97 &  -\\
$(200,0.4,5,4)$ & 3&	4	&4	&5	&0.997 &	0.997& 	0.997	& 0.998	&4	&4	&6	&6	&5	&8	&11	&18 & 4 & 1 & 1 & 0.992 & 0.982\\
$(200,0.8,5,2)$ & 4 & 5& -& -& 0.997 & 0.997 & - & - & 4 & 4& - & -& 2 & 3& -& -& 3&  1 & 0.99 & - & -
\\
\hline
\end{tabular}
}
\vspace{-3pt}
\end{table*}

\setlength{\tabcolsep}{3pt}
\begin{table*}[t]
\caption{Maximum CN degree for the LDPC codes used in different layers of the CMT. Under each algorithm, we enumerate the layer numbers and specify the maximum CN degree for that layer.}
 \label{table:CN_Degrees}
 \vspace{-2pt}
\centering
\resizebox{!}{1.4cm}{
\begin{tabular}{| c | c | c | c | c | c | c | c | c | c |c | c| c|c|c|c|c|c|}
\hline
CMT &\multirow{2}{*}{Ensemble}&\multicolumn{4}{c|}{PEG}&\multicolumn{4}{c|}{EC-PEG}&\multicolumn{4}{c|}{MC-PEG}&\multicolumn{4}{c|}{LC-PEG}\\
$\tCMT = (n_l,\tr,q,l)$ & & 1& 2& 3& 4& 1& 2& 3 & 4  & 1& 2 & 3 &4 &1&2&3&4\\
\hline
$(128,0.5,4,4)$& 16 & 8 &8&9&9	&10&13&12&14&	8&8&9&9	&8&8&9&9
\\
$(208,0.5,4,4)$ &16 &	8&9&8&8	&11&11&11&16&	8&9&8&9	&8&8&9&9
 \\
$(200,0.5,4,3)$ & 16	&9 &9 &9 &- &	12&11&18&-&	9&9&9&-&	8&9&9&-
\\
$(200,0.4,5,4)$ & 10	&7&7&7&7&	12&9&11&14&	7&7&7&8&7&8&7&7
\\
$(200,0.8,5,2)$ & 55 & 20&21&-&-&26&48&-&-&20&20&-&-&20&20&-&-\\
\hline
\end{tabular}
}
\vspace{-10pt}
\end{table*}

\vspace{-0.016cm}
\trim{In Figs. \ref{fig:LC-PEG-Full-CMT}, \ref{fig:LC-PEG-baselayer} and Table \ref{table:all_CMTs}, we demonstrate the performance of the LC-PEG algorithm and LP-sampling (LS) against a medium and a strong adversary. Fig. \ref{fig:LC-PEG-Full-CMT} and \ref{fig:LC-PEG-baselayer} correspond to CMT $\tCMT_1 = (128,0.5,4,4)$ where we have used
$\gamma = 4$, thus $\mu_j = \omega^{(j),PEG}_{\min} + 4$.
Table \ref{table:all_CMTs_parameters} lists $\omega^{(j),PEG}_{\min}$ for different $j$.
}
Additionally, for LP-sampling, we have used $\tnu^{(4)} = 0.993$, $\tnu^{(j)} = 1, j= 1,2,3$.
For the LC-PEG algorithm, we have used $d_v = 4$ for all layers, $\tr = 0.5$, $\ttg^{(4)} = 10$ and $\ttg^{(j)} = 8$ for $j = 1,2,3$, $T^{(j)}_{th} = 3$ for $j=1,2$, $T^{(j)}_{th} = 4$ for $j=3,4$,  $\hat{\theta}^{(j)} = 0.997$, $j=1,2,3,4$ (we tested with $T_{th} = 3,\; 4$ and $\hat{\theta} = 0.995,0.996,0.997,0.998$ and picked the codes that provide the lowest $P^{J^{\max}}_f(s)$) and $\hat{\mu}_j = \mu_j, j =1,2,3,4$.    
To demonstrate the effectiveness of the LC-PEG algorithm, we also plot the performance of an algorithm termed as the \emph{Minimum-Cycles} PEG (MC-PEG) algorithm. It is the same as the LC-PEG algorithm but instead of the CN selection steps in lines 10-13 of Algorithm \ref{alg:LC-PEG}, the MC-PEG algorithm selects a CN randomly from $\tK_{mincycles}$ as $\tcn^{sel}$. 

\vspace{-0.01cm}
We first look at the improvements provided by LP-sampling. Fig. \ref{fig:LC-PEG-Full-CMT} shows the performance of LP-sampling for a DA attack at different layers of the CMT constructed using the PEG,  LC-PEG and MC-PEG algorithms. We see that while the probability of failure for some layers worsens in comparison to random sampling, for the worst layer, which is the base layer, the probability
of failure improves for both the strong and medium adversary.  We generally find that the base layer is the worst layer so we focus on the base layer in the subsequent simulations.

We plot  $P^{(4)}_{f}(s)$ vs. $s$ for the PEG, MC-PEG, and  LC-PEG algorithms using LP-sampling in Fig. \ref{fig:LC-PEG-baselayer} left panel, where we see the following improvements. The first improvement is between the black and magenta curves due to using deterministic LDPC codes that produce larger stopping set sizes. The second improvement is due to using LP-sampling compared to random sampling. Compared to random sampling (magenta curve), LP-sampling with the original-PEG algorithm results in a lower probability of failure for the medium (red-solid curve) and strong adversary (red-dotted curve). 
The third improvement (between the red and light blue curves) comes from utilizing  the MC-PEG algorithm to reduce the number of small cycles as discussed in Remark \ref{remark:MC_PEG}. 
The final improvement comes from the informed CN selection in the LC-PEG algorithm to create tailored codes for LP-sampling as seen by comparing the dark and light blue curves.

In Fig. \ref{fig:LC-PEG-baselayer} right panel, we plot $P^{(4)}_f(s = 30)$ as a function of the parameter $\tnu^{(4)}$ for the original
PEG, MC-PEG and LC-PEG algorithms using LP-sampling.
From Fig. \ref{fig:LC-PEG-baselayer} right panel, we see that  $\tnu^{(4)}$ controls the trade-off between the probabilities of failure for the medium adversary and strong adversary. Thus, $\tnu^{(4)}$ can be chosen as a hyper-parameter based on the system specifications. 
We also see from Fig. \ref{fig:LC-PEG-baselayer} right panel that for all the values of $\tnu^{(4)}$, the LC-PEG algorithm outperforms
the PEG and MC-PEG algorithm for both the medium and strong adversary.

\trim{For completeness, we provide further examples of how our novel code constructions improve
the probability of failure for different CMT parameters. In Table \ref{table:all_CMTs}, we list $P^{(l)}_f(s)$ 
and compare various sampling strategies and LDPC code constructions.
 Similar to Fig. \ref{fig:LC-PEG-baselayer} left panel,  from Table \ref{table:all_CMTs}, we see that the novel co-design of the LC-PEG algorithm and LP-sampling results in the lowest probability of failure for the different CMT parameters. \editortwo{We see that even at a high rate of $0.8$, our techniques of LC-PEG algorithm and LP-sampling offer an improvement. %
 } }

  In Table \ref{table:CN_Degrees}, we compare the maximum CN degree for the LDPC codes used in different
 CMT layers for various construction techniques. We see that PEG based constructions have similar maximum CN degrees compared to the ensemble LDPC codes used in \cite{CMT}. Since the incorrect coding proof size is proportional to the maximum CN degree, we conclude that the new LDPC code constructions do not significantly impact the incorrect coding proof size to improve the probability of failure. \editortwo{Additionally for rate 0.8 codes, we see that the LC-PEG algorithm results in a significantly lower maximum CN degree compared to the ensemble LDPC codes thus also improving the incorrect coding proof size along with the probability of failure.
 }

\vspace{-0.2cm}
\section{Conclusion}\label{sec:conclusion}
\vspace{-0.1cm}
In this paper, we considered the problem of DA attacks pertinent to blockchains with
light nodes.
For various strengths of the malicious nodes, we demonstrated that, \editor{at short code lengths}, a suitable co-design of specialized LDPC codes and  the light node sampling strategy can result in a much lower probability of failure to detect DA attacks compared to schemes in prior literature.

\vspace{-0.2cm}
\appendix

\vspace{-0.1cm}

\subsubsection{Proof of Lemma \ref{eqn:p_f_str}}\hfill

$P^{(l)}_{(f,med)}(s) = \underset{{k \in \{1, 2, \ldots, \vert \tsj_l\vert\}}}{\max} P^{(l)}_f(s;\tSall^{(l)}_k) = \underset{{k \in \{1, 2, \ldots, \vert \tsj_l\vert\}}}{\max} \left(1 - \sum_{i: v^{(l)}_{i} \in \tSall^{(l)}_k}\tx_i\right)^s = \left[\max(1 - \tL^{(l)}\tx)\right]^s.$ Recall that $\tsj^{\infty}_{j}$ is set of all stopping sets of $H_j$. We have the following: $P^{(l)}_{(f,str)}(s) = \max_{\tSall \in \tsj^{\infty}_{l}}P^{(l)}_f(s;\tSall) = \max\left( \left[\max(1 - \tL^{(l)}\tx)\right]^s,\max_{\tSall \in \tsj^{\infty}_{l}, size(\tSall) \geq \mu_l}P^{(l)}_f(s;\tSall)\right)\\
     \leq \max\left( \left[\max(1 - \tL^{(l)}\tx)\right]^s, \left[1 - \beta^{(l)}\mu_l\right]^s\right)
     = \left(\max\left( \max(1 - \tL^{(l)}\tx), 1 - \beta^{(l)}\mu_l\right)\right)^s$. 
     
The second term in the maximum of $P^{(l)}_{(f,str)}(s)$ is because $\underset{\tSall \in \tsj^{\infty}_{l}, size(\tSall) \geq \mu_l}{\max}P^{(l)}_f(s;\tSall) \leq \left(1 - \beta^{(l)}\mu_l\right)^s$. \\[-1mm]
 
\subsubsection{Proof of Lemma \ref{lemma:p_f_med_str_any_layer}}\hfill

For $1 \leq j \leq l-1$, the $i^{th}$ column of $\tA^{(j)}$ (see Section \ref{sec:LP-sampling-anylayer}) corresponds to VN $v^{(l)}_i$ of the base layer 
and the non-zero positions in the $i^{th}$ column (two per column) correspond to the symbols of layer $j$ which are part of the Merkle proof of $v^{(l)}_i$. Thus, for a sampling strategy $\left(\tx\;,\;\beta^{(l)}\right)$ and $\tx^{(j)} = \tA^{(j)}\tx$, $1 \leq j \leq l$, it is easy to see that $\tx^{(j)}_k$ is the probability that $v^{(j)}_k$ is sampled.
Now, consider a stopping set $\psi$ that belongs to an intermediate layer $j$. Note that the Merkle proof 
for a base layer sample contains a single data and a single parity symbol from layer $j$ and is deterministic given the base layer sample. If both the symbols (VNs) exist in $\psi$, it is possible for a single base layer symbol to sample $\psi$ at two VNs. To avoid over-counting, we have defined the matrices $\piA^{(j)}$ in Section \ref{sec:LP-sampling-anylayer}. $\piA^{(j)}$ has the property that  $\piA^{(j)}_{ki}$ is 1 if the $i^{th}$ base layer symbol (i.e., $v^{(l)}_i$) samples, via its Merkle proof from layer $j$, the $k^{th}$ stopping set of $\tsj_j$ and zero otherwise. 
Thus, for a sampling strategy $\left(\tx\;,\;\beta^{(1)}\;,\ldots,\;\beta^{(l)}\right)$,
it is not difficult to see that $ P^{(j)}_{(f,med)}(s) = \left[\max(1 - \piA^{(j)}\tx)\right]^s$, $1 \leq j \leq l$. 

Now, let us consider the strong adversary. Since a Merkle proof contains one data and one parity symbol from every intermediate layer, all data (parity) symbols are sampled disjointly. As such, we can bound 
the probability of sampling a stopping set $\psi$ of size $\geq u_j$,  $1 \leq j <l$, by  $P^{(j)}_{f}(s=1;\psi) \leq 1 - \sum_{\tx^{(j)}_i : v^{(j)}_i \in \psi, v^{(j)}_i \text{is a data symbol}}\tx^{(j)}_i$ and $P^{(j)}_{f}(s=1;\psi) \leq 1 - \sum_{\tx^{(j)}_i : v^{(j)}_i \in \psi, v^{(j)}_i \text{is a parity symbol}}\tx^{(j)}_i$. Summing the two inequalities and dividing over $2$ yields

\noindent
$P^{(j)}_{f}(s=1;\psi) \leq 1 - \frac{1}{2}\sum_{\tx^{(j)}_i : v^{(j)}_i \in \psi}\tx^{(j)}_i \leq 1 - \frac{1}{2}\beta^{(j)}\mu_j$. Finally, use $P^{(j)}_{f}(s;\psi) = (P^{(j)}_{f}(s=1;\psi))^s$. \\[-1mm]

\subsubsection{Proof of Lemma \ref{lemma:soundess_weak}}\hfill

Soundness fails if the light nodes get back all the requested samples but no honest full node is able to fully decode the entire CMT. We consider two cases:\\
i) There is a DA attack at layer $j$: In this case, no honest full node will be able to decode layer $j$ of the CMT. Light nodes fail to detect this DA attack 
 using the overall greedy sampling strategy described in Remark \ref{remark:systematic_generator} with probability  $P^{(1)}_{f}(s) = \underset{\; \omega^{(j)} < \mu_j}{\max}\left[[1 - \tau(S^{(\tlambda s\;,\;j)}_{greedy},\omega^{(j)})]\left(1 - \frac{\omega^{(j)}}{n_j}\right)^{s - \tlambda s}\right]$. The term inside the maximum is the probability of failure using the overall greedy sampling strategy
 when the weak adversary hides a stopping set of size $\omega^{(j)}$.

ii) There is no DA attack: In this case, light nodes will accept the block. Soundness failure occurs when honest full nodes are not able to decode the entire CMT from the samples broadcasted by the light nodes. Let $P^{(2)}_f(s)$ be the probability of this event. To bound $P^{(2)}_f(s)$, we use the following property of the CMT which was proved in \cite{AceD}: the Merkle proof of $\eta$ fraction of distinct base layer coded symbols have at least $\eta$ fraction of distinct coded symbols from each layer of the CMT. Thus for $\eta_{rec} = \left(\max_{1 \leq j \leq l} \frac{n_j-\omega^{(j)}_{\min} + 1}{n_j}\right)$, if a full node has $\eta_{rec}$ fraction of distinct coded symbols from the base layer of the CMT, then it has at least $\eta_{rec}$ fraction or at least $\eta_{rec}n_j$ distinct coded symbols from layer $j$ of the CMT. Since $\eta_{rec}n_j \geq n_j - \omega^{(j)}_{\min}+1$, using these  $\eta_{rec}n_j$ distinct coded symbols, the full node will be able to successfully decode layer $j$, $\forall 1 \leq j \leq l$. Let $Z$ be the total number of distinct base layer coded symbols collected by a honest full node from the random portion of the light node's overall greedy sampling strategy.  Then, we have $ P^{(2)}_f(s) \leq P(Z \leq \eta_{rec}n_l)  \leq  {n_l \choose \eta_{rec}n_l} \frac{( \eta_{rec}n_l)^{Ms(1-\tlambda )}}{n^{Ms(1-\tlambda)}_l} \leq
     2^{[\tH(\eta_{rec},1-\eta_{rec})n_l - \tm s(1-\tlambda) \log(\frac{1}{\eta_{rec} })]}.$
The probability of soundness failure is smaller than the maximum of the above two cases. Moreover, in our system, for the same reasons as \cite{CMT}, soundness implies agreement (since each light node is connected to at least one honest full node and honest full nodes form a fully connected graph; see network model in Section \ref{sec:network_nodes}).  Thus, $\tpfsa \leq \max(P^{(1)}_f(s), P^{(2)}_f(s))$ completing the proof. \\[-1mm]

\subsubsection{Proof of Lemma \ref{lemma:soundness_med_strong}}\hfill

Again we consider the two cases described in the proof of Lemma \ref{lemma:soundess_weak}. For the first case, light nodes fail to detect the DA attack at layer $j$ using LP-sampling with probability $P^{(1)}_f(s) = \underset{1\leq j\leq l}{\max} P^{(j)}_{f,med}(s)$ and $P^{(1)}_f(s) = \underset{1\leq j\leq l}{\max} P^{(j)}_{f,str}(s)$ for the medium and the strong adversary, respectively. For the second case, let $Z$ be the total number of distinct base layer coded symbols collected by a honest full node when light nodes use LP-sampling. We have $    P^{(2)}_f(s) \leq P(Z \leq \eta_{rec}n_l) \leq  {n_l \choose \eta_{rec}n_l} \left(\sum_{i=1}^{\eta_{rec}n_l} \tx_{[i]}\right)^{Ms} \leq 2^{[\tH(\eta_{rec},1-\eta_{rec})n_l - \tm s \log\left(\frac{1}{\sum_{i=1}^{\eta_{rec}n_l}\tx_{[i]} }\right)]}$.
Similar to the proof of  Lemma \ref{lemma:soundess_weak}, soundness implies agreement and we have $\tpfsa \leq \max(P^{(1)}_f(s), P^{(2)}_f(s))$.

\begin{IEEEbiography}[{\includegraphics[width=1in,height=1.25in,clip,keepaspectratio]{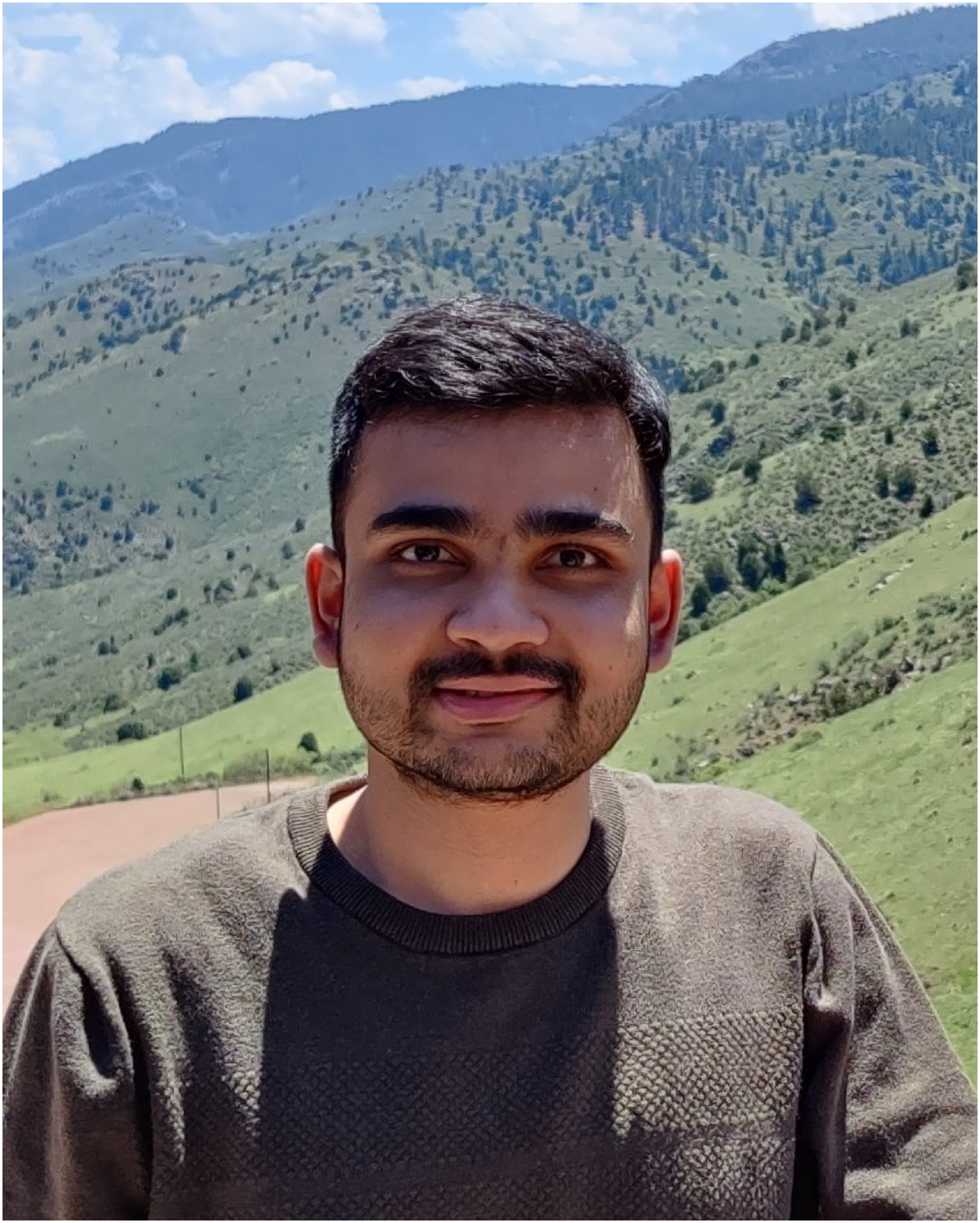}}]{Debarnab Mitra}
 is a Ph.D. candidate in the Electrical and Computer Engineering Department at the University of California, Los Angeles (UCLA). He received his B.~Tech (with honors) in Electrical Engineering from the Indian Institute of Technology Bombay in 2018 and his M.S. in Electrical and Computer Engineering from the University of California, Los Angeles in 2020.  Currently, he works at the Laboratory for Robust Information Systems (LORIS), and his focus is on coding schemes for blockchain systems. His research interests include coding and information theory, signal processing, graph theory, and blockchain systems. Debarnab is a receipt of the Best Poster Award from the IEEE North American School of Information Theory (NASIT), 2021. In 2020, he received the Distinguished Masters Thesis Award in Signals and Systems from the Electrical and Computer Engineering Department at UCLA.
 \end{IEEEbiography}

\begin{IEEEbiography}[{\includegraphics[width=1in,height=1.25in,clip,keepaspectratio]{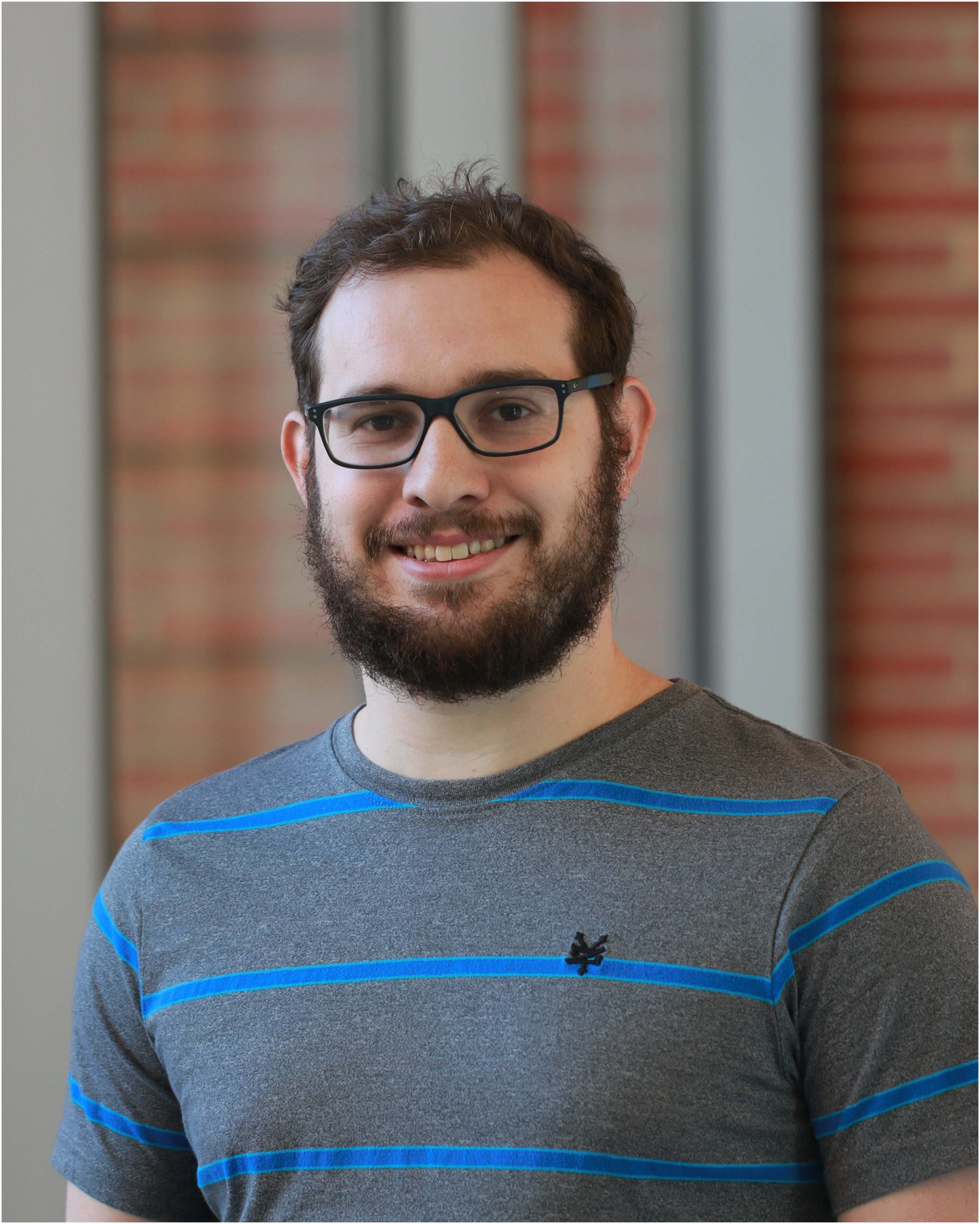}}]{Lev Tauz}
 is a Ph.D. candidate in the Electrical and Computer Engineering Department at the University of California, Los Angeles (UCLA). He received his B.S. (with honors)  in Electrical Engineering and Computer Science from the University of California, Berkeley in 2016 and his M.S. in Electrical and Computer Engineering from the University of California, Los Angeles in 2020. Currently, he works at the Laboratory for Robust Information Systems (LORIS), and is focused on coding techniques for distributed storage and computation. His research interests include distributed systems, error-correcting codes, machine learning, and graph theory. Lev is the recipient of the Memorable Paper Award from the 2021 Non-Volatile Memories Workshop (NVMW). 
 \end{IEEEbiography}

\begin{IEEEbiography}[{\includegraphics[width=1in,height=1.25in,clip,keepaspectratio]{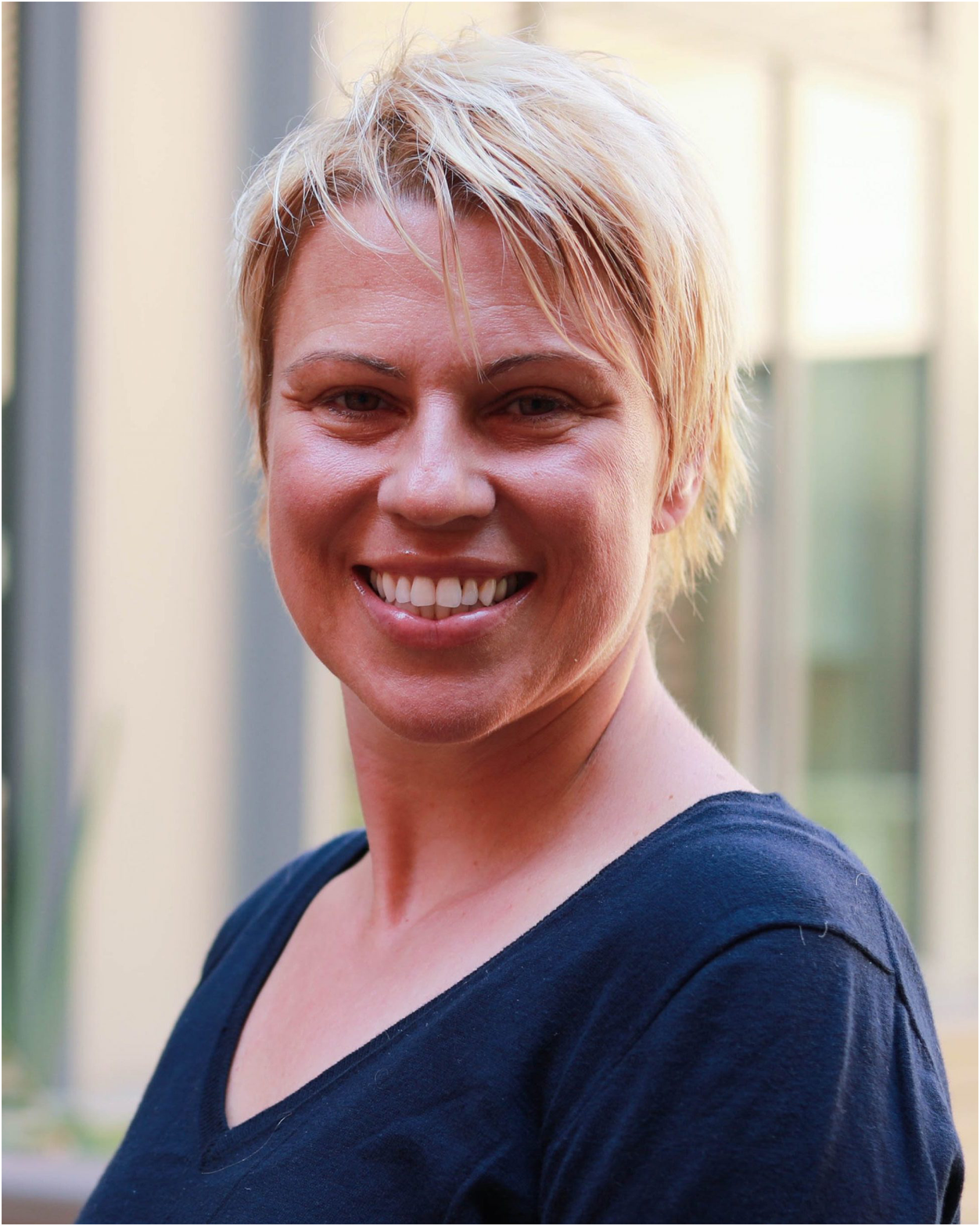}}]{Lara Dolecek} 
is a Full Professor with the Electrical and Computer Engineering Department and Mathematics Department (courtesy) at the University of California, Los Angeles (UCLA). She holds a B.S. (with honors), M.S. and Ph.D. degrees in Electrical Engineering and Computer Sciences, as well as an M.A. degree in Statistics, all from the University of California, Berkeley. She received the 2007 David J.~Sakrison Memorial Prize for the most outstanding doctoral research in the Department of Electrical Engineering and Computer Sciences at UC Berkeley. Prior to joining UCLA, she was a postdoctoral researcher with the Laboratory for Information and Decision Systems at the Massachusetts Institute of Technology. She received IBM Faculty Award (2014), Northrop Grumman Excellence in Teaching Award (2013), Intel Early Career Faculty Award (2013), University of California Faculty Development Award (2013), Okawa Research Grant (2013), NSF CAREER Award (2012), and Hellman Fellowship Award (2011). With her research group and collaborators, she received numerous best paper awards. Her research interests span coding and information theory, graphical models, statistical methods, and algorithms, with applications to emerging systems for data storage and computing. She currently serves as an Associate Editor for IEEE Transactions on Information Theory and as the Secretary of the IEEE Information Theory Society. Prof. Dolecek is 2021-2022 Distinguished Lecturer of the IEEE Information Theory Society. Prof. Dolecek has served as a consultant for a number of companies specializing in data communications and storage.

 \end{IEEEbiography}


\vspace{-0.2cm} 
\begin{thebibliography}{1}
\vspace{-0.1cm}
























 






 
 
 



































 






 
 
 









\bibitem{dataAvailOrg}
M.~Al-Bassam,  \emph{et al.}, ``Fraud and data availability proofs: Detecting invalid blocks in light clients," \emph{Int. Conf. on Financial Cryptography and Data Secur.}, Springer, Mar. 2021.

\bibitem{CMT} M.~Yu, \emph{et al.}, ``Coded merkle tree: Solving data availability attacks in blockchains," \emph{Int. Conf. on Financial Cryptography and Data Secur.,} Springer, Feb. 2020.

\bibitem{SSskewITW} D.~Mitra, \emph{et al.}, ``Concentrated stopping set design for coded merkle tree: Improving security against data availability attacks in blockchain systems," \emph{2020 IEEE Inf. Theory Workshop (ITW)}, Apr. 2021, full version: https://arxiv.org/abs/2010.07363.

\bibitem{Bitcoin} S.~Nakamato, ``Bitcoin: A peer to peer electronic cash system," 2008. [Online] Available: https://bitcoin.org/bitcoin.pdf.


\bibitem{supplychain} M.~J.~Casey and P.~Wong, ``Global supply chains are about to get better, thanks to
blockchain,” \emph{Harvard Business Review}, Mar.~2017. [Online] Available: https://hbr.org/2017/03/global-supply-chains-are-about-to-get-better-thanks-to-blockchain.

\bibitem{IoTsurvey}
X.~Wang, \emph{et al.}, ``Survey on blockchain for internet of things," \emph{Comput. Commun.}, vol.~136, pp.~10-29, 2019.

\bibitem{healthcare} M.~Mettler, ``Blockchain technology in healthcare: The revolution starts here,” \emph{IEEE 18th Int. Conf. on e-Health Network., Apps., and Services (Healthcom)}, Sept. 2016.

\bibitem{LowLatency}
T.~Rocket, \emph{et al.}, ``Scalable and probabilistic leaderless BFT consensus through metastability," \emph{arXiv:1906.08936}, Jun. 2019.

\bibitem{bitcoinsize} Online: https://www.blockchain.com/charts/blocks-size,  accessed:  May 5, 2022.

\bibitem{ethereumsize} Online: https://etherscan.io/chartsync/chaindefault,  accessed: May. 5, 2022.



\bibitem{DAnote} Online: https://github.com/ethereum/research/wiki/A-note-on-data-availability-and-erasure-coding

\bibitem{ModernCodingTheory} T.~Richardson, and R.~Urbanke, ``Modern coding theory," \emph{Cambridge: Cambridge University Press}, 2008.

\bibitem{SSelim} X.~Jiao, \emph{et al.}, ``Eliminating small stopping sets in irregular low-density parity-check codes," \emph{IEEE Commun. Lett.}, vol.~13, no.~6, pp.~435-437, Jun.~2009. 



\bibitem{SSNP-hard} K.~M.~Krishnan, and P.~Shankar, ``Computing the stopping distance of a Tanner graph is NP-hard," \emph{IEEE Trans. on Inf. Theory}, vol.~53, no.~6, pp.~2278-2280, Jun.~2007.

\bibitem{PEG} X.Y.~Hu, \emph{et al.}, ``Regular and irregular progressive edge-growth tanner graphs," \emph{IEEE Trans. on Inf. Theory}, vol.~51, no.~1, pp.~386-398, Jan.~2005.

\bibitem{SSemsemble} A.~Orlitsky, \emph{et al.}, ``Stopping set distribution of LDPC code ensembles," \emph{IEEE Trans. on Inf. Theory}, vol.~51, no.~3, pp.~929-953, Mar. 2005.

\bibitem{Cover} S.~Cao, \emph{et al.}, ``CoVer: Collaborative light-node-only verification and data availability for blockchains," \emph{IEEE Int. Conf. on Blockchain}, Nov. 2020.



\bibitem{Trifecta} Trifecta Team, ``Trifecta: The blockchain trilemma solved," http://pramodv.ece.illinois.edu/pubs/Whitepaper2019-9.pdf

\bibitem{AceD} P.~Sheng, \emph{et al.}, ``ACeD: Scalable data availability oracle,"  \emph{Financial Cryptography}, Springer, Mar.~2021.


\bibitem{DE-PEG} D.~Mitra, \emph{et al.}, ``Communication-efficient LDPC code design for data availability oracle in side blockchains," \emph{IEEE Inf. Theory Workshop (ITW)}, Oct. 2021.


\bibitem{WesselSS} T.~Tian, \textit{et al.}, ``Construction of irregular LDPC codes with low error floors," \emph{IEEE Int. Conf. on Commun.}, May 2003.
 


\bibitem{EMDcalculation} S.~Kim, \emph{et al.}, ``LDPC code construction with low error floor based on the IPEG algorithm," \emph{IEEE Commun. Lett.}, vol.~11, no.~7, pp.~607-609, Jul. 2007.

\bibitem{networkcodingstorage}
M.~Dai, \emph{et al.}, ``A low storage room requirement framework for distributed ledger in blockchain," \emph{IEEE Access}, vol.~6, pp.~22970-22975, Mar. 2018.

\bibitem{downsampling}
Q.~Huang, \emph{et al.}, ``Downsampling and transparent coding for blockchain" \emph{IEEE Trans. on Network Sci. and Eng.}, vol.~9, no.~4, pp.~2139-2149, Jul.-Aug.~2022.



\bibitem{SeF}
 S.~Kadhe, \emph{et al.}, ``SeF: A secure fountain architecture for slashing storage costs in blockchains," \emph{arXiv:1906.12140}, 2019.
 
 \bibitem{polyshard}
 S.~Li, \emph{et al.}, ``PolyShard: coded sharding achieves linearly scaling efficiency and security simultaneously," \emph{IEEE Trans. on Inf. Forensics and Secur.}, vol. 16, Jul. 2020.
 

 
\bibitem{erasurelowstorage}
D.~Perard, \emph{et al.}, ``Erasure code-based low storage blockchain node," \emph{IEEE Int. Conf. on Internet of Things (iThings) and IEEE Green Comput. and Commun. (GreenCom) and IEEE Cyber, Physical and Social Comput. (CPSCom) and IEEE Smart Data (SmartData)}, Jul. 2018. 

\bibitem{SnowWhite}
P.~Daian, \emph{et al.}, ``Snow white: Robustly reconfigurable consensus and applications to provably secure proof of stake," \emph{Financial Cryptography}, Sept. 2019.

\bibitem{PoSpace}
S.~Park, \emph{et al.}, ``Spacemint: A cryptocurrency based on proofs of space," \emph{ Financial Cryptography,} Springer, Feb.~2018.


\bibitem{ILPsearch}
A.~Sarıduman, \emph{et al.}, ``An integer programming-based search technique for error-prone structures of LDPC codes,"  \emph{AEU-Int. Journal of Electronics and Commun.}, vol.~8, no.~11, pp.~1097-1105, Nov. 2014.

\bibitem{LPcomplexity}
Y.~T.~Lee, and A.~Sidford, ``Efficient inverse maintenance and faster algorithms for linear programming," \emph{IEEE Annual Symp. on Foundations of Computer Science}, Oct.~2015.


\bibitem{zerocash}
E.~B.~Sasson,  \emph{et al.}, ``Zerocash: decentralized anonymous payments from bitcoin," \emph{IEEE Symp. on Secur. and Privacy}, May 2014.

\bibitem{InformationDispersal}
K.~Nazirkhanova, \emph{et al.}, ``Information dispersal with provable retrievability for rollups," \emph{arXiv:2111.12323}, Nov. 2021.

\bibitem{zk-STARKS}
E.~B.~Sasson, \emph{et al.}, ``Scalable, transparent, and post-quantum secure computational integrity," \emph{IACR Cryptol. ePrint Arch}, 2018.

\bibitem{RSoptimize}
P.~Santini, \emph{et al.}, ``Optimization of a Reed-Solomon code-based protocol against blockchain data availability attacks", \emph{IEEE Int. Conf. on Commun.}, May 2022.




\end{thebibliography}
\end{document}